\pdfoutput=1

\documentclass[12pt]{article}
\usepackage{amsfonts}
\usepackage{amsmath}
\usepackage{amssymb}
\usepackage{bigints}
\usepackage{booktabs}
\usepackage[nosort]{cite}
\usepackage{color}
\usepackage{dsfont}
\usepackage{float}
\usepackage{framed}
\usepackage{graphicx}
\usepackage{indentfirst}
\usepackage{mathrsfs}
\usepackage{multirow}
\usepackage{pdflscape}
\usepackage{setspace}
\usepackage{subdepth}
\usepackage{subfig}
\usepackage{titlesec}
\usepackage[dotinlabels]{titletoc}
\usepackage{wrapfig}
\usepackage[all]{xy}
\usepackage{young}
\usepackage[vcentermath]{youngtab}
\usepackage{relsize}
\usepackage{stackengine}
\usepackage{datetime}

\usepackage{hyperref}

\numberwithin{equation}{section}

\usepackage{verbatim}

\newcommand{\be}{\begin{equation}}
\newcommand{\ee}{\end{equation}}

\newcommand{\lbb}{\Lambda_{B \partial}}

\newmuskip\pFqmuskip
\newcommand*\pFq[6][8]{%
  \begingroup 
  \pFqmuskip=#1mu\relax
  \mathcode`\,=\string"8000
  \begingroup\lccode`\~=`\,
  \lowercase{\endgroup\let~}\pFqcomma
  {}_{#2}F_{#3}{\left[\genfrac..{0pt}{}{#4}{#5};#6\right]}%
  \endgroup
}
\newcommand{\pFqcomma}{\mskip\pFqmuskip}

\newmuskip\Ftmuskip
\newcommand*\Ft[6][8]{%
  \begingroup 
  \pFqmuskip=#1mu\relax
  \mathcode`\,=\string"8000
  \begingroup\lccode`\~=`\,
  \lowercase{\endgroup\let~}\pFqcomma
  F_2{\left[\genfrac..{0pt}{}{#4}{#5};#6\right]}%
  \endgroup
}

\def\({\left(} \def\){\right)}
\def\[{\left[} \def\]{\right]}

\def\sgn{\text{sgn}}

\def\ts{\tilde{s}}

\def\mF{\mathcal{F}}
\def\mS{\mathcal{S}}

\def\mE{\mathcal{E}}

\def\mO{\mathcal{O}}
\def\mC{\mathcal{C}}

\def\mI{\mathcal{I}}
\def\mA{\mathcal{A}}

\def\eps{\epsilon}

\def\mW{\mathcal{W}}

\def\mB{\mathcal{B}}

\newcommand{\bea}{\begin{eqnarray}}
\newcommand{\eea}{\end{eqnarray}}

\newcommand{\bml}{\begin{multline}}
\newcommand{\emll}{\end{multline}}

\usepackage[left=2.2cm,right=2.2cm,top=2.5cm,bottom=2.5cm]{geometry}
\titleformat{\section}{\normalfont\bfseries}{\thesection.}{4pt}{}
\titlespacing{\section}{0pt}{22pt}{6pt}

\titleformat{\subsection}{\normalfont\itshape}{\thesubsection.}{4pt}{}
\titlespacing{\subsection}{0pt}{18pt}{6pt}

\titleformat{\subsubsection}{\normalfont\itshape}{\thesubsubsection.}{4pt}{}
\titlespacing{\subsubsection}{0pt}{16pt}{6pt}

\def\ie{\begin{equation}\begin{aligned}}
\def\fe{\end{aligned}\end{equation}}

\def\tilde{\widetilde}

\def\bar{\overline}

\def\1{{\mathds 1}}

\DeclareFontShape{OT1}{cmr}{mx}{n}%
    {<->cmr10}{}
\newcommand{\mytitlefont}{\fontseries{mx}\selectfont}
\DeclareMathAlphabet{\titlemath}{OT1}{cmr}{mx}{n}

\begin{document}


\vfill


\begin{titlepage}

\begin{center}

~\\[2cm]

{\fontsize{20pt}{0pt} \mytitlefont All point correlation functions in SYK}

~\\[0.5cm]

{\fontsize{14pt}{0pt} David J.~Gross and Vladimir Rosenhaus}

~\\[0.1cm]

\it{Kavli Institute for Theoretical Physics}\\ \it{University of California, Santa Barbara, CA 93106}

~\\[0.8cm]

\end{center}

\noindent 

Large $N$ melonic theories are characterized by two-point function Feynman diagrams built exclusively out of melons. This leads to conformal invariance at strong coupling, four-point function  diagrams that are exclusively ladders, and higher-point functions that are built out of four-point functions joined together. We uncover an incredibly useful property of these theories: the six-point function, or equivalently, the three-point function  of the primary $O(N)$ invariant bilinears, regarded as an analytic function of the operator dimensions, fully determines all correlation functions, to leading nontrivial order in $1/N$, through simple Feynman-like rules. The result is applicable to any theory, not necessarily melonic, in which higher-point correlators are built out of four-point functions. We explicitly calculate the bilinear three-point function for $q$-body SYK, at any $q$. This leads to the bilinear four-point function, as well as all higher-point functions, expressed in terms of higher-point conformal blocks, which we discuss. We find universality of correlators of operators of large dimension, which we simplify through a saddle point analysis. We comment on the implications for the AdS dual of SYK. 

\vfill

\end{titlepage}

\tableofcontents

~\\[.1cm]

\section{Introduction}
Strongly coupled quantum field theories are often prohibitively difficult to study,  yet, in the rare cases that one succeeds, they  reveal a  wealth of  phenomena. This has been evidenced over the past decade with the remarkable integrability results in maximally supersymmetric $\mathcal{N} = 4$ Yang-Mills \cite{Beisert:2010jr}. 
The integrability of $\mathcal{N} = 4$ implies that the theory is, in principle, solvable at large $N$. However, in practice the solution is neither simple nor direct. Like any matrix model, the large $N$ dominant Feynman diagrams are planar, and there are no known general techniques to sum planar diagrams. It would  be incredibly useful to have simpler large $N$ models, with  diagrammatic structures that allow for full summation. 

Melonic models are of this type. These have arisen in a number of independent contexts, including:  models of Bose fluids\cite{Pat}, models of spin glasses \cite{SY}, and tensor models  \cite{Gurau:2009tw}. The specific theory we will focus on is the SYK model \cite{Kitaev}: a $0+1$ dimensional model of Majorana fermions with $q$-body interactions. Through a simple extension, our results are  applicable to any melonic theory. In fact, as we will discuss later, our results extend to an even broader class of theories, provided they have the diagrammatic structure that higher-point correlators are built out of four-point functions. In this paper we  solve SYK: we give expressions for the connected piece of the fermion $2 p$-point correlation function, for any $p$, to leading nontrivial order in $1/N$. 

What are the features of melonic theories that make them solvable?
At the level of the two-point function, it is the fact that, at leading order in $1/N$, all Feynman diagrams are iterations of melons nested within melons. This self-similarity leads to an integral equation determining the two-point function, which in turn has a conformal solution at strong coupling. At the level of the four-point function, all leading large $N$ Feynman diagrams are ladders, with an arbitrary number of rungs: summing all ladder diagrams is no more difficult than summing a geometric series, provided one uses the appropriate basis. 

The focus of this paper is the six-point function, and higher. As input, we need to know the conformal two-point and four-point functions, but it is irrelevant to us how they were obtained or which diagrams contributed to them. The essential property we do need is that higher-point correlation functions have the diagrammatic structure of four-point functions that are glued together, as shown in Fig.~\ref{FigIntro1} for the six-point function and Fig.~\ref{FigIntro3} for the eight-point function. Computing the six-point function involves gluing together three four-point functions. As the four-point function is a sum of conformal blocks, this amounts to evaluating a conformal integral, though a nontrivial one. In all higher-point functions, the six-point function acts like an interaction vertex. As a result, the structure of the six-point function fully determines all higher-point correlation functions. 

To be slightly more specific, since 
SYK has an $O(N)$ symmetry after disorder averaging,  it is convenient to work with 
the primary, $O(N)$ invariant, fermion bilinear operators, $\mathcal{O}_i$. These are the analogs of the single-trace operators in gauge theories. The fermion six-point function determines the three-point function of the bilinears, and hence the OPE coefficients $c_{123}$ appearing in  $\mO_1 \mO_2 \sim c_{123}\,  \mO_3$. The essential point is to regard the $c_{1 2 3}$ as analytic functions of the dimensions of the $\mO_i$. All higher-point correlation functions will be expressed in terms of contour integrals involving  $c_{i j k}$ . We stress that $c_{123}$ are the OPE coefficients of the single-trace operators. Somehow, their analytic structure, combined with the fermion four-point function,  is encoding the OPE coefficients of the double-trace operators, as well as all others.  
We finish this introduction with a heuristic sketch of the main result, followed by an outline of the paper.

\subsection{Outline of computation}
\begin{figure}[t]
\centering
\includegraphics[width=6.5in]{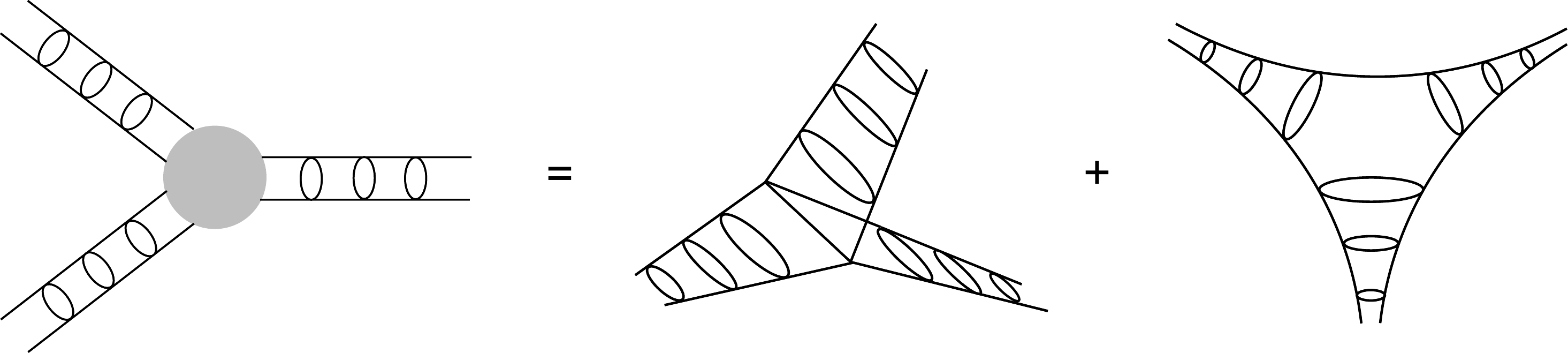}
\caption{The connected fermion six-point function, to leading nontrivial order in $1/N$, is given by a sum of Feynman diagrams, of the kind shown on the right. This consists of three fermion four-point functions, ladders, that are glued together. There are two classes of diagrams, as shown on the right; only the second is planar. This figure, as well as all others, is for $q=4$ SYK, and the lines denote the full propagators: they should be dressed with melons.} \label{FigIntro1}
\end{figure}

We will focus on the three-point and four-point functions of the primary $O(N)$ invariant bilinear operators, schematically of the form, $\mO = \sum_i \chi_i \partial_{\tau}^{2n+1} \chi_i$. These arise from a limit of the fermion six-point function and eight-point functions, respectively. 
 
The fermion six-point function consists of a sum of two classes of diagrams,``contact'' and planar, as shown in Fig.~\ref{FigIntro1}. Summing these gives the conformal three-point function $\langle \mO_1(\tau_1) \mO_2(\tau_2) \mO_3(\tau_3)\rangle$ of the $\mO_i$ of dimension $h_i$. Up to a constant, $c_{123}$, the form of the  three-point function is fixed by conformal symmetry. This constant is of course the same one that appears in the OPE, $\mO_1 \mO_2 \sim c_{123}\, \mO_3$.  In \cite{GR2} we computed the contact diagram  exactly, whereas the planar diagram was evaluated in the large $q$ limit, in which the computation simplifies. In Sec.~\ref{sec:3pt} we revisit the three-point function, and compute the planar diagrams exactly at finite $q$. The form of $c_{123}$ involves  generalized hypergeometric functions, of type ${}_4 F_3$, at argument one.

\begin{figure}[t]
\centering
\includegraphics[width=3.5in]{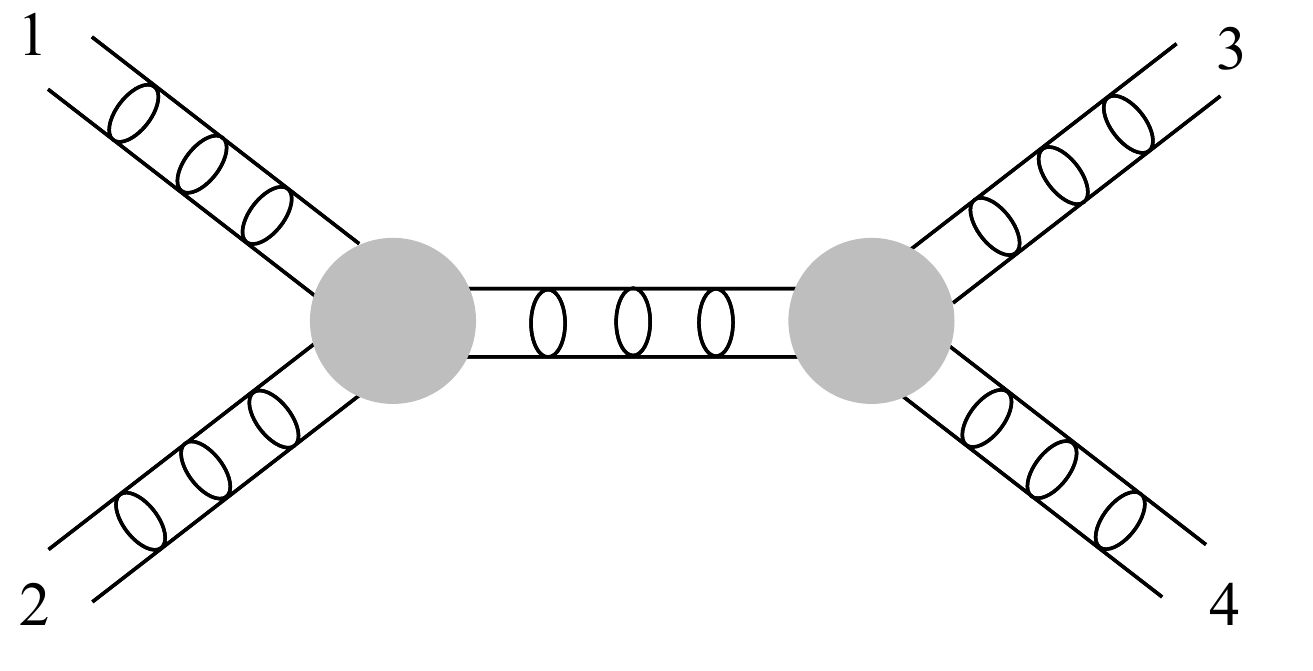}
\caption{ The fermion eight-point function is composed of Feynman diagrams such as the one shown. It is built out of two six-point functions; the shaded circle is defined by Fig.~\ref{FigIntro1}. } \label{FigIntro3}
\end{figure}

In Sec.~\ref{sec:4pt} we turn to the fermion eight-point function. While the six-point function involves gluing together three fermion four-point function, the eight-point function involves gluing together five four-point functions, as shown in Fig.~\ref{FigIntro3}. While this at first appears significantly more involved, it is actually quite simple, and builds off of the analytic structure of the three-point function structure constants, $c_{123}$. The essential step is to use the representation of a CFT four-point function in terms of a contour integral over a complete basis of $SL_2$ conformal blocks. Specifically, for any CFT$_1$, let $\mF_{1234}^H(x)$ denote a conformal block, with the subscript labeling the four external operators $\mA_i$, the superscript  labeling the exchanged operator, and $x$ denoting the conformal cross-ratio of the four times. It is a familiar fact that the four-point function can be expanded as a sum of conformal blocks, 
\be \label{B1}
\langle \mA_1\cdots \mA_4\rangle = \sum_H c_{12 H} c_{3 4 H} \mF_{1234}^H(x)~,
\ee
where the sum is over all exchanged primaries. One may just as well write the four-point function as a contour integral,~\footnote{ We are being slightly imprecise here, in that what  should really enter this expression is the conformal block plus its shadow; we will be more explicit in the main body of the paper. }
\be \label{B2}
\langle \mA_1 \cdots \mA_4\rangle = \int_{\mC} \frac{d h}{2\pi i} f(h) \mF_{1234}^h(x)~,
\ee
with some appropriately chosen $f(h)$, where the contour consists of a line running parallel to the imaginary axis, $h = \frac{1}{2} + i s$, as well as circles around the positive even integers, $h = 2n$. The distinction between these two expansions is that  the former sums over conformal blocks corresponding to physical operators in the theory, whereas  the latter sums over the blocks that form a complete basis. If one closes the contour in the latter, one recovers the sum in the former.

Let us write the SYK fermion four-point function in the form of such a contour integral, 
\be \label{introSYK4pt}
\sum_{i j} \langle \chi_i(\tau_1) \chi_i(\tau_2) \chi_j(\tau_3) \chi_j(\tau_4) \rangle= \int_{\mC} \frac{ d h}{2\pi i} \tilde{\rho}(h) \mF_{\Delta}^h(x)~. 
\ee
Closing the contour yields the standard conformal block expansion, with OPE coefficients $\sum_i \chi_i \chi_i \sim \sum_{h_n} c_{h_n} \mO_{h_n}$, given by, 
\be
c_{h_n}^2 = - \text{Res } \tilde{\rho}(h)\Big|_{h = h_n}~.
\ee
The main step in evaluating the contribution to the SYK four-point function $\langle \mO_1 \cdots \mO_4 \rangle$ shown in Fig.~\ref{FigIntro3}, is to use the above contour integral representation  for the intermediate fermion four-point function. After some manipulation, we will find these diagrams are, 
\be \label{exchangeW}
\langle \mO_1(\tau_1) \cdots \mO_4(\tau_4)\rangle_s = \int_{\mC}\frac{d h}{2\pi i}\frac{\tilde{\rho} (h)}{c_h^2}  \,c_{1 2 h} c_{ 3 4 h}\, \mF_{1234}^h(x)~.
\ee

This result is simple and intuitive, following Feynman-like rules: there are cubic interactions $c_{ 1 2 h}$ and $c_{3 4 h}$,  the conformal block of $\mO_h$, $\mF_{1234}^h$, acts as the CFT analog of a propagator, and $h$-space acts as the CFT analog of Fourier space.

If one closes the contour in (\ref{exchangeW}), one is left with the standard representation of a CFT four-point function as a sum of conformal blocks. The analytic structure of the integrand is such that the only blocks that appear are those corresponding to single-trace and double-trace operators, as should be the case. In fact, the argument leading to (\ref{exchangeW}) is general, and is valid for any cubic level interactions of four-point functions, not necessarily the ones specific to SYK that were depicted in Fig.~\ref{FigIntro1}. 

The expression (\ref{exchangeW}) is just for the $s$-channel diagrams. We must also include the $t$-channel and $u$-channel diagrams, which follow from the $s$-channel ones by a simple permutation of operators. In adding these three contributions, we will over-count the diagram which has no exchanged melons, shown later in Fig.~\ref{FigIntro4}, which  must then be explicitly subtracted off. 

\subsection*{Outline}
The paper is organized as follows: Sec.~\ref{ladders} reviews the SYK model and the fermion four-point function. The bilinear three-point function is computed in Sec.~\ref{sec:3pt} and the bilinear four-point function is computed in Sec.~\ref{sec:4pt}.  
Higher-point functions are studied in Sec.~\ref{sec:ppt}. 
The correlation functions of the bilinears, in the limit that all of them have large dimension, reduce to the correlators of generalized free field theory of fermions in the singlet sector. This provides a good way of studying their asymptotic behavior, via saddle point, and is discussed in Sec.~\ref{FREE}. In Sec.~\ref{sec:BULK}, we make some comments on what the correlators teach us about the bulk dual of SYK, and  discuss the relation between exchange Feynman diagrams in SYK and exchange Witten diagrams. We end in Sec.~\ref{sec:dis} with a brief discussion.

In Appendix.~\ref{blocks} we review conformal blocks, the shadow formalism,  and Mellin space. Appendix.~\ref{largeQ} discusses the SYK correlation functions in the large $q$ limit, and  Appendix.~\ref{FFT} discusses the generalized free field limit. In Appendix.~\ref{FermionP} we discuss the relation between the fermion correlation functions and the bilinear correlation functions. In Appendix.~\ref{sec:con} we study additional contact Feynman diagrams that must be included in the computation of correlation functions if $q$ is sufficiently large. In Appendix.~\ref{app:Witten} we express  exchange and contact Witten diagrams as sums of conformal blocks. In Appendix.~\ref{KK} we show that the spectrum of large $q$ SYK can be  reproduced by placing an AdS$_2$ brane inside of AdS$_3$, however this does not reproduce the necessary  cubic couplings.

\section{SYK Ladders} \label{ladders}
\subsection{SYK basics}
The SYK model  describes $N\gg1$ Majorana fermions satisfying $\{\chi_i, \chi_j\} = \delta_{i j}$, with  action,  $S_{top} + S_{SYK}^{\rm int}$, where,
\be \label{Stop}
S_{top} =   \frac{1}{2} \sum_{i = 1}^N \int d\tau\, \chi_i \, \frac{d}{d \tau}\, \chi_i~,
\ee
is the action for free Majorana fermions, and the interaction is, 
\be  \label{SSYK}
S_{SYK}^{\rm int} = \frac{(i)^{\frac{q}{2}}}{q!}\sum_{i_1, \ldots, i_q=1}^N\,\int d\tau \, J_{i_1\, i_2\, \ldots i_q}\,  \chi_{i_1}\chi_{i_2}\, \cdots \chi_{i_q}~,
\ee
where the coupling $J_{i_1, \ldots, i_q}$ is totally antisymmetric and, for each $i_1, \ldots, i_q$, is chosen from a Gaussian ensemble, with variance,
\be \label{disA}
\frac{1}{(q-1)!} \sum_{i_2, \ldots, i_q=1}^{N}\langle J_{i_1 i_2 \ldots i_q}  J_{i_1 i_2 \ldots i_q}\rangle= J^2~.
\ee
One can consider SYK for any even $q\geq 2$, with  $q=4$ being the prototypical case.

 In the UV, at zero coupling, the total action is  (\ref{Stop}), and the fermions have a two-point function given by $\frac{1}{2} \sgn(\tau)$. In the infrared, for $J |\tau| \gg 1$, the fermion two-point function is, at leading order in $1/N$, 
\be
G(\tau) = b\frac{\sgn(\tau)}{|J\tau|^{2\Delta}}~,
\ee
where $b$ is given by,
\be \label{psiDelta}
\psi (\Delta) \equiv 2 i \sqrt{\pi}\, 2^{-2\Delta} \frac{\Gamma(1- \Delta)}{\Gamma(\frac{1}{2} + \Delta)} ~, \ \ \ \ \ b^{q} = - \frac{1}{\psi(\Delta) \psi(1-\Delta)} = \frac{1}{2\pi}(1 - 2\Delta)\tan\pi \Delta~,
\ee
and the IR dimension of the fermions is $\Delta = 1/q$. 

While SYK appears conformally invariant at the level of the two-point function, the conformal invariance is broken at the level of the four-point function \cite{Kitaev, PR, MS}, resulting in SYK being ``nearly'' conformally invariant in the infrared.  There is a variant of SYK, cSYK \cite{GR3}, which is conformally invariant at strong coupling, and in fact, for any value of the coupling. The action for cSYK is  $ S_0 + S_{SYK}^{\rm int}$, 
where $S_{SYK}^{\rm int}$ is given by (\ref{SSYK}), while  $S_0$ is the bilocal action, 
\be  \label{S0}
S_0 =b^q\sum_{i = 1}^N \int d\tau_1 d\tau_2\, \chi_i(\tau_1)\, \frac{\! \sgn(\tau_1 - \tau_2)}{\, |\tau_1 - \tau_2|^{2(1-\Delta)}}\,  \chi_i(\tau_2)~.
\ee
The distinction between SYK and cSYK is in the kinetic term, $S_{top}$ versus $S_0$. 
As a result, for SYK the coupling $J$ is dimension-one, while for cSYK it is dimensionless. 

At strong coupling, the correlation functions of all bilinear, primary, $O(N)$ singlet operators $\mO_n$, schematically of the form $\mO_n =\sum_{i=1}^N \chi_i \partial_{\tau}^{1+2n}\chi_i$, are the same for SYK and for cSYK, for $n\geq 1$. The distinction between SYK and cSYK appears in the correlators involving $\mO_0$ (the ``$h=2$'' operator); it is these that break conformal invariance in SYK. Our results for the correlation functions of the $\mO_n$ that will be presented in the body of the paper are for cSYK at strong coupling, or, equivalently, for all the $\mO_n$ in SYK at strong coupling, with the exception of those correlators involving $\mO_0$.~\footnote{In particular, the fermion four-point function in SYK, has a block coming from $\mO_0$ that breaks conformal invariance and, at finite temperature, scales as $\beta J$. We will not be including this contribution. It  would give rise to terms in the higher-point functions that  scale as powers of $\beta J$, and are straightforward to compute, using the $\mO_0$ block in the fermion four-point function.} Since cSYK is conformally invariant for all $J$, it is trivial to extend the results to cSYK correlators at finite $J$.

\subsection{Fermion four-point function: summing ladders}
\begin{figure}[t]
\centering
\includegraphics[width=5in]{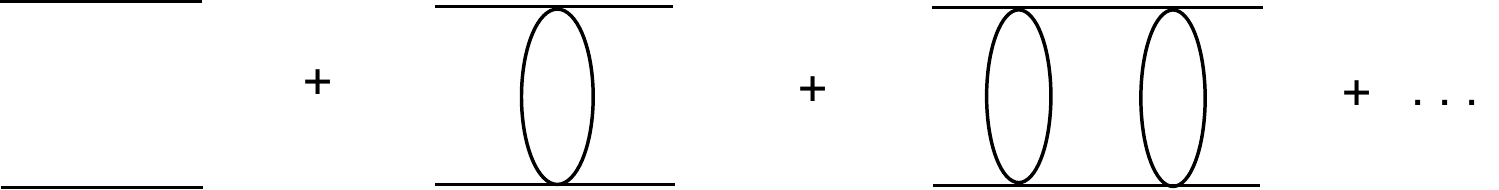}
\caption{ The fermion four-point function, at order $1/N$, is a sum of ladder diagrams.  There are also crossed diagrams, which are not shown. } \label{Figladder}
\end{figure} 

 The SYK four-point function to order $1/N$, is given by,
\be \label{4ptSum}
\frac{1}{N^2}\sum_{i, j=1}^N\langle \chi_i(\tau_1) \chi_i(\tau_2) \chi_j(\tau_3) \chi_j(\tau_4)\rangle = G(\tau_{12}) G(\tau_{34}) + \frac{1}{N}\mF(\tau_1, \tau_2, \tau_3, \tau_4)~,
\ee
where $\tau_{12} \equiv \tau_1 - \tau_2$ and $\mF$ is given by the sum of ladder diagrams, as shown in  Fig.~\ref{Figladder}. Due to the restored $O(N)$ invariance the leading behavior in $1/N$ is completely captured by $\mF$.
 The first diagram in Fig.~\ref{Figladder}, although disconnected, is suppressed by $1/N$ as it requires setting the indices to be equal, $i=j$.  This diagram is denoted by $\mF_0$,
\be \label{F0}
\mF_0 = - G(\tau_{13}) G(\tau_{24})+ G(\tau_{14}) G(\tau_{23})~.
\ee
Letting $K$ denote the kernel that adds a rung to the ladder, 
\be
K(\tau_1, \ldots \tau_4) = - (q-1) J^2 G(\tau_{13}) G(\tau_{24}) G(\tau_{34})^{q-2}~, 
\ee
and then summing the ladders yields, schematically, $\mF =(1 + K + K^2 + \ldots) \mF_0 = (1- K)^{-1} \mF_0$.  To write this explicitly, one should decompose $\mF_0$ in terms of a complete basis of eigenvectors of the kernel $K$. 

The eigenvectors of the kernel are conformal three-point functions  involving two fermions and a scalar of dimension $h$,~\footnote{In the current context the subscript on $\mO_h$ denotes that the operator has dimension $h$. This is different from another usage of subscript, $\mO_n$, which  denotes the operator in SYK, which in the weak coupling limit has dimension $2\Delta + 2n +1$. Finally, we will also sometimes use the shorthand $\mO_1$ to mean $\mO_{h_1}$. }
\be \label{Ochichi}
\langle \mO_h(\tau_0) \chi(\tau_1) \chi(\tau_2)\rangle = c_{ h} \frac{b}{J^{2\Delta} }\frac{\, \sgn(\tau_{12})}{|\tau_{12}|^{2\Delta - h} |\tau_{01}|^{h} |\tau_{02}|^{h}}~,
\ee
and have corresponding eigenvalues \cite{Kitaev}, 
\be \label{ghSYK}
k_c(h) = - (q-1) \frac{\psi(\Delta)}{\psi(1-\Delta)}  \frac{\psi(1- \Delta - \frac{h}{2})}{\psi(\Delta - \frac{h}{2})}~,
\ee
where $\psi(\Delta)$ was defined in (\ref{psiDelta}). For our purposes, one should regard the right side of (\ref{Ochichi}) as defining what we mean by the left side.  It is manifest that $k_c(h) = k_c(1-h)$, and moreover, that the singularities of $k_c(h)$ in the right-half complex plane are at $h = 2\Delta+ 2n + 1$, for integer $n$.
The three-point function involving the shadow of $\mO_h$, $\langle \mO_{1- h} \chi \chi\rangle$ is also an eigenfunction of the kernel, with the same eigenvalue, $k_c(h)$. As a result, $\Psi_h$, defined as \cite{MS}, 
\be \label{PsiIntr}
2\, c_{h} c_{1-h} \frac{b\, \sgn(\tau_{12})\, b\, \sgn(\tau_{34})}{ |J \tau_{12}|^{2 \Delta} |J \tau_{34}|^{2\Delta}}  \Psi_h(x) = \int d\tau_0 \, \langle \chi(\tau_1) \chi(\tau_2) \mO_h(\tau_0)\rangle \langle  \chi(\tau_3) \chi(\tau_4) \mO_{1-h}(\tau_0)\rangle~,
\ee
is also an eigenfunction of the kernel. Moreover, (\ref{PsiIntr}) can be seen to be an eigenfunction of  the $SL(2,R)$ Casimir, and is simply the sum of a conformal block and its shadow, see Appendix~\ref{blocks}. The conformal cross-ratio of times, $x$, is defined as, 
\be \label{xX}
x= \frac{\tau_{12} \tau_{34}}{\tau_{13} \tau_{24}}~.
\ee

\begin{figure}[t]
\centering
\includegraphics[width=2.4in]{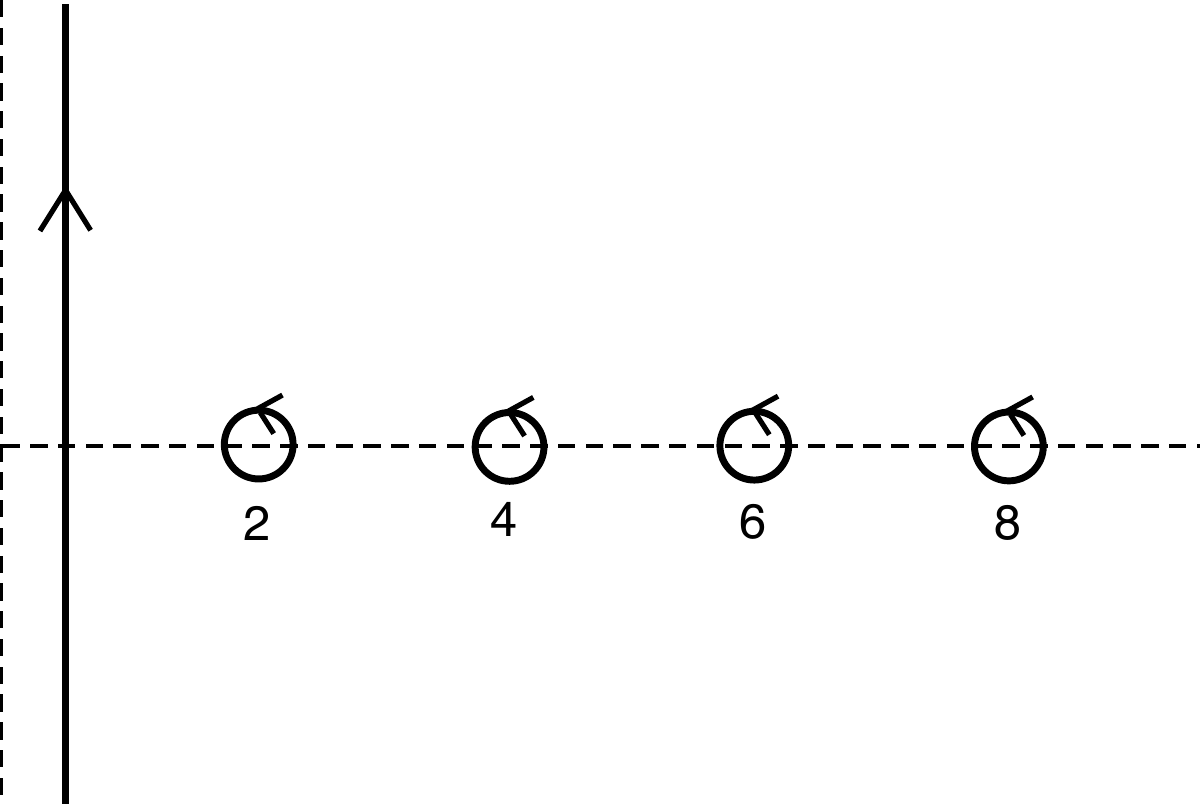}
\caption{The contour of integration $\mC$ in the complex $h$-plane.  } \label{figcontour}
\end{figure}

The necessary  range of $h$ in order to form a complete basis is dictated by representation theory of the conformal group. In even spacetime  dimensions, one only needs the continuous series, $h = \frac{d}{2} + i s$, where $d$ is the dimension and $ - \infty \!<\!s\!<\!\infty$. In odd dimensions, the  case relevant for SYK,  one must also include the discrete series, $h = 2n$ for $n\geq 1$. The eigenfunctions are orthonormal with respect to the Plancherel measure, 
\be \label{muh}
\mu(h) = \frac{2h - 1}{\pi \tan \frac{\pi h}{2}}~.
\ee
The measure has poles at $h = 2n$; indeed, the complete basis includes the discrete series specifically in order to cancel off these poles \cite{Gadde:2017sjg}. 
We can  now write $\mF_0$, as well as $\mF$, in terms of the complete basis of $\Psi_h$ \cite{MS},
\bea \label{4pt0Contour}
\mF_0(\tau_1, \ldots, \tau_4) &=&  G(\tau_{12}) G(\tau_{34})\int_{\mC}\frac{d h}{2\pi i}\, \rho^0 (h) \Psi_h(x)~, \\\label{4ptContour}
\mF(\tau_1, \ldots, \tau_4) &=&  G(\tau_{12}) G(\tau_{34})\int_{\mC}\frac{d h}{2\pi i}\, \rho(h) \Psi_h(x)~,
\eea 
where, 
\bea
 \rho^0 (h) = \mu(h) \frac{\alpha_0 }{2} k_c(h)~, \ \ \ \ \ \ \ 
  \rho (h) = \mu(h) \frac{\alpha_0 }{2} \frac{ k_c(h)}{1- k_c(h)}~,
\eea
and $\alpha_0$ is a constant,
\be
\alpha_0 = \frac{2\pi \Delta}{(1- \Delta) (2- \Delta) \tan \pi \Delta}~,
\ee
and the contour of integration $\mC$ in (\ref{4pt0Contour}) consists of the line $h = 1/2 + i s$ with $s$ running from $- \infty$ to $\infty$, as well as circles going counterclockwise around $h = 2n$ for $n\geq 1$, see Fig.~\ref{figcontour}.

A property of the measure $\mu(h)$ that we will use, which follows immediately from its definition is, 
\be \label{mu1mh}
\mu(1-h) = - \tan^2 \frac{\pi h}{2}\, \mu(h)~.
\ee
As $k_c(1-h) = k_c(h)$, both $\rho^0(h)$ and $\rho(h)$ satisfy an analogous relation.

\begin{figure}[t]
\centering
\includegraphics[width=4.5in]{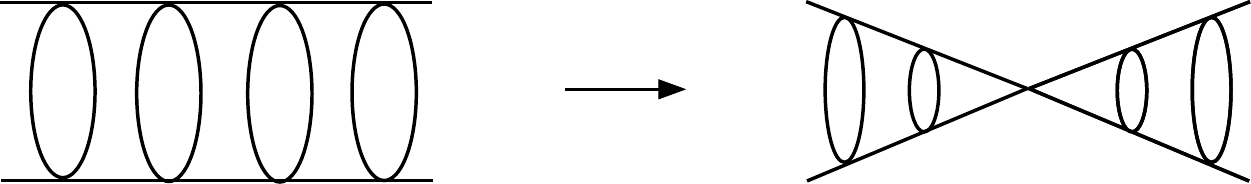}
\caption{ A pictorial representation of the four-point function, split into a product of two three-point functions $\langle \chi \chi \mO\rangle$, see \cite{GR1}, using the shadow formalism. See Eq.~\ref{PsiIntr}. } \label{figladderSplit}
\end{figure}

The fermion four-point function $\mF$, written as a contour integral over $h$, is of the form that was expected on general grounds, as mentioned in the introduction, (\ref{introSYK4pt}). This form of the four-point function will be very useful in our later studies of higher-point correlation functions. Inserting into $\mF$ the representation of $\Psi_h$ given in (\ref{PsiIntr}), we can pictorially view the four-point function as shown in Fig.~\ref{figladderSplit}.

\subsubsection*{Closing the contour}
In order to write the four-point function as a sum of conformal blocks of the operators in the theory, we simply need to close the contour of integration in (\ref{4ptContour}). 

First, consider the case of $0<x<1$. We split the contour into the line piece and the sum of poles, 
\be
\int_{\mC}\frac{d h}{2\pi i} \rho (h) \Psi_h(x) = \int_{h = \frac{1}{2}+i s} \frac{d h}{2\pi i} \rho(h) \Psi_h(x) + \sum_{n>0} \text{Res } \rho(h)\,  \Psi_h(x) \Big|_{h = 2n}~.
\ee
Focusing first on the line piece of the contour,  we write $\Psi_h$ in terms of a sum of a conformal block and its shadow, see Appendix.~\ref{blocks},
\be  \label{GGPsi}
\!\!\! G(\tau_{12}) G(\tau_{34})\! \int_{\frac{1}{2}+i s}\frac{d h}{2\pi i} \rho (h) \Psi_h(x) = \frac{b^2}{2 J^{4 \Delta}} \int_{\frac{1}{2} + i s} \frac{d h}{2\pi i} \rho(h)\[ \beta(h, 0) \mF_{\Delta}^h(x) + \beta(1\!-\! h, 0) \mF_{\Delta}^{1-h}(x)\]~,
\ee
where $\mF_{\Delta}^h$ is the conformal block with external fermions of dimension $\Delta$ and an exchanged scalar of dimension $h$,  while,
\be
\beta(h,0) = \sqrt{\pi} \frac{\Gamma(\frac{h}{2})^2 \Gamma(\frac{1}{2}- h)}{\Gamma(\frac{1-h}{2})^2 \Gamma(h)}~.
\ee
In  (\ref{GGPsi}), let us change integration variables for the second term, $h \rightarrow 1- h$, use the reflection relation (\ref{mu1mh}) for the measure, as well as,
\be
\beta(h,0) \(1 -  \tan^2 \frac{\pi h}{2}\) = 2 \frac{\Gamma(h)^2}{\Gamma(2h)}~, 
\ee
to write, 
\be \label{line11}
G(\tau_{12})G(\tau_{34})\int_{\frac{1}{2}+i s}\frac{d h}{2\pi i} \rho (h) \Psi_h(x) =  \frac{b^2}{J^{4 \Delta}}\int_{ \frac{1}{2}+i s}\frac{d h}{2\pi i}\,  \rho(h) \frac{\Gamma(h)^2}{\Gamma(2h)} \mF_{\Delta}^h(x)~.
\ee
Turning now to the sum over the discrete series, we  rewrite this as, 
\be \label{line22}
G(\tau_{12}) G(\tau_{34}) \sum_{n>0} \text{Res}\, \rho(h)\,  \Psi_h(x) \Big|_{h = 2n} = \frac{b^2}{J^{4 \Delta}} \sum_{n>0} \text{Res}\, \rho(h)\,\frac{\Gamma(h)^2}{\Gamma(2h)} \mF_{\Delta}^h(x)\Big|_{h = 2n} ~,
\ee
where we have used that $\beta( 1- 2n, 0) = 0$ for $n>0$ and $\beta(h, 0) = 2 \Gamma(h)^2/ \Gamma(2h)$ for $h = 2n$. Recombining the continuous and discrete series terms gives,
\be
\mF(\tau_1, \ldots, \tau_4) = \frac{b^2}{J^{4 \Delta}}\int_{\mC}\frac{d h}{2\pi i} \rho(h) \frac{\Gamma(h)^2}{\Gamma(2h)}\, \mF_{\Delta}^h(x)~.
\ee
Finally, we close the line piece of the contour to the right, giving a sum over the poles at the $h$ for which $k_c(h) = 1$,~\footnote{The poles at $h = 2n$ coming from measure $\mu(h)$ are outside of the closed contour, as a result of the piece of the  contour made up of the circles at $h=2n$.}
\be \label{Fxl1}
\mF(\tau_1, \ldots, \tau_4) = \frac{b^2}{J^{4 \Delta}} \sum_{h_n} c_n^2\, \mF_{\Delta}^{h_n}(x)~, ~\ \ \ \ \ 0<x<1~,
\ee
where  $h_n$ are the single-trace operator dimensions, $k_c(h_n) = 1$, and we have defined \cite{MS}~\footnote{We have suppressed the  $1/N$ scaling of $c_n \sim 1/\sqrt{N}$. In order to not carry around factors of $1/N$, we will generally suppress them. A connected $p$-point correlation function scales as $\langle \mO_1\cdots \mO_p\rangle \sim 1/N^{\frac{(p-2)}{2}}$. }
\be \label{OPEres}
c_n^2 \equiv - \text{Res}\, \rho(h)\Big|_{h = h_n}\,  \frac{\Gamma(h_n)^2}{\Gamma(2h_n)}  =  \alpha_0 \frac{(h_n - 1/2)}{ \pi \tan( \pi h_n/2)} \frac{\Gamma(h_n)^2}{\Gamma(2 h_n)} \frac{1}{k_c'(h_n)}~.
\ee
One can identify the $c_n$ as the OPE coefficients $\frac{1}{N} \sum_{i=1}^N\chi(0) \chi(\tau) \sim\frac{1}{\sqrt{N}} \sum_n c_n \mO_n$. We will sometimes use the short-hand, $c_{h_1}$ (or $c_1$) to denote $c_n$ for $h_1$ that is given by $h_1 = 2\Delta + 2n +1$ at weak coupling. 

This is the expression for the fermion four-point function when the conformal cross-ratio $x$ in the range $0<x<1$.  For the case of $x>1$, we return to (\ref{4ptContour}) and simply close the line piece of the contour to the right, giving, 
\be
\mF(\tau_1, \ldots, \tau_4) = G(\tau_{12})G(\tau_{34}) \sum_{h_n} \text{Res }\rho(h)\Big|_{h = h_n}\, \Psi_{h_n}(x)~,  \ \ \ \ x>1~.
\ee

We conclude with a comment on the singularity structure in $h$-space of the ladder diagrams. One can see that the first diagram in the sequence of ladders, $\mF_0$, is, in $h$-space, proportional to $k_c(h)$. Similarly, a  diagram with $n$ rungs is proportional to $k_c(h)^{n+1}$. Summing any finite number of ladder diagrams gives a polynomial in $k_c(h)$ which, like $k_c(h)$, will have singularities at $h = 2\Delta + 2n +1$. Correspondingly, upon closing the contour to return to physical space, the finite sum of ladder diagrams will be expressed in terms of conformal blocks of exchanged operators of dimension $2\Delta + 2n + 1$: the free-field dimensions of the primaries, schematically of the form $\sum \chi_i \partial_{\tau}^{2n+1} \chi_i$. It is only when one sums an infinite number of ladder diagrams, such as the geometric sum $k_c(h)( 1+ k_c(h) + k_c(h)^2 + \ldots)$, as in $\rho(h)$ appearing in $\mF$, that the singularities of the expression are no longer where $k_c(h)$ is singular, but rather where $k_c(h) = 1$. Correspondingly, the expansion of $\mF$ is in terms of conformal blocks at the infrared dimensions of the primaries, the $h$ for which  $k_c(h) = 1$. 

\section{Bilinear Three-Point Function} \label{sec:3pt}
In this section we  compute, to leading nontrivial order in $1/N$,  the fermion six-point function, and correspondingly the three-point function   $\langle \mO_1 \mO_2 \mO_3\rangle$ of the bilinear $O(N)$ invariant primaries, $\mO_i$, of dimension $h_i$. 

The six-point function of the fermions can be written as, 
\be \label{Six} 
\frac{1}{N^3} \sum_{i, j, l = 1}^N \langle \chi_i (\tau_1) \chi_i (\tau_2) \chi_j (\tau_3) \chi_j (\tau_4) \chi_l (\tau_5) \chi_l (\tau_6)\rangle = \ldots  + \frac{1}{N^2} \mS(\tau_1, \ldots, \tau_6) + \ldots~,
\ee
where $\mS$ is the lowest order term in $1/N$ that contains fully connected diagrams. 
 There are two classes of diagrams contributing to $\mS$: the ``contact'' diagrams, whose sum we denote by $\mS_1$, and the planar diagrams, whose sum we denote by $\mS_2$, 
\be \label{SS12}
\mS = \mS_1 + \mS_2~.
\ee
We  study the contact diagrams in Sec.~\ref{contact}, and the planar diagrams in Sec.~\ref{planar}.  

From the fermion six-point function, we will extract the three-point function of the bilinear primary $O(N)$ singlets, 
\be \label{OOO}
\langle \mO_{1}(\tau_1) \mO_{2}(\tau_2) \mO_{3}(\tau_3)\rangle =  \frac{1}{\sqrt{N}}\frac{ c_{123}}{|\tau_{12}|^{h_{1} +h_{2}  - h_{3}}| \tau_{23}|^{h_{2} + h_3 - h_{1}} |\tau_{13}|^{h_{1} + h_{3} - h_{2}}}~,
\ee 
where $c_{123}$ will have two contributions, 
 \be\label{cnmk}
c_{123} = c_{123}^{(1)} + c_{ 123}^{(2)}~, 
\ee
coming from the contact and the planar diagrams, respectively. We compute $c_{123}^{(1)}$ in Sec.~\ref{contact} and $c_{123}^{(2)}$ in Sec.~\ref{planar}.

\subsection{Contact diagrams} \label{contact}
\begin{figure}[t]
\centering
\includegraphics[width=6.5in]{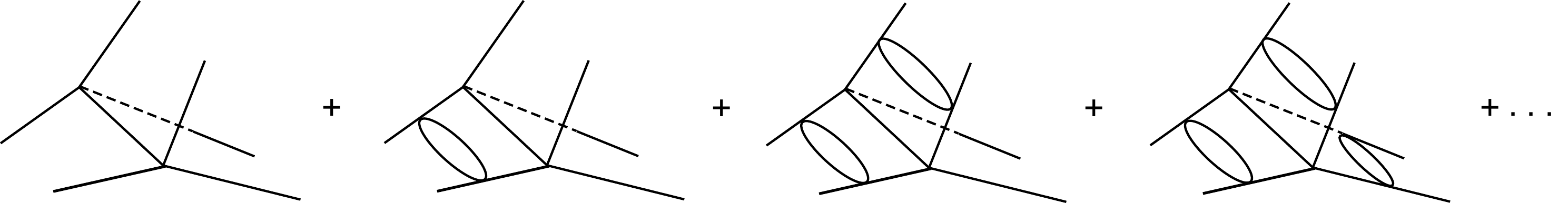}
\caption{ The first set of diagrams (``contact'' diagrams) contributing to the six-point function at order $1/N^2$. 
}
\label{Fig6pt0}
\end{figure}

The ``contact diagrams'' are composed of three fermion four-point functions glued to two interaction vertices connected by $q-3$ propagators, see Fig.~\ref{Fig6pt0} and the first diagram on the right in Fig.~\ref{FigIntro1},
\be \label{S1}
\mS_1 = (q-1)(q-2) J^2 \int d\tau_a d\tau_b\, G(\tau_{ab})^{q-3} \mF(\tau_1, \tau_2, \tau_a, \tau_b) \mF(\tau_3, \tau_4, \tau_a, \tau_b)\mF(\tau_5, \tau_6, \tau_a, \tau_b)~.
\ee
The fermion four-point function $\mF$ is a sum of conformal blocks, and the functional  form of each block is fixed by conformal invariance. It will be most convenient to write the blocks in terms of the differential operator $\mC_n(\tau_{12}, \partial_2)$, which sums the contributions of all descendants associated with the primary $\mO_n$, acting on a conformal three-point function, see Appendix~\ref{blocks},
\be \label{mFmC}
\mF(\tau_1, \tau_2, \tau_a, \tau_b) = \sum_n c_n\, \mC_{n}(\tau_{12}, \partial_2)\, \langle \mO_n(\tau_2) \chi(\tau_a) \chi(\tau_b)\rangle~,
\ee
where the three-point function was given in (\ref{Ochichi}).
Using this form for each of the four-point functions appearing in $\mS_1$ gives,
\be \label{mS_1}
\mS_1 = \sum_{n_1, n_2, n_3} \prod_{i=1}^3 c_{ n_i} \mC_{ n_i}(\tau_{2i-1, 2i}, \partial_{2i})\,\, \langle\mO_{n_1}(\tau_2)\mO_{n_2} (\tau_4)\mO_{n_3}( \tau_6)\rangle_1~,
\ee
where, 
\be \label{OOO}
\langle\mO_{n_1}(\tau_1)\mO_{n_2} (\tau_2)\mO_{n_3}( \tau_3)\rangle_1=  (q-1)(q-2) J^2 \int d\tau_a d\tau_b\, G(\tau_{ab})^{q-3}\, \prod_{i=1}^3  \langle \mO_{n_i}(\tau_i) \chi(\tau_a) \chi(\tau_b)\rangle~.
\ee

Explicitly writing out the integrand in the expression for the three-point function of bilinears gives, 
\be
\!\!\!\!\!\!\! \langle\mO_{n_1}(\tau_1)\mO_{n_2} (\tau_2)\mO_{n_3}( \tau_3)\rangle_1=c_{n_1}c_{n_2}c_{n_3} (q-1)(q-2) b^{q}\,   I_{1 2 3 }^{(1)}(\tau_1, \tau_2, \tau_3) ~,
\ee
where
\be \label{I1}
I_{1 2 3 }^{(1)}(\tau_1, \tau_2, \tau_3) = \int d\tau_a d\tau_b \frac{|\tau_{ab}|^{h_1 + h_2 + h_3-2}}{|\tau_{1a}|^{h_1}|\tau_{1 b}|^{h_1}|\tau_{2a}|^{h_2}|\tau_{2 b}|^{h_2}|\tau_{3 a}|^{h_3} |\tau_{3 b}|^{h_3}}~.
\ee
In order to evaluate the integral, it is convenient to change  integration variables to the two cross-ratios, 
\be
A= \frac{\tau_{a1} \tau_{23}}{\tau_{a 2} \tau_{13}}~, \ \ \ \ \ B = \frac{\tau_{b 2} \tau_{13}}{\tau_{b 1} \tau_{2 3}}~,
\ee
resulting in a conformal thee-point function,
\be
I_{123}^{(1)}(\tau_1, \tau_2, \tau_3) = \frac{\mI_{123}^{(1)}}{|\tau_{12}|^{h_1 + h_2 - h_3}|\tau_{13}|^{h_1 +h_3 - h_2}|\tau_{23}|^{h_2 + h_3 - h_1}}
\ee
with  coefficients $\mI_{123}^{(1)}$,
\be
\mI_{123}^{(1)} = \int d A d B \frac{| 1- A B|^{h_1 +h_2 + h_3 - 2}}{|A|^{h_1}  |B|^{h_2} |(1-A)(1-B)|^{h_3}}~.
\ee
In \cite{GR2}, we evaluated this integral by noticing that, after a change of variables $B \rightarrow 1/B$, it is of the form of a Selberg integral. Equivalently, one may notice that if the integration range in the integral were $A\in (0,1)$ and $B\in (0,1)$, then the result would be proportional to a generalized hypergeometric function at argument one, 
\be
\pFq{3}{2}{1-h_1~,\, , 1-h_2~,\, , 2- h_1 - h_2 - h_3}{ 2 - h_1 - h_3~,\,, 2-h_2 - h_3}{1}~.
\ee
Breaking the integral in $\mI_{123}^{(1)}$ up into regions for which the integrand is analytic, identifying the integral in each region as a particular ${}_3 F_2$ at argument one, all of which in this case simplify to products of ratios of gamma functions, and then adding the contributions, we recover  the result of \cite{GR2},
\be \label{I1exact}
\!\!\mathcal{I}_{123}^{(1)}\! =\!\frac{  \sqrt{\pi}\, 2^{h_1 + h_2+ h_3 -1}\, \Gamma(1\!-\!h_1) \Gamma(1\!-\!h_2) \Gamma(1\!-\!h_3)}{\Gamma\(\frac{3 - h_1 - h_2 - h_2}{2}\)}\! \[\rho(h_1, h_2, h_3)\! +\! \rho(h_2, h_3, h_1)\!+\! \rho(h_3, h_1, h_2) \]~,
\ee
where we defined,
\be
\rho(h_1, h_2, h_3) = \frac{\Gamma(\frac{h_2 +h_3 - h_1}{2})}{\Gamma(\frac{2-h_1-h_2 +h_3}{2})\Gamma(\frac{2-h_1-h_3+h_2}{2})}\( 1+ \frac{\sin(\pi h_2)}{\sin(\pi h_3) - \sin(\pi h_1 + \pi h_2)}\)~.
\ee
The contribution of the contact diagrams to the three-point function is thus, 
\be \label{c1231}
c_{1 2 3}^{(1)} = c_1 c_2 c_3\,(q-1)(q-2) b^{q}\,  \mI_{123}^{(1)}~.
\ee

\subsection{Planar diagrams} \label{planar}
\begin{figure}[t]
\centering
\includegraphics[width=6in]{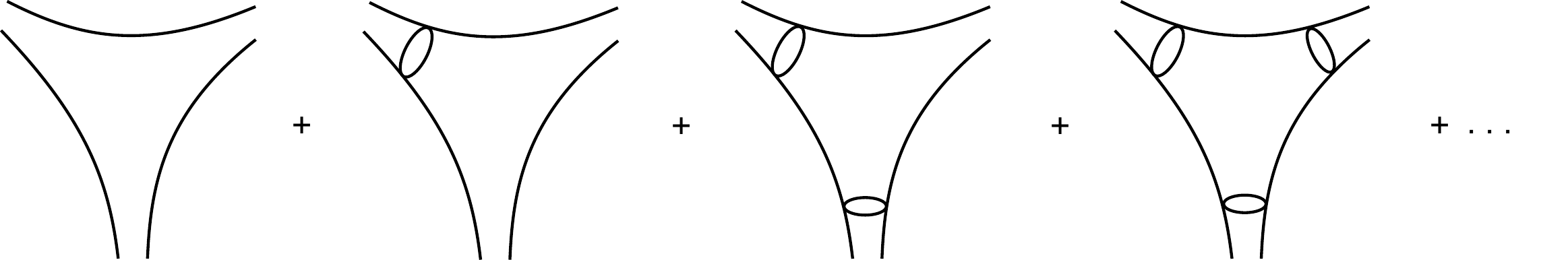}
\caption{The second set of diagrams  (planar diagrams)  contributing to the six-point function at order $1/N^2$.} \label{Fig6pt}
\end{figure}
Turning now to the planar diagrams, these similarly consist of three fermion four-point functions glued together, though now in a smooth way, see Fig.~\ref{Fig6pt} as well as the second diagram on the right in Fig.~\ref{FigIntro1},
\be \label{mS2}
\!\mS_2 =\! \int\! d\tau_a d\tau_{\bar{a}}d\tau_b d\tau_{\bar{b}} d\tau_c d\tau_{\bar{c}}\,\,  \mF(\tau_1, \tau_2, \tau_a, \tau_{\bar{b}})\, D(\tau_{\bar{b} b}) \mF(\tau_3, \tau_4, \tau_c, \tau_{\bar{a}})\, D(\tau_{\bar{a} a})\, \mF(\tau_5, \tau_6, \tau_b, \tau_{\bar{c}})\, D(\tau_{\bar{c} c})~,
\ee
where we have let $D(\tau)$ denote the inverse of the propagator, 
\be
\int d\tau_0\, D(\tau_{10}) G(\tau_{0 2}) = \delta(\tau_{12})~.
\ee
The $D(\tau)$ are needed in (\ref{mS2}) in order to strip off some of the propagators on the external legs of the four-point functions which would, otherwise, be overcounted. In the infrared,  $D(\tau)$ is simply, $D(\tau) = -\Sigma(\tau) = -J^2 G(\tau)^{q-1}$. 

Applying the same logic as with the contact diagrams, and writing the fermion four-point function in the form (\ref{mFmC}), gives the planar diagram contribution to the three-point function of the $\mO$, 
\begin{multline}
\langle\mO_{1}(\tau_1)\mO_{2} (\tau_2)\mO_{3}( \tau_3)\rangle_2 = \int\! d\tau_a d\tau_{\bar{a}}d\tau_b d\tau_{\bar{b}} d\tau_c d\tau_{\bar{c}}\, \langle \mO_{1}(\tau_1) \chi(\tau_a)\chi( \tau_{\bar{b}}) \rangle\, D(\tau_{\bar{b} b})\\
 \langle \mO_{2}(\tau_2) \chi(\tau_c) \chi(\tau_{\bar{a}})\rangle D(\tau_{\bar{a} a})\, \langle \mO_{3}(\tau_3)\chi(\tau_b) \chi(\tau_{\bar{c}})\rangle D(\tau_{\bar{c} c})~.
\end{multline}
Explicitly writing out the terms appearing  in the integrand gives, 
\begin{multline}
\langle\mO_{1}(\tau_1)\mO_{2} (\tau_2)\mO_{3}( \tau_3)\rangle_2 \\ =c_1 c_2 c_3 \,b^{3 q} \int\! d\tau_a d\tau_{\bar{a}}d\tau_b d\tau_{\bar{b}} d\tau_c d\tau_{\bar{c}}\, 
 \frac{\,\sgn(\tau_{a \bar{b}} \tau_{c \bar{a}} \tau_{b \bar{c}} \tau_{\bar{a} a} \tau_{\bar{b} b} \tau_{\bar{c} c})\,  |\tau_{a \bar{b}}|^{h_1-2 \Delta}|\tau_{c \bar{a}}|^{h_2-2\Delta} |\tau_{b\bar{c}}|^{h_3-2\Delta}}{|\tau_{\bar{a} a} \tau_{\bar{b} b} \tau_{\bar{c} c}|^{2(1-\Delta)}   |\tau_{1 a } \tau_{1 \bar{b}}|^{h_1} |\tau_{2 c} \tau_{2 \bar{a}}|^{h_2} |\tau_{3 b} \tau_{3 \bar{c}}|^{h_3}}~.
\end{multline}
This form exhibits all the symmetries that  are manifest of the Feynman diagrams. 
The integrals over $\tau_{\bar{a}}, \tau_{\bar{b}}, \tau_{\bar{c}}$ are conformal three-point integrals, and are simple to evaluate, see Appendix.~B of \cite{GR2}. Defining,
\be \label{eq:xi}
\xi(h) =  \frac{1}{\sqrt{\pi}} \frac{\Gamma(\frac{2\Delta+1}{2})}{\Gamma(1-\Delta)} \frac{\Gamma(\frac{1-h}{2})}{\Gamma(\frac{h}{2})}\frac{\Gamma(\frac{2 - 2 \Delta + h}{2})}{\Gamma(\frac{1 + 2\Delta - h}{2})}~,
\ee
gives  
\be
\langle\mO_{1}(\tau_1)\mO_{2} (\tau_2)\mO_{3}( \tau_3)\rangle_2  = c_1 c_2 c_3\, \xi(h_1) \xi(h_2) \xi(h_3)\, I_{123}^{(2)}(\tau_1, \tau_2, \tau_3)~,
\ee
where \cite{GR2}, 
\be \label{I2}
\!\!\!\!\!\!\!\!\! I_{1 23}^{(2)}(\tau_1, \!\tau_2, \!\tau_3) \!= \!\!\! \int\!\! d\tau_a d\tau_b d\tau_c \frac{  -\sgn(\tau_{1a} \tau_{1b} \tau_{2a} \tau_{2c} \tau_{3b} \tau_{3c}) \! |\tau_{ab}|^{h_1  \!- \!1} |\tau_{ca}|^{h_2 \!- \!1} |\tau_{b c}|^{h_3  \!- \!1}}{|\tau_{1a}|^{h_1 \!- \!1+2\Delta} |\tau_{1b}|^{h_1 +1-2\Delta} |\tau_{2 c}|^{h_2  \!- 1\!+2\Delta} |\tau_{2a}|^{h_2+1-2\Delta} |\tau_{3 b}|^{h_3  \!-1 \!+2\Delta} |\tau_{3 c}|^{h_3+1-2\Delta}}.
\ee
 In making the choice of, for instance, evaluating the $\tau_{\bar{a}}$ integral instead of the $\tau_a$ integral, some of the symmetries are no longer manifest. 
 
 To proceed with evaluating the remaining three integral,  we change integration  variables to the cross-ratios $A, B, C$, defined as, 
\be
A = \frac{\tau_{a1} \tau_{32}}{\tau_{a2} \tau_{31}}~, \,\,  B = \frac{\tau_{13} \tau_{ab}}{\tau_{1a} \tau_{3b}}~,\,\, C = \frac{\tau_{2a} \tau_{3c}}{\tau_{23}\tau_{ac}}~.
\ee
This change of variables transforms $I_{123}^{(2)}$ into a form that is manifestly a conformal three-point function, 
\be
I_{123}^{(2)}(\tau_1, \tau_2, \tau_3) = \frac{\mI_{123}^{(2)}}{|\tau_{12}|^{h_1 + h_2 - h_3}|\tau_{13}|^{h_1 +h_3 - h_2}|\tau_{23}|^{h_2 + h_3 - h_1}}~,
\ee
with a coefficient,
\be \label{911v1}
\mI_{123}^{(2)} = \int d A d B d C \frac{\sgn(C(1-B) (1-C))\,\, |1 -  A B C|^{h_3-1}}{|A|^{h_1 } |1-A|^{1 - h_1 - h_2+h_3} |B|^{1-h_1} |1-B|^{h_1+1-2\Delta} |C|^{1-2\Delta+h_3} |1-C|^{h_2-1+2\Delta}}~.
\ee

To evaluate this integral we note the following: if the integration range were over $A\in (0,1), B\in (0,1), C\in (0,1)$, then this would be of the form of a generalized hypergeometric function at argument equal to one, 
\be \label{4F31}
 \pFq{4}{3}{1\!-\!h_1~, h_1~,  2\Delta\! -\! h_3~, 1\!-\!h_3}{1\!+\!h_2\!-\! h_3~, 2\Delta~,  2\!-\!h_2\!-\!h_3}{1}~.
 \ee
In order to account for the other regions of integration, one should consider each region  separately and perform simple changes of variables combined with ${}_2 F_1$ connection identities and  Euler's integral transform, 
\be \nonumber
\pFq{p}{q}{a_1, \dots, a_{p}}{b_1, \ldots,  b_{q} }{z} = \frac{\Gamma(b_q)}{\Gamma(a_p)\Gamma(b_q-a_p) }\int_0^1 dt\, t^{a_p - 1} (1-t)^{b_q-a_p -1}\,  \pFq{p-1}{q-1}{a_1, \ldots, a_{p-1}}{b_1, \ldots, b_{q-1}}{ t  z}~.
\ee

A faster method is the following. Consider the more general integral, which is  a function of an additional variable $z$, 
\be \label{911vz}
\mI_{123}^{(2)}(z) = \int d A d B d C \frac{\sgn(C(1-B) (1-C))\,\, |1 -  z A B C|^{h_3-1}}{|A|^{h_1 } |1-A|^{1 - h_1 - h_2+h_3} |B|^{1-h_1} |1-B|^{h_1+1-2\Delta} |C|^{1-2\Delta+h_3} |1-C|^{h_2-1+2\Delta}}~.
\ee 
The generalized hypergeometric function $_{4}F_3$  satisfies a fourth-order differential equation. Since the piece of this  integral coming from  the region  $A\in (0,1), B\in (0,1), C\in (0,1)$ is a ${}_4 F_3$, of the type (\ref{4F31}), it must be the case  that the integrand satisfies the appropriate differential equation. Breaking the integral up into regions in which the integrand is analytic, the integrand in each region should also satisfy the same differential equation. As there are four solutions to the differential equation defining ${}_4 F_3$, the integral (\ref{911vz}) should take a form that is a superposition of these, with some coefficients, $\bar{\alpha}_i$,~\footnote{It is conceivable that, as result of boundary terms, this is not true. However, we have also evaluated the integral (\ref{911v1}) explicitly, by breaking it up into regions, as outlined in the previous paragraph, and found the same answer as the one quoted below, though in a less nice form. }
\bea \nonumber
\mI_{123}^{(2)}(z) &=& \bar{\alpha}_1\,\, \pFq{4}{3}{1\!-\!h_1~, h_1~,  2\Delta\! -\! h_3~, 1\!-\!h_3}{1\!+\!h_2\!-\! h_3~, 2\Delta~,  2\!-\!h_2\!-\!h_3}{z} \\[8pt] \nonumber
&+& \bar{\alpha}_2\,  z^{h_3- h_2}\, \pFq{4}{3}{1\!-\!h_1\! -\! h_2\! +\! h_3~,  h_1\! -\! h_2\! +\! h_3~, 2\Delta \!-\! h_2~, 1\!-\!h_2}{2\!-\!2h_2~, 1\!-\! h_2\! +\! h_3~, 2\Delta\! -\! h_2\! +\! h_3}{z} \nonumber
\\[8pt]  &+& \bar{\alpha}_3\, z^{1- 2\Delta}\, \pFq{4}{3}{2\! -\! h_1\! -\!2 \Delta~, 1\!+\! h_1\! -\! 2\Delta, 1\!-\!h_3~, 2\!-\! h_3\! -\! 2\Delta}{2 \!+\! h_2 \!-\! h_3\!-\! 2\Delta~, 3\! -\! h_2\! -\! h_3\!-\! 2\Delta~, 2\!-\! 2\Delta}{z}  \nonumber
\\[8pt] &+&  \bar{\alpha}_4\, z^{h_2 + h_3 - 1}\, \pFq{4}{3}{h_2\! +\! h_3\! -\! h_1~, h_1\! +\! h_2\! +\! h_3\! -\!1~, h_2\! -\!1\!+\!2\Delta~, h_2}{2 h_2~, h_2 \!+\! h_3\! -\!1\! +\! 2\Delta~, h_2\! +\! h_3}{z}~. \label{Ansatz}
\eea
To fix the coefficient $\bar{\alpha}_1$ we simply set $z=0$ in (\ref{911vz}): the integrals decouple, and are trivial to evaluate, see Appendix.~B of \cite{GR2} for relevant equations. 
Similarly to fix $\bar{\alpha}_2$, we change integration variables $A\rightarrow A/ (z B C)$, and then take small $z$ and evaluate the integral. To fix $\bar{\alpha}_3$ we change variables $B\rightarrow B/ ( z A C)$, and for $\bar{\alpha}_4$ we change variables $C\rightarrow C/ ( z A B)$. 
It is convenient to define $\alpha_i$, which is related to $\bar{\alpha}_i$ through the coefficients $\xi(h)$ that arose earlier in performing the first three of six integrals,
\be
\alpha_i = \xi(h_1) \xi(h_2) \xi(h_3)\, \bar{\alpha}_i~.
\ee
The result for the $\alpha_i$ is the following, 
\bea \nonumber
\alpha_1 &=& - \frac{\Gamma(\frac{2\Delta+1}{2})^2}{\Gamma(1-\Delta)^2} \prod_{i=1}^3 \frac{\Gamma(\frac{1- h_i}{2})}{\Gamma(\frac{h_i}{2})}\, \, \frac{\Gamma(\frac{3-h_2 - 2\Delta}{2}) \Gamma(\frac{2 +h_2 - 2\Delta}{2})}{\Gamma(\frac{h_2 + 2\Delta}{2}) \Gamma( \frac{1-h_2 + 2\Delta}{2})} \frac{\Gamma(\frac{h_3 - h_2}{2})\Gamma(\frac{h_2+h_3-1}{2})}{\Gamma(\frac{2-h_2 - h_3}{2})\Gamma(\frac{1+h_2- h_3}{2})} \frac{\Gamma(\frac{h_1 + h_2 - h_3}{2})}{\Gamma(\frac{1-h_1 - h_2 + h_3}{2})}~, \\[10pt] \nonumber
\alpha_2 &=& - \frac{\Gamma(\frac{2\Delta+1}{2})^3}{\Gamma(1-\Delta)^3}\frac{\Gamma(\frac{1-h_1}{2})}{\Gamma(\frac{h_1}{2})}\frac{\Gamma(\frac{1-h_2}{2})^2\, \Gamma(\frac{2h_2 -1}{2})}{\Gamma(\frac{h_2}{2})^2\, \Gamma(\frac{2-2h_2}{2})}\frac{\Gamma(\frac{3 - h_2 - 2\Delta}{2})}{\Gamma(\frac{h_2 + 2\Delta}{2})} \frac{\Gamma(\frac{2+ h_3 - 2\Delta}{2})}{\Gamma(\frac{1- h_3 + 2\Delta}{2})} \\ \nonumber
&{ } &\, \, \, \, \cdot \frac{\Gamma(\frac{h_2 - h_3}{2})\Gamma(\frac{h_2 -h_3 + 2 - 2\Delta}{2})}{\Gamma(\frac{1 - h_2 +h_3}{2})\Gamma(\frac{h_3-h_2 +1 +2 \Delta}{2})} \frac{\Gamma(\frac{h_1 - h_2 + h_3}{2})}{\Gamma(\frac{1-h_1+h_2 - h_3}{2})}~,   \\[10pt]\nonumber
\alpha_3 &=& - \frac{\Gamma(\frac{2\Delta+1}{2})^3\, \Gamma(\Delta)}{\Gamma(1-\Delta)^3\, \Gamma(\frac{3 - 2\Delta}{2})} \prod_{i=1}^3\frac{\Gamma(\frac{1-h_i}{2})\Gamma(\frac{2+h_i - 2\Delta}{2})\Gamma(\frac{ 3- h_i - 2\Delta}{2})}{\Gamma(\frac{h_i}{2})\Gamma(\frac{1-h_i + 2\Delta}{2})\Gamma(\frac{h_i + 2\Delta}{2})}\\ \nonumber
&&\,\, \, \cdot \frac{\Gamma(\frac{h_3 - h_2 +2\Delta}{2}) \Gamma(\frac{h_2 + h_3 - 1+2\Delta}{2})}{\Gamma(\frac{3+h_2 - h_3 - 2\Delta}{2})\Gamma(\frac{ 4 - h_2 - h_3 - 2\Delta}{2})}\frac{\Gamma(\frac{h_1 + h_2 - h_3}{2})}{\Gamma(\frac{1-h_1 - h_2 + h_3}{2})} ~, \\[10pt]\nonumber
\alpha_4 &=& - \frac{\Gamma(\frac{2\Delta+1}{2})^3}{\Gamma(1- \Delta)^3}\frac{\Gamma(\frac{1- h_1}{2})}{\Gamma(\frac{h_1}{2})}\frac{\Gamma(\frac{1- 2h_2}{2})}{\Gamma(h_2)}\frac{\Gamma(\frac{2+ h_2 - 2\Delta}{2})}{\Gamma(\frac{1 - h_2 + 2\Delta}{2})} \frac{\Gamma(\frac{ 2 + h_3 - 2\Delta}{2})}{\Gamma(\frac{1- h_3 + 2\Delta}{2})}\\ 
&& \,\, \, \cdot \frac{\Gamma(\frac{1- h_2 - h_3}{2})}{\Gamma(\frac{h_2 + h_3}{2})} \frac{\Gamma(\frac{3- h_2 - h_3 - 2\Delta}{2})}{\Gamma(\frac{h_2 + h_3 + 2\Delta}{2})}\frac{\Gamma(\frac{h_1 + h_2 - h_3}{2})\Gamma(\frac{-h_1 + h_2 +h_3}{2})\Gamma(\frac{h_1 + h_2 +h_3 -1 }{2})}{\Gamma(\frac{ 1- h_1 - h_2 + h_3}{2})\Gamma(\frac{1+h_1 - h_2-h_3}{2})\Gamma(\frac{ 2- h_1 -h_2 - h_3}{2})}~.
\eea
This completes the evaluation of the planar diagram contribution to the three-point function. The result is,
\be \label{c1232}
c_{123}^{(2)} = c_1 c_2 c_3\, \xi (h_1)\xi(h_2) \xi(h_3)\, \mI_{123}^{(2)}~,
\ee
where $\mI_{123}^{(2)}$ is a sum of four generalized hypergeometric functions with argument one, $ \mI_{123}^{(2)} = \mI_{123}^{(2)} (z=1)$  given by (\ref{Ansatz}).  Although it is not manifest, $c_{123}^{(2)}$ must be symmetric under all permutations of the $h_i$. 
 In Appendix.~\ref{largeQ} we study $c_{123}^{(2)}$ in the large $q$ limit in which it somewhat simplifies. 

\subsubsection*{Universality}
The full three-point function coefficient is a sum of the contact diagram and the planar diagram contributions, $c_{123} = c_{123}^{(1)} + c_{123}^{(2)}$. It is instructive to write this as, 
\be \label{c123I}
c_{123} = c_1 c_2 c_3 \mI_{123}~. 
\ee
There are two distinct contributions. The product of OPE coefficients  $c_i$ of two fermions turning into an $\mO_i$ reflects the sum of the ladder diagrams; this sum determines the dimensions $h_i$ of the $\mO_i$. The contribution $\mI_{123}$ comes from gluing the ladders. It is universal in the sense that it is determined by an integral whose parameters are the  fermion dimension $\Delta$ and the dimensions $h_i$.

\section{Bilinear Four-Point Function} \label{sec:4pt}

\subsection{Cutting melons and $2p$-point functions}
\begin{figure}[t]
\centering
\subfloat[]{
\includegraphics[width=1.6in]{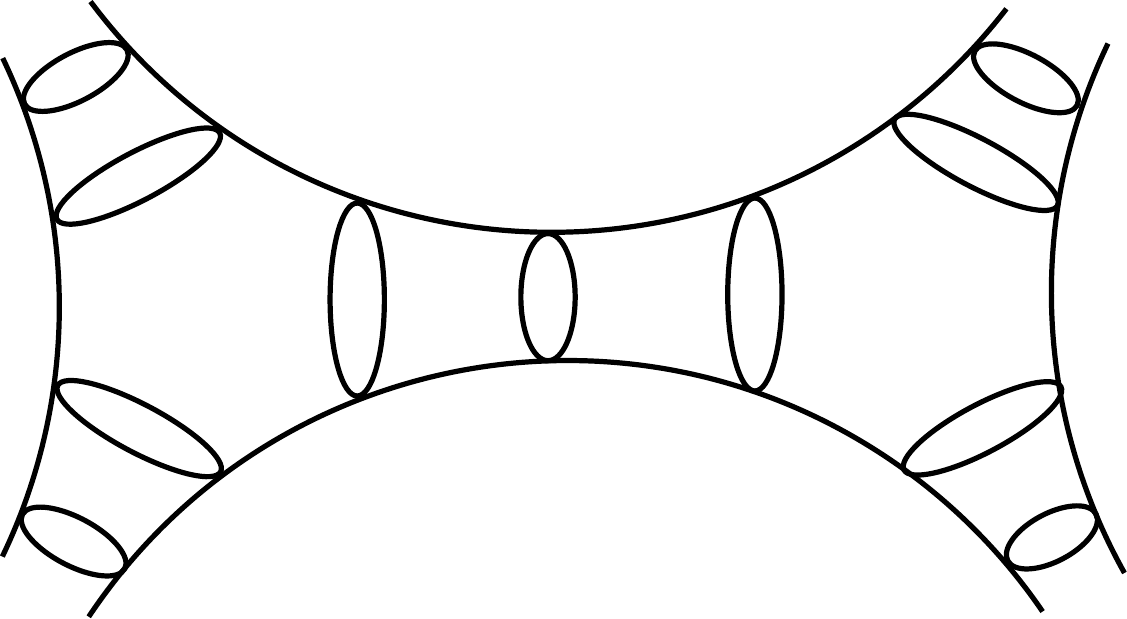}
} \ \ \ \ 
\subfloat[]{
\includegraphics[width=1.8in]{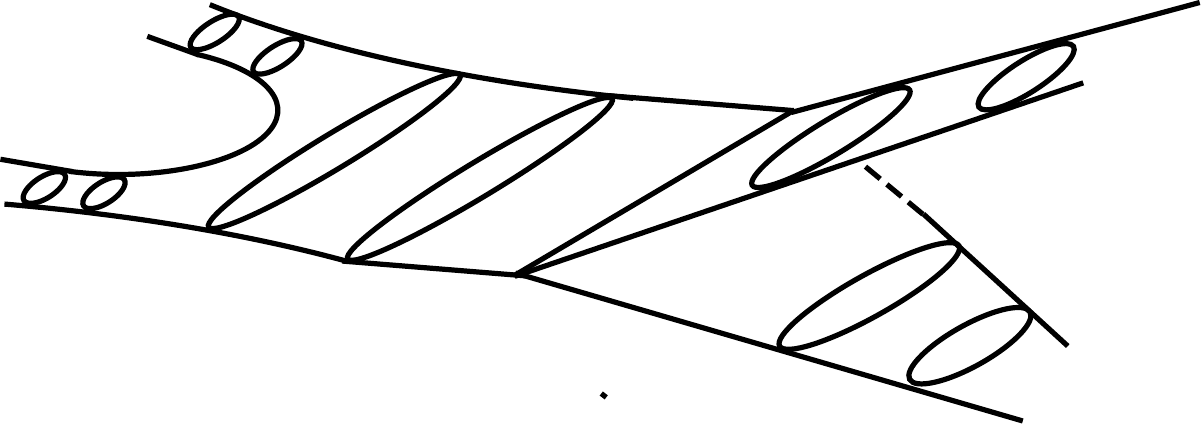}
} \ \ \ \ 
\subfloat[]{
\includegraphics[width=1.8in]{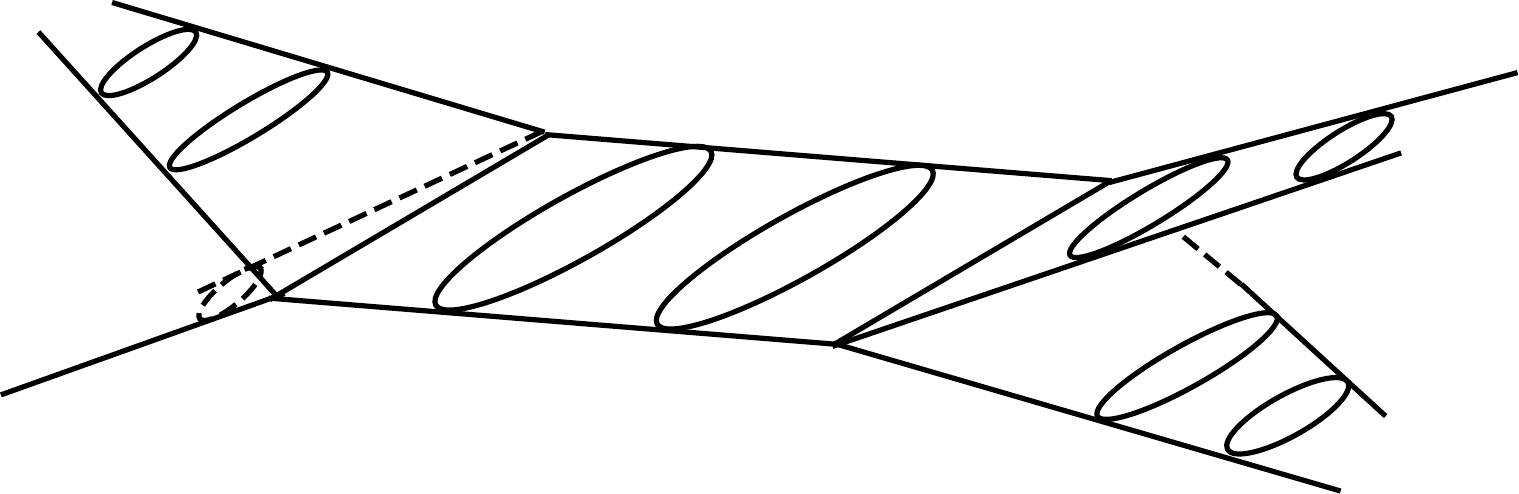}
}
\caption{The two different kinds of vertices between three ladders, as was shown in Fig.~\ref{FigIntro1}, lead to three kinds of contributions to the eight-point function.}  \label{fig:eight}
\end{figure} 
We begin by classifying which Feynman diagrams will appear, at leading nontrivial order in $1/N$, for a $2p$-point correlation function of fermions. As noted in \cite{Witten:2016iux},  for any large $N$ theory, this is found by drawing all diagrams contributing to the vacuum energy and successively considering all cuts of the propagators. A single cut gives a diagram contributing to the two-point function. Two cuts gives a contribution to the four-point function, and so on.

The diagrams contributing to the two-point function consist entirely of melons. This is true of SYK, as well as  variants of SYK \cite{GR1, Gu:2016oyy, Fu:2016vas, Davison:2016ngz, Murugan:2017eto, Song:2017pfw} and their extensions, and of  certain tensor models \cite{Gurau:2009tw, Bonzom:2011zz, Witten:2016iux, Klebanov:2016xxf,  Klebanov:2017nlk, Bulycheva:2017ilt, Choudhury:2017tax} and their extensions. A cut of a melon diagram gives a ladder diagram, contributing to the four-point function. Starting with the four-point function, we have two nonequivalent options of which lines we may cut. We may either cut a melon along a rail, giving  a planar diagram contribution to the six-point function, or we may cut a melon that is along a rung, giving a ``contact'' diagram contribution. 
Proceeding to the eight-point function, there are now four possible cuts: two from a cut of the planar  six-point diagram, and two from a cut of the contact six-point diagram. In particular, for the planar diagram, a cut of a melon along a rail leads to a planar diagram contribution to the eight-point function, as in Fig.~\ref{fig:eight} (a), while a cut of a melon along a rung leads to a mixed planar/contact eight-point diagram, as in Fig.~\ref{fig:eight} (b). For the contact six-point  diagram, a cut of a melon along a rail also leads to a mixed planar/contact eight-point diagram, while a cut of a melon along a rung leads to a contact/contact eight-point diagram, as in Fig.~\ref{fig:eight} (c). 
The same structure will  persist for higher-point functions.

\subsection{Outline}
\begin{figure}[t]
\centering
\includegraphics[width=5.5in]{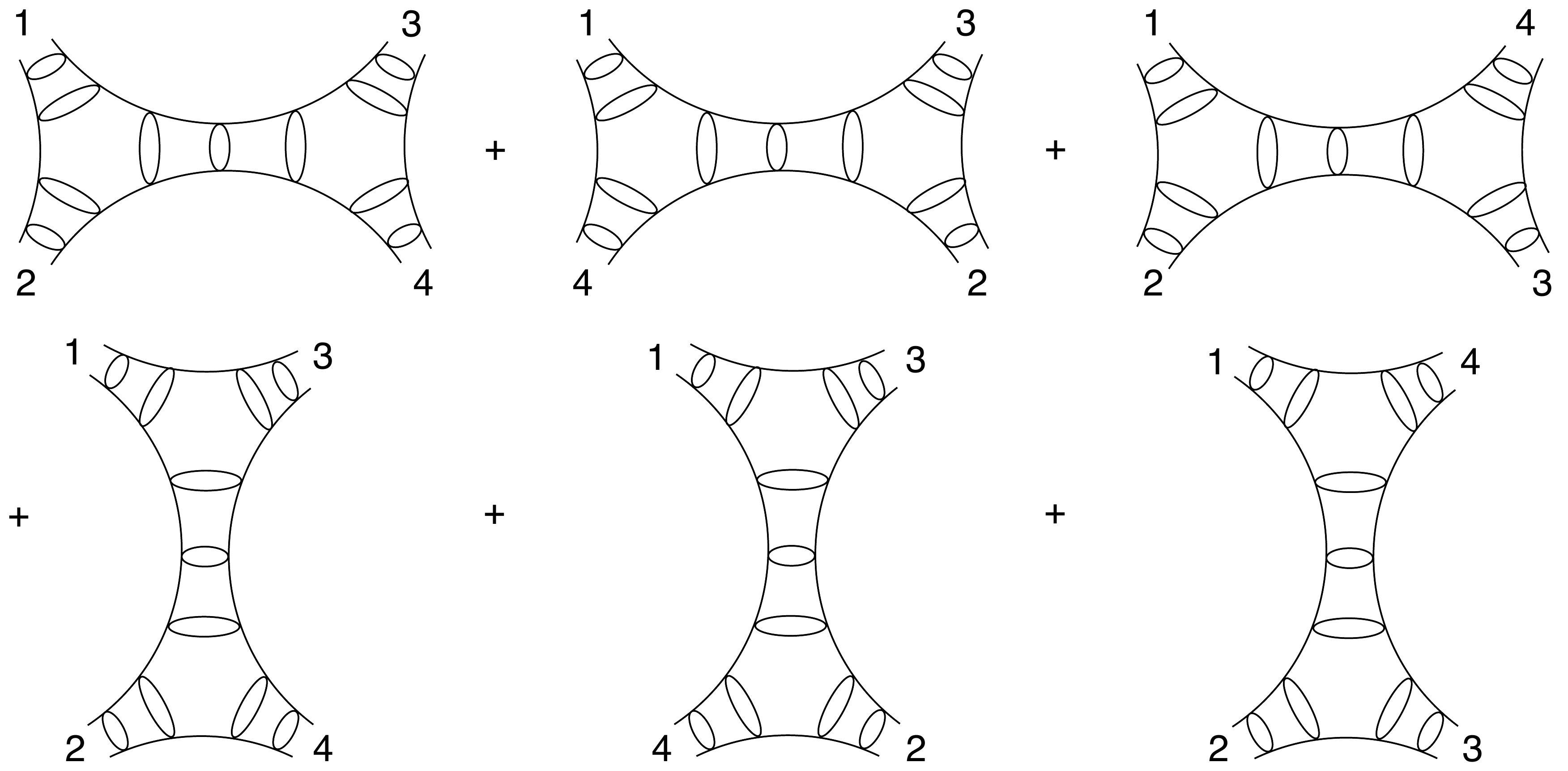}
\caption{Some of the diagrams contributing to the eight-point function. We must include diagrams with melons exchanged in both directions (first and second line). } \label{Fig8pt}
\end{figure}

Having established the basic structure of the Feynman diagrams contributing to the eight-point function, we now list more precisely all the diagrams that will need to summed.  Let $\mE_s(\tau_1, \ldots, \tau_8)$ denote the Feynman diagram shown previously in Fig.~\ref{FigIntro3}, and let $\langle \mO_1(\tau_1) \cdots \mO_4(\tau_4)\rangle_s$ denote its contribution to the four-point function of the $\mO_i$.  In addition, let $\mE_s^0(\tau_1, \ldots, \tau_8)$, and correspondingly $\langle \mO_1(\tau_1) \cdots \mO_4(\tau_4)\rangle_s^0$, denote similar Feynman diagrams, but only the planar one, and with no exchanged melons, as will be illustrated later in Fig.~\ref{FigIntro4}. Then, the four-point function of the $\mO_i$ is, 
\bea \nonumber
\langle \mO_1(\tau_1) \cdots \mO_4(\tau_4) \rangle&=&\Big(\langle \mO_1(\tau_1) \cdots \mO_4(\tau_4) \rangle_s + (2\leftrightarrow 3) + (2\leftrightarrow 4)\Big) \\
&-& \frac{1}{2}\Big(\langle \mO_1(\tau_1) \cdots \mO_4(\tau_4) \rangle_s^0 + (2\leftrightarrow 3) + (2\leftrightarrow 4)\Big)~. \label{4ptAll}
\eea
Finally,  there is an additional diagram, which is discussed in Appendix.~\ref{sec:con}, and 
consists of four fermion four-point functions glued to the same melon.

Let us explain why (\ref{4ptAll}) is correct. If we, for the moment, focus on only the planar diagrams, then all the diagrams that need to be summed are shown in Fig.~\ref{Fig8pt}. The three classes of diagrams in the first line are  the three different  channels. The diagrams in the second line must be included as well - these are similar to the diagrams on the first line, except now the exchanged melons are going in the other direction. For the diagrams in which there are no exchanged melons, the top and bottom diagrams are the same, and we should only include one of them. 
One can see that  $\mE_s(\tau_1, \ldots, \tau_8)$  corresponds  to the sum of the first and third diagrams on the top line of Fig.~\ref{Fig8pt}. The reason it corresponds to two sets of diagrams is because the fermion four-point function is antisymmetric under interchange of the last two (or the first two) fermions: in summing the ladder diagrams, there were two sets of diagrams, coming from adding rungs to the two terms in $\mF_0$ in (\ref{F0}). The sum of the three terms on the first line of (\ref{4ptAll}) accounts for all six terms in Fig.~\ref{Fig8pt}. The second line of (\ref{4ptAll}) compensates for the double counting of diagrams in which no melons are exchanged. Finally, in addition to the diagrams shown in Fig.~\ref{Fig8pt}, there are diagrams in which the cubic vertex is contact rather than planar, such as those in Fig.~\ref{fig:eight}; these have already been taken into account in (\ref{4ptAll}), as the shaded circle in the diagram in Fig.~\ref{FigIntro3} includes both such vertices, see Fig.~\ref{FigIntro1}. 

We now turn to computing $\langle \mO_1(\tau_1) \cdots \mO_4(\tau_4)\rangle_s$.

\subsection{Splitting and recombining conformal blocks} \label{Sec42}
\begin{figure}[t]
\centering
\includegraphics[width=1.5in]{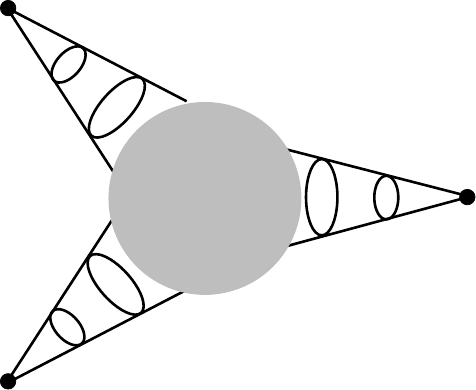}
\caption{The three-point function of bilinears. This looks like the fermion six-point function, with fermions brought together in pairs. } \label{Fig6to30}
\end{figure}

The eight-point function of the fermions can be written as, 
\be \label{Eight} 
\frac{1}{N^4} \sum_{i_1,\ldots, i_4} \langle \chi_{i_1} (\tau_1) \chi_{i_1} (\tau_2) \chi_{i_2} (\tau_3) \chi_{i_2}(\tau_4) \chi_{i_3} (\tau_5) \chi_{i_3} (\tau_6)\chi_{i_4} (\tau_7) \chi_{i_4} (\tau_8)\rangle = \ldots  + \frac{1}{N^3} \mE(\tau_1, \ldots, \tau_8) + \ldots~,
\ee
where $\mE$ is the lowest order term in $1/N$ that contains fully connected diagrams. 

In this section we will study the contribution to the eight-point function that was shown in Fig.~\ref{FigIntro3}, denoted by $\mE_s(\tau_1, \ldots, \tau_8)$. 
This consists of two six-point functions glued together. We can write a general expression for the six-point function, containing a piece $\mS^{\text{core}}$ which encodes the details of the interactions, attached to three external four-point functions,
\be \label{SFFF}
\!\mS(\tau_1, \ldots, \tau_6)\! =\!\! \int d\tau_{a_1} \cdots d\tau_{a_6} \mF(\tau_1, \tau_2, \tau_{a_1}, \tau_{a_2}) \mF(\tau_3, \tau_4, \tau_{a_3}, \tau_{a_4})\mF(\tau_5, \tau_6, \tau_{a_5}, \tau_{a_6}) \mS^{\text{core}}(\tau_{a_1}, \ldots, \tau_{a_6})~.
\ee
Pictorially, $\mS^{\text{core}}$ is the shaded circle that appeared before in Fig.~\ref{FigIntro3}. For SYK, $\mS^{\text{core}}$ is pictorially defined in Fig.~\ref{FigIntro1}. More explicitly, we  found in Sec.~\ref{sec:3pt} that $\mS^{\text{core}}$ is, 
\be \nonumber
\mS^{\text{core}} = (q-1)(q-2)J^2 G(\tau_{a_1 a_2})^{q-3} \delta(\tau_{a_1 a_3} )\delta( \tau_{a_1 a_5}) \delta(\tau_{a_2 a_4}) \delta(\tau_{a_2 a_6}) +D(\tau_{a_2 a_5}) D(\tau_{a_4 a_1})D(\tau_{a_6 a_3})~,
\ee
however the explicit form of $\mS^{\text{core}}$ is not relevant for the argument that follows. 

Employing the same logic as used perviously in the derivation of the three-point function of bilinears from the six-point function of fermions, and utilizing the conformal block structure of $\mF$ given in (\ref{mFmC}), we may write for the three-point function, see Fig.~\ref{Fig6to30},
\begin{multline}
\langle \mO_1(\tau_1) \mO_2(\tau_2) \mO_3(\tau_3) \rangle = \int d\tau_{a_1} \cdots d\tau_{a_6}  \mS^{\text{core}}(\tau_{a_1}, \ldots, \tau_{a_6})
\\  \label{OOO}
\langle \mO_1(\tau_1) \chi(\tau_{a_1} )\chi(\tau_{a_2})\rangle \langle\mO_2(\tau_2)\chi( \tau_{a_3})\chi( \tau_{a_4})\rangle \langle\mO_3(\tau_3) \chi(\tau_{a_5})\chi( \tau_{a_6})\rangle~. 
\end{multline}

\begin{figure}[t]
\centering
\includegraphics[width=5.5in]{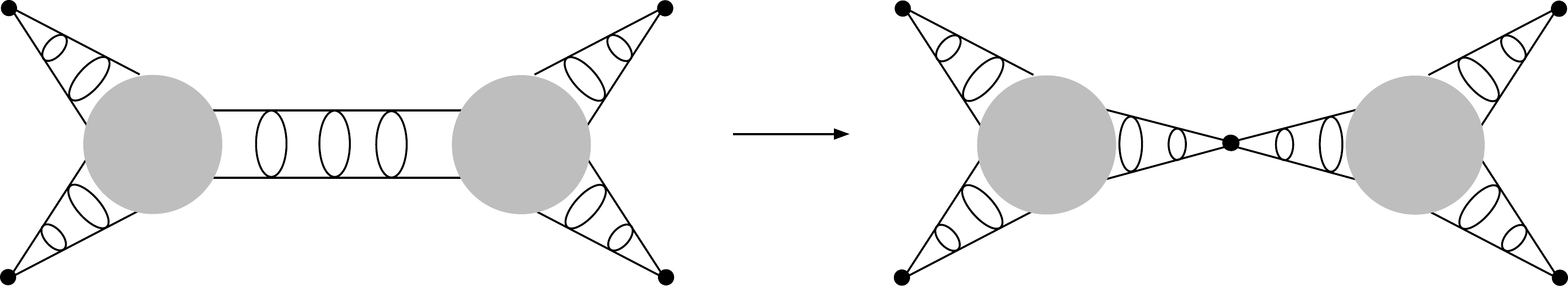}
\caption{An important step in computing the bilinear four-point function is to use the split representation for the intermediate four-point function, as was shown previously in Fig.~\ref{figladderSplit}.} \label{8split}
\end{figure}

With this building block, we construct $\mE_s$. As shown in Fig.~\ref{FigIntro3}, gluing together two six-point functions gives, 
\bea \nonumber
\mE_{s}(\tau_1, \ldots, \tau_8)\!\! &=&\!\!\!\! \int\!d\tau_{a_1} \cdots d\tau_{a_8} d\tau_{b_1}\cdots d\tau_{b_4} 
\mS^{\text{core}}(\tau_{a_1}, \ldots, \tau_{a_4}, \tau_{b_1}, \tau_{b_2}) \mS^{\text{core}}(\tau_{a_5}, \ldots,\tau_{a_8}, \tau_{b_3}, \tau_{b_4}) \\ \nonumber
&&\!\!\!\! \mF(\tau_1, \tau_2, \tau_{a_1}, \tau_{a_2}) \mF(\tau_3, \tau_4, \tau_{a_3}, \tau_{a_4}) 
 \mF(\tau_{b_1}, \ldots,  \tau_{b_4}) \mF(\tau_5, \tau_6, \tau_{a_5}, \tau_{a_6}) \mF(\tau_7, \tau_8, \tau_{a_7}, \tau_{a_8})
\eea
Again using (\ref{mFmC}), the four-point function of the $\mO$ is thus,
\begin{multline} \label{OOOO}
\hspace{-.5cm}\langle \mO_1(\tau_1) \cdots \mO_4(\tau_4)\rangle_s =\! \int\!d\tau_{a_1} \cdots d\tau_{a_8} d\tau_{b_1}\cdots d\tau_{b_4} 
\mS^{\text{core}}(\tau_{a_1}, \ldots, \tau_{a_4}, \tau_{b_1}, \tau_{b_2}) \mS^{\text{core}}(\tau_{a_5}, \ldots,\tau_{a_8}, \tau_{b_3}, \tau_{b_4}) \\ 
 \hspace{-.3cm}\langle\mO_1(\tau_1 ) \chi(\tau_{a_1})\chi(\tau_{a_2})\rangle \langle \mO_2 (\tau_2) \chi( \tau_{a_3}) \chi( \tau_{a_4}) \rangle
 \mF(\tau_{b_1}, \ldots,  \tau_{b_4}) \langle \mO_3(\tau_3)\chi( \tau_{a_5})\chi( \tau_{a_6})\rangle \langle \mO_4(\tau_4)\chi( \tau_{a_7})\chi( \tau_{a_8})\rangle~.
\end{multline}
The fermion four-point function is a sum of conformal blocks, hypergeometric functions, and this integral is clearly challenging to evaluate directly in position space. The crucial step  is to use the more elementary representation of the four-point function, in terms of the complete basis $\Psi_h(x)$ of eigenfunctions of the conformal Casimir, as given in (\ref{4ptContour}),
\be \label{Fin2}
\mF(\tau_{b_1}, \ldots, \tau_{b_4}) =\frac{1}{2} \int_{\mC}\frac{d h}{2\pi i}\frac{ \rho (h) }{c_h c_{1- h}} \int d\tau_0\, \langle \chi(\tau_{b_1}) \chi(\tau_{b_2}) \mO_{h}(\tau_0)\rangle \langle \mO_{1-h}(\tau_0) \chi(\tau_{b_3}) \chi(\tau_{b_4})\rangle~,
\ee
where we have made use of the representation (\ref{PsiIntr}) of $\Psi_h(x)$ in terms of a product of a three-point function involving  $\mO_h$ and  a three-point function involving the shadow $\mO_{1- h}$.

With this representation of the fermion four-point function, upon comparing with the expression (\ref{OOO}) for the three-point function of $\mO$, we may write (\ref{OOOO}) as an integral involving a three-point function of the external ingoing $\mO_1$ and $\mO_2$ and the exchanged $\mO_h$, along with a three-point function involving the shadow $\mO_{1-h}$ and the external outgoing $\mO_3$ and $\mO_4$,  see Fig.~\ref{8split},
\be \label{exchangeF}
\langle \mO_1(\tau_1) \cdots \mO_4(\tau_4)\rangle_s =\frac{1}{2} \int_{\mC}\frac{d h}{2\pi i}\frac{ \rho (h) }{c_h c_{1- h}} \int d\tau_0 \langle \mO_1(\tau_1) \mO_2(\tau_2) \mO_h(\tau_0)\rangle \langle \mO_{1- h}(\tau_0) \mO_3(\tau_3) \mO_4(\tau_4)\rangle~.
\ee
The integral over $\tau_0$ over this product of three-point functions give a sum of a conformal block and its shadow, now for external operators $\mO_1, \ldots, \mO_4$, see Appendix.~\ref{blocks},
\be \label{OOOOb}
\langle \mO_1(\tau_1) \cdots \mO_4(\tau_4)\rangle_s =\frac{1}{2} \int_{\mC}\frac{d h}{2\pi i}\frac{ \rho (h) }{c_h c_{1- h}}\, c_{1 2 h}\, c_{3 4\, 1-h}\[ \beta(h, h_{34}) \mF_{1234}^h(x) + \beta(1-h, h_{12}) \mF_{1234}^{1-h} (x)\]~.
\ee
Here $\mF_{1234}^h(x)$ is the conformal block for external operators $\mO_1, \ldots, \mO_4$ and exchanged operator $\mO_h$, 
\be \label{CPW2}
\mF_{1234}^h(x) = \Big|\frac{\tau_{24}}{\tau_{14}}\Big|^{h_{12}}\Big|\frac{\tau_{14}}{\tau_{13}}\Big|^{h_{34}} \frac{1}{|\tau_{12}|^{h_1 +h_2}|\tau_{34}|^{h_3 + h_4}}x^{h}\, {}_2 F_1 (h- h_{12}, h+ h_{34}, 2h, x)~,
\ee
while, 
\be \label{betaD2}
\beta(h,\Delta) =\sqrt{\pi} \frac{\Gamma(\frac{h+\Delta}{2}) \Gamma(\frac{h-\Delta}{2})}{\Gamma(\frac{1-h+\Delta}{2})\Gamma(\frac{1-h-\Delta}{2})}\frac{\Gamma(\frac{1}{2}- h)}{\Gamma(h)}~.
\ee
Also, to be clear, $c_{3 4\, 1-h}$ denotes the coefficient of the three-point function of operators of dimensions $h_3$, $h_4$, and $1-h$: $\langle \mO_3\mO_4 \mO_{1- h}\rangle$.
The contour $\mC$ consists of a line parallel to the imaginary axis, $h = \frac{1}{2} + i s$, as well as the circles around $h=2n$ for $n\geq 1$. We consider each piece separately. Starting with the contribution from the line,  and changing variables $h\rightarrow 1-h$ for the second term in (\ref{OOOOb}), we get, 
\begin{multline} \label{OOOOline}
\langle \mO_1(\tau_1) \cdots \mO_4(\tau_4)\rangle_s \supset \frac{1}{2} \int_{\frac{1}{2} + is}\frac{d h}{2\pi i}\rho (h)\frac{c_{1 2 h} c_{ 3 4 h}}{c_h^2}\,\\
 \[\frac{c_{3 4\, 1-h}}{c_{3 4 h} }\frac{c_h}{c_{1-h}} \beta(h, h_{34}) - \frac{c_{12\, 1-h}}{c_{1 2 h }}\frac{c_h}{c_{1-h}}\tan^2 \frac{\pi h}{2} \beta(h, h_{12}) \]  \mF_{1234}^h(x)~.
\end{multline}
We now use the following relation between the coefficient $c_{ 1 2 h}$ of the  three-point function $ \langle \mO_1 \mO_2 \mO_h\rangle$ and that of $c_{1 2\, 1-h}$, involving the shadow, $\langle \mO_1 \mO_2 \mO_{1-h}\rangle$,
\be \label{c123Rel}
\frac{c_{1 2\, 1 - h}}{c_{1-h}} \frac{\Gamma(\frac{1- h}{2})^2}{\Gamma(\frac{1-h + h_{12}}{2})\Gamma(\frac{1-h - h_{12}}{2})} = \frac{ c_{1 2 h}}{c_h} \frac{\Gamma(\frac{h}{2})^2}{\Gamma(\frac{h+ h_{12}}{2}) \Gamma(\frac{h-h_{12}}{2})}~.
\ee
Using the explicit form of the $c_{123}$ for SYK found in Sec.~\ref{sec:3pt}, one can verify that  this relation is satisfied. However, it should be true more generally. 
The contribution of the line integral (\ref{OOOOline}) now simplifies to become, 
\be
\langle \mO_1(\tau_1) \cdots \mO_4(\tau_4)\rangle_s \supset \int_{\frac{1}{2} + i s}\frac{d h}{2\pi i}\frac{\rho (h)}{c_h^2}  \frac{\Gamma(h)^2}{\Gamma(2h)}\,c_{1 2 h} c_{ 3 4 h}\, \mF_{1234}^h(x)~.
\ee
Now consider the portion of the contour integral (\ref{OOOOb}) consisting of the circles wrapping $h = 2n$. Noting that for $h=2n$, $\beta(1-h, h_{12})$ vanishes,  and as a result of (\ref{c123Rel}), 
\be
\beta(h, h_{34})\frac{c_{3 4\, 1- h}}{c_{ 3 4\, h}} \frac{c_h}{c_{1-h}}  = \(1 + \frac{1}{\cos \pi h}\) \frac{\Gamma(h)^2}{\Gamma(2h)}~.
\ee
For $h=2n$, the factor in parenthesis becomes $2$, and so the integrand for the portion of the contour consisting of the circles is the same as for the line piece of the contour. Recombining the two gives a single  expression, 
\be \label{RESULT}
\langle \mO_1(\tau_1) \cdots \mO_4(\tau_4)\rangle_s = \int_{\mC}\frac{d h}{2\pi i}\frac{\rho (h)}{c_h^2}  \frac{\Gamma(h)^2}{\Gamma(2h)}\,c_{1 2 h} c_{ 3 4 h}\, \mF_{1234}^h(x)~.
\ee
This is one of our main results. It is simple and intuitive.

\subsection{Combining ingredients and comments}

\subsubsection*{Universality}
It is instructive to recall the form of the three-point function, as written in (\ref{c123I}), $c_{123} = c_1 c_2 c_3 \mI_{123}$, which separates the $c_i$, which arise from summing the ladders, from $\mI_{123}$ which arises from gluing the ladders. With this, the $s$-channel piece of the four-point function takes the form, 
\be \label{RESULTv2}
\langle \mO_1(\tau_1) \cdots \mO_4(\tau_4)\rangle_s =c_1 c_2 c_3 c_4 \int_{\mC}\frac{d h}{2\pi i}\rho (h) \frac{\Gamma(h)^2}{\Gamma(2h)}\,\mI_{1 2 h} \mI_{ 3 4 h}\, \mF_{1234}^h(x)~.
\ee
The four-point function, as well as all higher-point correlation functions, are analytic functions of the fermion dimension $\Delta$ and the $\mO_i$ dimensions $h_i$. As one flows from weakly coupled cSYK to strongly coupled cSYK, the $h_i$ change, or, as one changes the order of the interaction, $q$, the fermion dimension $\Delta = 1/q$ changes. To the extent that $h_i$ and $\Delta$ are close for these different theories, Eq.~\ref{RESULTv2} shows that the four-point functions will also be close, and, through a simple generalization, so will all correlation functions.~\footnote{The statement is true to the extent that one can neglect the additional contact diagrams discussed in Appendix.~\ref{sec:con}.} A useful case is when all the operators have large dimensions,   $h_i \gg 1$, as in this limit the  anomalous dimensions at strong coupling are small, $h_i \approx 2\Delta + 2n + 1$. This allows for the study of this universal sector of the theory through study of weakly coupled cSYK, which is just generalized free field theory, and will be discussed in Sec.~\ref{FREE}. 


\subsubsection*{Closing the contour}
Closing the contour in (\ref{RESULT}) will turn the integral over conformal blocks in $h$-space into a sum over conformal blocks.  To do this, we need to look at the singularity structure of the integrand, for $h$ in the right-half complex plane. For simplicity, we assume none of the $h_i$ are equal. 

The first term in the integrand, $\rho(h)$, has poles at the dimensions of the single-trace operators, the $h=h_n$ for which $k_c(h_n) = 1$. Next, let us look at the other term, involving the three-point function coefficients, $c_{h 1 2 }/c_h$. The contact contribution $c_{h 12}^{(1)}/c_h$, see Eq.~\ref{c1231}, has poles at $h = h_1 + h_2 + 2 n$, as well as at $h = 2n+1$.~\footnote{It may naively appear that there are also poles at $h=2n$, but in fact there aren't.} The planar contribution $c_{h 1 2}^{(2)}/c_h$, see Eq.~\ref{c1232}, has poles at $h = h_1 + h_2 + 2n$ as well as $h = 2n + 1$, and $h = 3 - 2\Delta+ 2n$.~\footnote{It is most convenient to look at $c_{1 2 3}^{(2)}$ found in Sec.~\ref{planar} as a function of $h_1$ (as it is symmetric under permutations, we are free to do this). Then, all of the poles in $h_1$ arise form the gamma functions in the $\alpha_i$; the generalized hypergeometric functions, as functions of $h_1$, do not have any poles.} 
The poles at $h= 2n+1 $ and $h = 3- 2\Delta+2n$ are  irrelevant, since $\rho(h\! =\! 2n\!+\!1)\!=\! 0$ and $\rho(h\!=\! 3\!-\! 2 \Delta\!+\!2n)\! =\!0$.~\footnote{In fact, this is a bit subtle. One may notice that even though $\rho(h) = 0$ for $h = 2n+1$ or $h = 3- 2\Delta+2n$, one would still have a pole at these $h$, because the product  $c_{12h} c_{34 h}$ gives rise to a double pole at these values. However, this  divergence is an artifact of an earlier step, in which we exchanged the order of the $h$ contour integral and the time integrals. More simply stated, what we should really do is instead of the contour $\mC$ in the fermion four-point function in (\ref{Fin2}), we should use a contour $\mC'$ which excludes $h = 2n+1$ and $h = 3- 2\Delta+2n$; since the integrand vanishes at these values of $h$, this is a justified replacement. }

Therefore, as  expected, $\langle \mO_1 \cdots \mO_4\rangle_s$ is a sum of single-trace and double-trace conformal blocks, 
\bea \label{OOOOs3}
\hspace{-1cm}\langle \mO_1(\tau_1) \cdots \mO_4(\tau_4)\rangle_s  &=&\sum_{h= h_n} c_{12 h} c_{3 4 h}\, \mF_{1234}^h(x)\\ \nonumber
& +& \sum_{n=0}^{\infty} - \text{Res }\Big[\frac{c_{12h}}{c_h}\Big]_{h = h_1 + h_2 +2n}\, \Big[ \rho (h)\frac{\Gamma(h)^2}{\Gamma(2h)}\frac{c_{3 4 h}}{c_h} \mF_{1234}^{h}(x)\Big]_{h = h_1 + h_ 2+2n}\\  \nonumber
&+&\sum_{n=0}^{\infty} - \text{Res }\Big[\frac{c_{3 4 h}}{c_h}\Big]_{h = h_3 + h_4 +2n}\, \Big[\rho (h) \frac{\Gamma(h)^2}{\Gamma(2h)}\frac{c_{1 2  h}}{c_h} \mF_{1234}^{h}(x)\Big]_{h = h_3 + h_ 4+2n}~.
\eea
In Appendix.~\ref{largeQ} we write the terms on the second and third line more explicitly, and also study their large $q$ limit. 

Let us recall why we expect that the four-point function of bilinears, at order $1/N$,  is composed of single-trace and double-trace conformal blocks.  On general grounds the OPE is of the form \cite{Heemskerk:2009pn}, 
\be
\mO_1 \mO_2 \sim \frac{1}{\sqrt{N}} c_{12 h} \mO_h + d_{ 1 2 [12]_n}^0 [\mO_1 \mO_2]_n + \frac{1}{N} d_{1 2 [i j]_n}^1 [\mO_i \mO_j]_n + \ldots, 
\ee
where $[\mO_i \mO_j]_n$ denotes a double-trace operator, schematically of the form, $\mO_i \partial^{2n } \mO_j$, and the dots denote terms that are higher order in $1/N$. 
If we look at the four-point function, and apply the OPE to $\mO_1 \mO_2$ and to $\mO_3 \mO_4$ then we schematically get, for the $1/N$ piece,
\be \nonumber
\hspace{-.5cm}\langle \mO_1 \cdots \mO_4\rangle \sim \frac{1}{N} \Big(c_{1 2 h} c_{3 4 h} \langle \mO_h \mO_h\rangle +d^{0}_{12 [12]_n} d^{1}_{34 [12]_n}\langle [\mO_1 \mO_2]_n  [\mO_1 \mO_2]_n\rangle + d^{0}_{34 [34]_n} d^{1}_{12 [34]_n}\langle [\mO_3 \mO_4]_n  [\mO_3 \mO_4]_n\rangle\Big)~.
\ee
This structure is precisely reflected in the actual result, (\ref{OOOOs3}).

\subsubsection*{Cross-channel}
As stated in Eq.~\ref{4ptAll}, in addition to the sum of the $s$-channel Feynman diagrams, given by (\ref{RESULT}), we must also include the $t$-channel and $u$-channel diagrams. The sum of the $t$- channel diagrams is simply (\ref{RESULT}), but with $h_2 \leftrightarrow h_3$, and $\tau_2 \leftrightarrow \tau_3$ and correspondingly for the cross-ratio $x\rightarrow 1/x$. The sum of the $u$-channel diagram is (\ref{RESULT}), but with $h_2 \leftrightarrow h_4$, and $\tau_2 \leftrightarrow \tau_4$ and correspondingly $x \rightarrow 1- x$. 

It is straight forward to combine these three contributions into a single expression suited to performing the OPE. To do this, one should use (\ref{OOOOb}), which has an integral over the conformal block plus its shadow, $\mB_{1234}^h(x)= \beta(h, h_{34}) \mF_{1234}^h(x) + \beta(1-h, h_{12}) \mF_{1234}^{1-h} (x)$. The range of $h$ is  $\frac{1}{2} - i \infty < h < \frac{1}{2} + i \infty $ and $h = 2n$. Since these form a complete basis, one could expand $\mB_{1234}^h( 1-x)$ and $\mB_{1234}^h(1/x)$ in terms of the basis of $\mB_{1234}^{\tilde{h}}(x)$. This would be analogous to the computation in \cite{Hogervorst:2017sfd}, though slightly different since there one has a  linear combination  of the block plus shadow block that is different from $\mB_{1234}^h$.

Combining these three channels is actually unnecessary for us, since, as we will see later, in the bulk computation of the four-point function, there are three types of Witten diagrams, $s$, $t$, and $u$ channel, related to the SYK $s$, $t$, $u$ channel Feynman diagrams.~\footnote{It may be of interest to do this calculation anyway, in order to compute the $1/N$ corrections to the OPE coefficients. One should note, however, that we are only computing the connected piece of the bilinear four-point function. In order to compute the $1/N$ anomalous dimensions of operators, one needs to compute the disconnected diagrams as well: in particular, one needs the loop corrections to the fermion four-point function.}

\subsubsection*{Subtracting a planar} \label{Sec43}
\begin{figure}[t]
\centering
\includegraphics[width=1.5in]{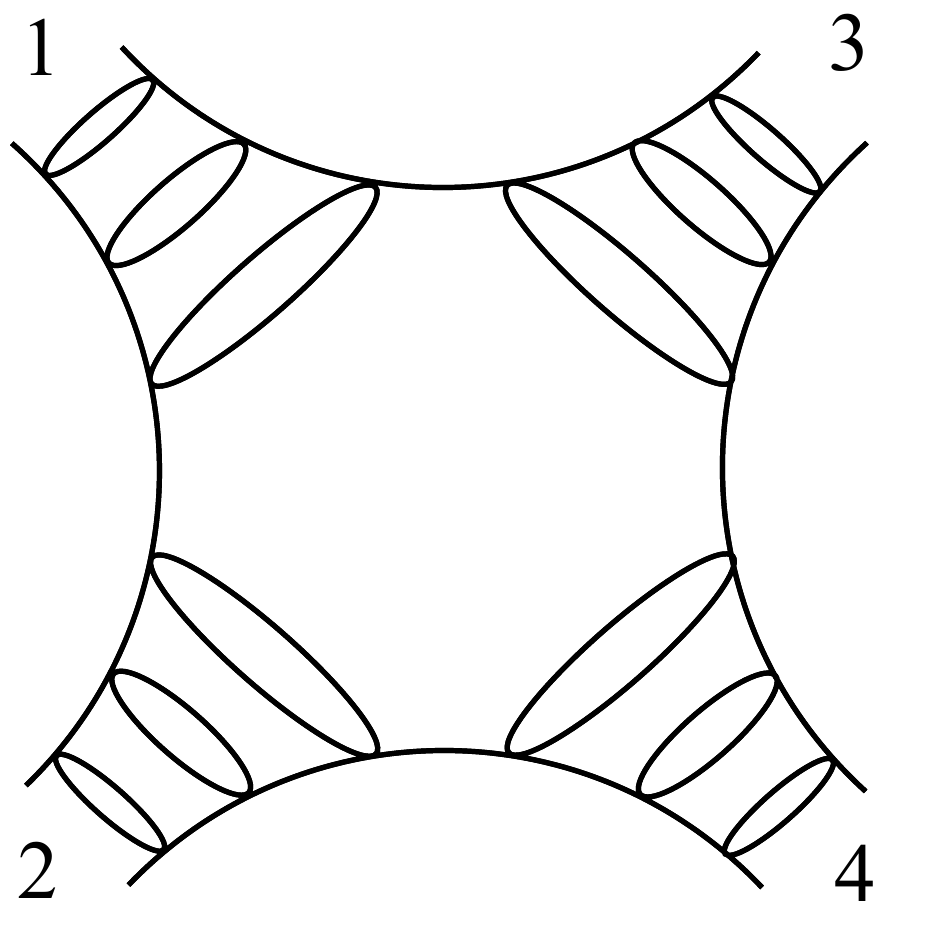}
\caption{A contribution to the eight-point function. This was included in both lines shown before in Fig.~\ref{Fig8pt}, and so must be subtracted due to double counting. } \label{FigIntro4}
\end{figure} 
The first term on the second line of (\ref{4ptAll}) is the diagram shown in Fig.~\ref{FigIntro4}. This is similar to the sum of the $s$-channel exchange diagrams we already computed, the only difference being that it only sums planar diagrams, and that instead of the full fermion four-point function $\mF$ appearing in the exchange, one has the free fermion four-point function, $\mF_0$. This allows us to immediately write the answer, 
\be \label{OOOOs0}
\langle \mO_1(\tau_1) \cdots \mO_4(\tau_4)\rangle_s^0 = \int_{\mC}\frac{d h}{2\pi i}\frac{\rho^0 (h)}{c_h^2}  \frac{\Gamma(h)^2}{\Gamma(2h)}\,c_{1 2 h}^{(2)} c_{ 3 4 h}^{(2)}\, \mF_{1234}^h(x)~.
\ee
Closing the contour yields a sum of both single-trace and double-trace conformal blocks. The single-trace blocks are for operators of dimension $2\Delta +2n +1$, which serve to cancel the same blocks that arise from expanding the exchange diagrams in the cross-channel. The double-trace blocks are again for operators of the type $[\mO_1 \mO_2]_n$ and $[\mO_3 \mO_4]_n$. 

\subsubsection*{Mellin space}

It is sometimes useful to represent the four-point function in Mellin space, see Appendix~\ref{app:Mellin} for our conventions. 
In order to find the Mellin transform of $\langle \mO_1 \cdots \mO_4\rangle_s$, denoted by $M_s(h_i, \gamma_{12})$, it is most convenient to use the form of the expression in (\ref{exchangeF}). The integral appearing there is denoted by $\mB_{1234}^h$ in Appendix~\ref{blocks}, see Eq.~\ref{SbS}, and its Melin transform, $\tilde{M}_{1234}^{h}(\gamma_{12})$, is given in (\ref{Mtilde}). Therefore $M_s (h_i, \gamma_{12})$ is the contour integral,  
\be
M_s(h_i, \gamma_{12}) = \frac{1}{2} \int_{\mC}\frac{d h}{2\pi i}\frac{ \rho (h) }{c_h c_{1- h}}\, c_{1 2 h}\, c_{3 4\, 1-h}\, \tilde{M}_{1234}^h(\gamma_{12})~.
\ee
Similarly, the Mellin transform of $\langle \mO_1\cdots \mO_4\rangle_s^0$ is, 
\be
M_s^0(h_i, \gamma_{12}) = \frac{1}{2} \int_{\mC}\frac{d h}{2\pi i}\frac{ \rho^0 (h) }{c_h c_{1- h}}\, c_{1 2 h}^{(2)}\, c_{3 4\, 1-h}^{(2)}\, \tilde{M}_{1234}^h(\gamma_{12})~.
\ee

Due to the complexity of $c_{12 3}$, these expressions are not in themselves especially enlightening. In Sec.~\ref{FREE} we will study the limit of $h_ i \gg 1$, in which the full four-point function, as well as its Mellin transform, significantly simplify.

\section{Higher-Point Correlation Functions} \label{sec:ppt}
\begin{figure}[t]
\centering
\includegraphics[width=4in]{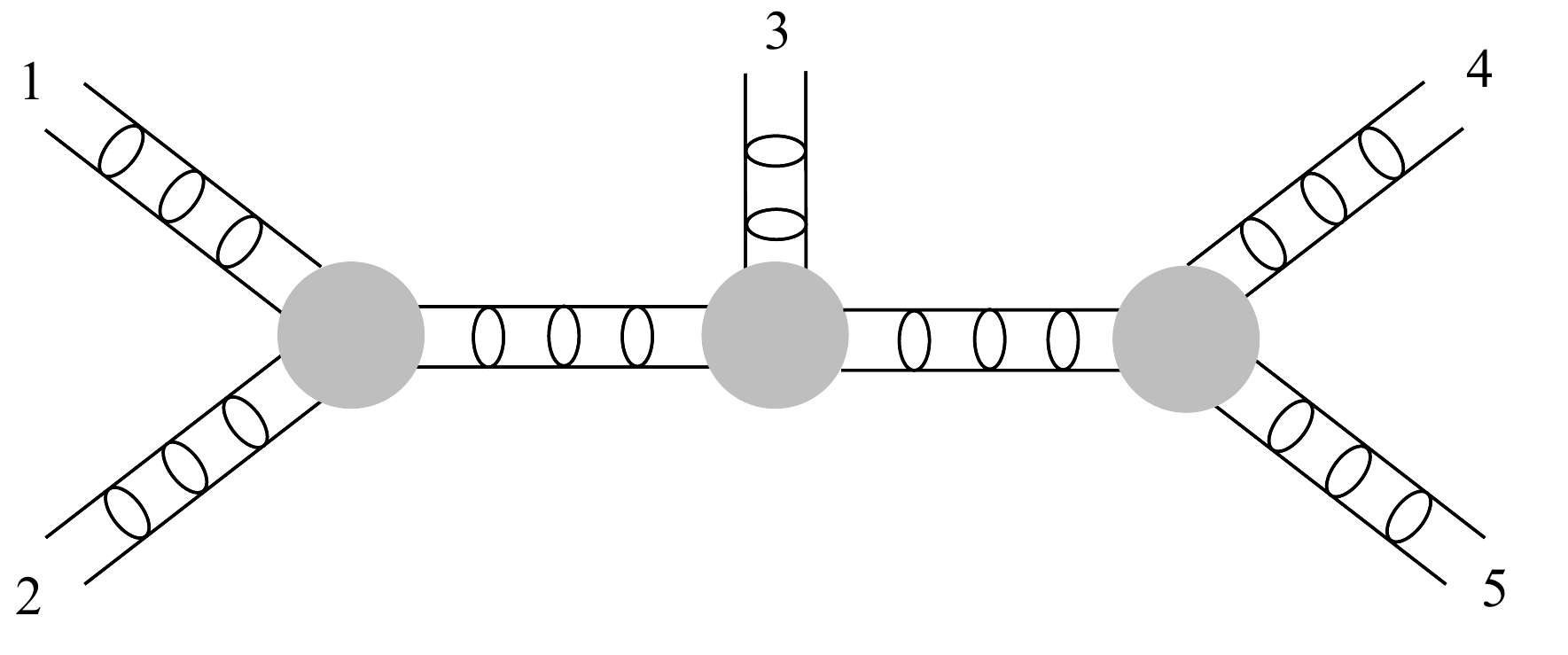}
\caption{ A contribution to the ten-point function.} \label{Fig5pt}
\end{figure} 
In the previous section we computed the bilinear four-point function. It is straightforward to generalize to higher-point functions. These will be expressed in terms of contour integrals involving the $\rho(h)$ from summing ladders in Sec.~\ref{ladders}, the $c_{1 2 3}$ computed in Sec.~\ref{sec:3pt}, and  higher-point conformal blocks.

For instance, consider a fermion ten-point function. An example of a class of diagrams that contribute is shown in Fig.~\ref{Fig5pt}. 
To compute such diagrams, we use the same method as in the previous section, writing the intermediate fermion four-point functions (of which there are now two) in the form given by Eq.~\ref{Fin2}. 
The contribution to the bilinear five-point function is then, 
\begin{multline} \label{5Os1}
\langle \mO_1(\tau_1) \cdots \mO_5(\tau_5) \rangle_s = \frac{1}{4} \int_{\mC}\frac{d h_a}{2\pi i}\frac{ \rho (h_a) }{c_{h_a} c_{1- h_a}}  \int_{\mC}\frac{d h_b}{2\pi i}\frac{ \rho (h_b) }{c_{h_b} c_{1- h_b}} \\
\int d\tau_a  d\tau_b\, \langle \mO_1(\tau_1) \mO_2(\tau_2) \mO_{h_a}(\tau_a) \rangle \langle \mO_{1- h_a}(\tau_a) \mO_3(\tau_3) \mO_{h_b}(\tau_b)\rangle \langle \mO_{1- h_b}(\tau_b) \mO_4(\tau_4) \mO_5(\tau_5)\rangle~.
\end{multline}
The integrals over $\tau_a, \tau_b$ will be evaluated in the next section; the result is a sum of five-point conformal blocks and their shadows. After changing variables, $h_a \rightarrow 1- h_a$ and $h_a \rightarrow 1-h_b$ on some of the terms, similar to what was done in the case of the bilinear four-point function, we find,
\be \label{5Os2}
\!\!\! \langle \mO_1(\tau_1) \cdots \mO_5(\tau_5) \rangle_s =\!\!\!  \int_{\mC}\frac{d h_a}{2\pi i}\frac{ \rho (h_a) }{c_{h_a}^2}\frac{\Gamma(h_a)^2}{\Gamma(2 h_a)}   \int_{\mC}\frac{d h_b}{2\pi i}\frac{ \rho (h_b) }{c_{h_b}^2}\frac{\Gamma(h_b)^2}{\Gamma(2 h_b)}\, c_{1 2\, h_a} c_{h_a\, 3\, h_b} c_{h_b\, 4  5}\, \mF_{12345}^{h_a, h_b}(x_1, x_2)~,
\ee
where $\mF_{12345}^{h_a, h_b}(x_1, x_2)$ is the five-point conformal block, depending on the two cross-ratios of times,
\be \label{twoX}
x_1 = \frac{\tau_{12} \tau_{34}}{\tau_{13} \tau_{24}}~, \ \ \ \ x_2 = \frac{\tau_{23} \tau_{45}}{\tau_{24} \tau_{35}}~.
\ee

The prescription for writing a general connected $p$-point correlation function $\langle \mO_{h_1} \cdots \mO_{h_p}\rangle$, to leading nontrivial order in $1/N$,  is clear. One draws all Feynman-like skeletons, in which the lines are ladders and there are ``cubic interactions'' $c_{1 2 3 }$ (where $c_{1 2 3}$ is the coefficient of $\langle \mO_1 \mO_2 \mO_3\rangle$ found in Sec.~\ref{sec:3pt}). For each internal line, one has a contour integral, 
\be
\int_\mC \frac{dh_a}{2 \pi i} \frac{\rho(h_a)}{c_{h_a}^2} \frac{\Gamma(h_a)^2}{\Gamma(2h_a)}~.
\ee
The integrand consists of the ``cubic interactions'' $c_{123}$, and a $p$-point conformal block. 
One writes down such an expression for each of the skeleton diagrams. 
One should then subtract diagrams with no exchanged melons in some channels, which were over-counted; these have the same rules but with a $\rho^0$ and a $c_{123}^{(2)}$ (as was discussed in the four-point function case, Eq.~\ref{OOOOs0}). Finally, if $q$ is sufficiently large, there are additional contact diagrams one must add, which consist of four or more ladders meeting at a melon; these are discussed in Appendix~\ref{sec:con}. From the  correlation functions $\langle \mO_{h_1} \cdots \mO_{h_p}\rangle$, one can obtain the $2p$-point fermion correlation function, as discussed in 
Appendix~\ref{FermionP}.

\subsection{Five-point conformal blocks} \label{BigBlock}
In conformal field theories, the functional form of the building blocks of correlation functions is fully fixed by conformal invariance. As discussed in Appendix~\ref{blocks}, the OPE takes the form, 
\be
\mO_1(\tau_1) \mO_2(\tau_2) = \sum_h c_{12 h}\, \mC_{12 h}(\tau_{12}, \partial_2) \mO_h~,
\ee
where $\mC_{12h}(\tau_{12}, \partial_2)$ accounts for descendants of $\mO_h$, and is fully determined by the functional form of the three-point function. The conformal blocks are in turn fully determined by the $\mC_{12 h}(\tau_{12}, \partial_2)$. 
For instance, the four-point block is, 
\be
\mF_{1234}^h(x) = \mC_{1 2 h}(\tau_{12}, \partial_2)\, \mC_{3 4 h}(\tau_{34}, \partial_4)\,\langle \mO_h(\tau_2) \mO_h(\tau_4)\rangle~,
\ee
and the five-point block is, 
\be
\mF_{12345}^{h_a, h_b}(x_1, x_2) = \mC_{1 2 h_a}(\tau_{12}, \partial_2)\, \mC_{4 5  h_b}(\tau_{45}, \partial_4)\,\langle \mO_{h_a}(\tau_2) \mO_3(\tau_3) \mO_{h_b}(\tau_4)\rangle~.
\ee
To determine the higher-point conformal blocks, one simply continues to successively apply the OPE. See  \cite{Cho:2017oxl} for a recent study, in the context of Virasoro blocks. 

An alternative way to obtain an explicit form for the higher-point $SL_2$ conformal blocks is to simply evaluate the integrals that appear in the higher-point correlation function. For instance, the expression that appeared in the five-point function is,
\be \nonumber
\mB_{12345}^{a, b} = \int d\tau_a d\tau_b\, \langle \mO_1(\tau_1) \mO_2(\tau_2) \mO_{h_a}(\tau_a) \rangle \langle \mO_{1- h_a}(\tau_a) \mO_3(\tau_3) \mO_{1-h_b}(\tau_b)\rangle \langle \mO_{h_b}(\tau_b) \mO_4(\tau_4) \mO_5(\tau_5)\rangle~,
\ee
where we have changed $h_b \rightarrow 1- h_b$, relative to (\ref{5Os1}), in order to make the expression more symmetric. Through a change of variables, we rewrite this so that it is a function of the two cross-ratios $x_1, x_2$ defined in (\ref{twoX}), 
\be
\mB_{12345}^{a, b} =\frac{1}{|\tau_{12}|^{h_1 + h_2}|\tau_{45}|^{h_4 + h_5}|\tau_{34}|^{h_3}} \Big|\frac{\tau_{23}}{\tau_{13}}\Big|^{h_{12}} \Big|\frac{\tau_{24}}{\tau_{23}}\Big|^{h_3} \Big|\frac{\tau_{35}}{\tau_{34}}\Big|^{h_{45}} \mC_{12345}^{a, b}~,
\ee
where, 
\be
\mC_{12345}^{a, b}  = |x_1|^{1-h_a} |x_2|^{1-h_b} \int d\tau_a d\tau_b \frac{| 1- \tau_a x_1 - \tau_b x_2|^{h_a + h_b +h_3 - 2}}{|\tau_a|^{h_a - h_{12}}|\tau_a - 1|^{h_a + h_{12}}|\tau_b|^{h_b + h_{45}}|\tau_b - 1|^{h_b - h_{45}}}~.
\ee
 Let us assume $0<x_1, x_2<1$. From the integral definition of the Appell function $F_2$ we notice that, 
if our integral were in the range $0< \tau_a, \tau_b<1$, then $\mC_{12345}^{a, b}$ would be proportional  to, 
\be \label{511}
 x_1^{1-h_a} x_2^{1-h_b}  \Ft{3}{2}{ 2\!- \!h_a\!-\! h_b\! -\! h_3~, 1\!+\!h_{12}\! -\!h_a~, 1\!-\!h_{45}\! -\! h_b}{2\! -\! 2 h_a~, 2\!-\! 2 h_b}{x_1~,x_2}~.
\ee
The differential equation defining the Appell function $F_2$ has a total of four solutions, which follow from (\ref{511}). Our integral $\mC_{12345}^{a, b}$ should be a linear combination of these. We set the coefficients by studying the integral $\mC_{12345}^{a, b}$ in various limits, similar to what we did for the integral appearing in the three-point function in Sec.~\ref{planar}. The result is expressed in terms of the   five-point conformal blocks,
\begin{multline}
\mF_{12345}^{h_a, h_b}(x_1, x_2) = \frac{1}{|\tau_{12}|^{h_1 + h_2}|\tau_{45}|^{h_4 + h_5}|\tau_{34}|^{h_3}} \Big|\frac{\tau_{23}}{\tau_{13}}\Big|^{h_{12}} \Big|\frac{\tau_{24}}{\tau_{23}}\Big|^{h_3} \Big|\frac{\tau_{35}}{\tau_{34}}\Big|^{h_{45}}  \\
x_1^{h_a} x_2^{h_b}  \Ft{3}{2}{h_a\! +\! h_b\! -\! h_3~,h_a\! +\! h_{12}, h_b\!-\!h_{45}}{2 h_a~,  2 h_b}{x_1~,x_2}~, 
\end{multline}
and is given by, 
\begin{multline} \nonumber
\!\!\!\!\!\! \mB_{12345}^{a, b} =\beta(h_a, h_b + h_3 -1)\beta(h_b, h_3 - h_a)\, \mF_{12345}^{h_a, h_b}(x_1, x_2) +\beta(1-h_a, h_{12}) \beta(h_b, h_a + h_3-1)\mF_{12345}^{1- h_a, h_b}(x_1, x_2) \\
+ \beta(h_a, h_b+h_3 - 1)\beta(1-h_b, h_{45}) \mF_{12345}^{h_a, 1-h_b}(x_1, x_2)+ \beta(1-h_a, h_{12}) \beta(1-h_b, h_{45})\mF_{12345}^{1-h_a, 1-h_b}(x_1, x_2)~,
\end{multline}
where $\beta(h, \Delta)$ is defined in Appendix.~\ref{blocks}, see Eq.~\ref{betaD}.  We established which of the four terms in this expression is identified as the five-point conformal block by looking at the small $\tau_{12}, \tau_{45}$ behavior. 

One could, in this way, compute six-point blocks and higher, though we will stop here.

\section{Generalized Free Field Theory} \label{FREE}
In the previous sections we gave a prescription for determining all correlation functions in SYK, $\langle \mO_{h_1} \cdots \mO_{h_p}\rangle$. The operators $\mO_h$ have small anomalous dimensions when the dimension $h$ is large, $h\gg 1$. As we showed, the correlators of these are determined from the weak coupling limit of cSYK: generalized free field theory of fermions, and can be found through Wick contraction. This provides significant simplification. 

In this section, we study the generalized free field theory of $N$ fermions of dimension $\Delta$, in the singlet sector. In Sec.~\ref{wick} we compute the correlation functions of the primary $O(N)$ invariant fermion bilinears. Then in Sec.~\ref{A3} and Sec.~\ref{A4} we use saddle point analysis to simplify the three-point and four-point functions, respectively, in the limit of large $h_i$. 

\subsection{Wick contractions and generating function} \label{wick}

The fermion bilinear, primary, $O(N)$ invariant operators are given by, 
\be \label{Ofree}
\mO_n =\frac{1}{\sqrt{N}} \sum_{i=1}^N \sum_{r=0}^{n} d_{n\, r}\, \partial_{\tau}^{r}\, \chi_i\, \partial_{\tau}^{n-r} \chi_i~,
\ee
where $d_{n\, r}$ is,
\be
d_{n\, r} = \frac{(-1)^r}{\Gamma(n-r+1) \Gamma(\Delta+n-r)\, \Gamma(r+1)\, \Gamma(\Delta+r)}~.
\ee
Due to fermion antisymmetry, only correlation functions of $\mO_n$ involving odd $n$ are nonzero. As a result, throughout this paper $\mO_n$ has been used to denote what in the current language is $\mO_{2n +1}$; for the purposes of this section, the current definition is more convenient.  

\subsubsection*{Wick Contractions}
The correlation functions of the $\mO_n$ follow trivially by Wick contractions. The connected piece of a $p$-point correlation function is,
\begin{multline}
\langle \mO_{n_1}(\tau_1) \cdots \mO_{n_p}(\tau_p)\rangle = 
\frac{1}{N^{\frac{p-2}{2}}}\sum_{r_1,\ldots, r_p} d_{n_1 r_1}\cdots d_{n_p r_p}  \(\partial_{p}^{n_{p} - r_{p}} \partial_1^{r_1} G(\tau_{1\, p})\)\\ 
\times \(\partial_1^{n_1 - r_1} \partial_2^{ r_2} G(\tau_{12})\) \(\partial_2^{n_2 - r_2} \partial_3^{r_3} G(\tau_{23})\)\cdots\(\partial_{p-1}^{n_{p\!-\!1} - r_{p\!-\!1}} \partial_p^{r_p} G(\tau_{p-\!1\, p})\)
+ \text{perm}~.
\end{multline}
Using that $d_{n r} = (-1)^n d_{n\, n-r}$, one can see that the addition of permutations gives a factor $(1-(-1)^{n_1})\cdots (1-(-1)^{n_p})$ multiplying the term we explicitly wrote. Making use of the derivative of the two-point function, 
\be \label{Gderiv}
\partial_1^p\, G(\tau_{12}) = G(\tau_{12}) \frac{\Gamma(2\Delta+p)}{\Gamma(2\Delta)}\frac{(-1)^p}{\tau_{12}^p}~,
\ee
the $p$-point function becomes, 
\begin{multline} \label{pPOINT}
\langle \mO_{n_1}(\tau_1) \cdots \mO_{n_p}(\tau_p)\rangle =\frac{1}{N^{\frac{p-2}{2}}}(\prod_i \delta_{n_i =\text{odd}})\, \frac{-(-2)^{p}}{\Gamma(2\Delta)^p}\, G(\tau_{12}) G(\tau_{23}) \cdots G(\tau_{p-1\, p})\,\, G(\tau_{1 p})\\
 \sum_{r_1,\ldots, r_p} d_{n_1 r_1}\cdots d_{n_p r_p}\,(-1)^{r_2 + \ldots r_{p-1}}\,
 \frac{\Gamma(2\Delta+n_p -r_p+r_1)}{\tau_{1p}^{n_p - r_p+r_1}} \\
  \frac{\Gamma(2\Delta+n_1 -r_1+r_2 )}{\tau_{12}^{n_1 - r_1+r_2}}  \frac{\Gamma(2\Delta+n_2 -r_2+r_3 )}{\tau_{23}^{n_2 - r_2+r_3}}\cdots  \frac{\Gamma(2\Delta+n_{p-1} -r_{p-1}+r_p )}{\tau_{p-1\, p}^{n_{p-1}- r_{p-1}+r_p}}~.
\end{multline}

\subsubsection*{Generating function}
It is convenient to introduce a generating $\mO(\tau, x)$ which includes all the $\mO_n(\tau)$, see  for instance \cite{Sleight:2016dba},
\be
\mO(\tau, x) = \sum_{n=0}^{\infty} \mO_n(\tau)\, x^{n}~.
\ee
Using the explicit definition of the $\mO_n$ in terms of fermions the generating $\mO(\tau,x)$ becomes,
\be
\mO(\tau,x) = \frac{1}{\sqrt{N}}\sum_{i = 1}^N D(x,\tau)\chi_i(\tau)\, D(-x, \tau) \chi_i(\tau)~,
\ee
where we have defined, 
\be 
D(x, \tau) =\sum_{r=0}^{\infty} \frac{(-x)^{r}}{r!\,  \Gamma(r+\Delta)} \partial_{\tau}^{r}= (x \partial_{\tau})^{\frac{1-\Delta}{2}} J_{\Delta-1}(2 \sqrt{x \partial_{\tau} }) ~.
\ee

We now compute the correlation functions of $\mO(\tau, x)$. The two-point function is, 
\be \label{OO2}
\langle O(\tau_1, x_1) O(\tau_2,x_2) \rangle = H(x_1, \tau_1, -x_2, \tau_2) H(-x_1, \tau_1, x_2, \tau_2) - H(x_1, \tau_1, x_2, \tau_2) H(-x_1, \tau_1, -x_2, \tau_2)
\ee
where
\be
H(x_1, \tau_1, x_2, \tau_2) = D(x_1, \tau_1) D(x_2,\tau_2) G(\tau_{12})~.
\ee
Using the definition of $D(x_i, \tau_i)$ and acting with the derivatives on $G(\tau_{12})$, and then using the integral definition of the Gamma function, performing the sum, and evaluating the resulting integral, we get,
\be
H(x_1, \tau_1, x_2, \tau_2) = \frac{G(\tau_{12})}{\Gamma(2\Delta)} \(\frac{x_1 x_2}{\tau_{12}^2}\)^{\frac{1-\Delta}{2}}\, e^{\frac{x_1 - x_2}{\tau_{12}}}\, J_{\Delta-1}\(2 \sqrt{\frac{x_1 x_2}{\tau_{12}^2}}\)~.
\ee
If we insert $H$ into (\ref{OO2}),  and  Taylor expand,  we recover the two-point functions $\langle \mO_n(\tau_1) \mO_n(\tau_2)\rangle$. In the  $\Delta=0$ limit these are, 
\be \label{2pt}
\langle \mO(x_1,\tau_1) \mO(x_2, \tau_2) \rangle   = \frac{1}{\Gamma(2\Delta)^2}\sum_{n=1}^{\infty} \(\frac{x_1\, x_2}{\tau_{12}^2}\)^{2n+1}\, \(\frac{N_n}{(2n!)}\)^2~, \ \  \ \ \ \ N_n^2 =\frac{2^{4n+1}}{(2n+1)}\frac{\Gamma(2n+\frac{1}{2})}{\sqrt{\pi}\Gamma(2n)}~.
\ee
In the large $n$ limit these simplify to,
\be \label{2ptBig}
\(\frac{N_n}{(2n!)}\)^2 \approx \(\frac{e}{n}\)^{4n}~, \ \ \ \ \ \ n\gg1~.
\ee

A $p$-point correlation function of the $\mO(\tau, x)$ is a simple generalization of the two-point function, 
\bea \label{pPoint}
&&\hspace{-1cm}\langle \mO(\tau_1, x_1) \mO(\tau_2, x_2) \cdots \mO(\tau_p, x_p)\rangle =\\  \nonumber&& \hspace{-1cm}H(x_1, \tau_1, -x_p, \tau_p) H(-x_1, \tau_1, x_2,\tau_2)H(-x_2, \tau_2, x_3, \tau_3) \cdots H(-x_{p-1},\tau_{p-1}, x_p, \tau_p) + \text{perm}~.
\eea
To obtain the correlators of the $\mO_{n_i}$, one should Taylor expand the right-hand side, extracting the  coefficient of the $x_i^{n_i}$ term. Upon Taylor expansion, each of the permutations  gives the same contribution, up to a sign, and serves to   ensure that the correlation function $\langle \mO_{n_1} (\tau_1)\cdots \mO_{n_p}(\tau_p)\rangle$ is nonzero only for odd $n_i$.

\subsection{Asymptotic three-point function} \label{A3}
We would like to find the form of the three-point function $\langle \mO_{n_1}(\tau_1) \mO_{n_2}(\tau_2) \mO_{n_3}(\tau_3)\rangle$ in the limit that $n_1, n_2, n_3 \gg 1$. 
This is simplest to do through study of the correlator \\
$\langle \mO(\tau_1, x_1) \mO(\tau_2, x_2) \mO(\tau_3, x_3)\rangle $.~\footnote{
In Sec.~\ref{planar} we found the three-point function by evaluating  Feynman diagrams, obtaining an expression in terms of a  generalized hypergeometric function at argument $1$; see Eq.~\ref{s1232} for the expression in the current context. However, since some of the arguments of this hypergeometric function are negative, written as a single sum,  it includes both positive and negative terms, which makes its asymptotic analysis, in this form, difficult.} Writing this out explicitly, 
\begin{multline} \label{3pt2}
\langle \mO(x_1, \tau_1) \mO(x_2, \tau_2) \mO(x_3, \tau_3) \rangle =\frac{\sgn(\tau_{12}\tau_{13}\tau_{23})}{\Gamma(2\Delta)^3} \Big|\frac{x_1 x_2}{\tau_{12}^2} \Big|^{\frac{1-\Delta}{2}} \Big|\frac{-x_1 x_3}{\tau_{13}^2}\Big|^{\frac{1- \Delta}{2}} \Big|\frac{x_2 x_3}{\tau_{23}^2}\Big|^{\frac{1-\Delta}{2}} \\ \exp\( x_1 \frac{\tau_{23}}{\tau_{12} \tau_{13}} - x_2 \frac{\tau_{13}}{\tau_{12} \tau_{23}} + x_3 \frac{\tau_{12}}{\tau_{13}\tau_{23}}\)J_{\Delta-1}\! \(2  \sqrt{\frac{x_1 x_2}{\tau_{12}^2}}\)\! J_{\Delta-1}\! \(2  \sqrt{\frac{-x_1 x_3}{\tau_{13}^2}}\) J_{\Delta-1}\(2  \sqrt{\frac{x_2 x_3}{\tau_{23}^2}}\)\, \\
+\, \text{7 perm.}
\end{multline}
If we were to expand  this, it would give, for $\Delta\rightarrow 0$, 
\begin{multline} \label{OOOs}
\langle \mO(x_1, \tau_1) \mO(x_2, \tau_2) \mO(x_3, \tau_3) \rangle = \\
\frac{1}{{\Gamma(2\Delta)^3}}\sum \frac{x_1^{2n_1+1}}{(2n_1)!}\frac{x_2^{2n_2+1}}{(2n_2)!}\frac{x_3^{2n_3+1}}{(2n_3)!} \frac{8\, s_{n_1 n_2 n_3}^{(2)}}{|\tau_{12}|^{2(n_1 + n_2-n_3)+1}   |\tau_{23}|^{2(n_2 + n_3-n_1)+1} |\tau_{31}|^{2(n_3 + n_1-n_2)+1}}~,
\end{multline}
where $s_{n_1 n_2 n_3}^{(2)}$ is a triple sum,  see Eq.~\ref{s1232}. 

More directly, we can extract the desired correlator through a triple contour integral over the unit circle,
\be
\langle \mO_{n_1}(\tau_1) \mO_{n_2}(\tau_2) \mO_{n_3}(\tau_3) \rangle = \prod_{i=1}^3 \oint \frac{d x_i}{(2\pi i) x_i^{n_i+1}} \langle \mO(x_1, \tau_1) \mO(x_2, \tau_2) \mO(x_3, \tau_3) \rangle~.
\ee
To work with (\ref{3pt2}), we  use the following representation of the Bessel function, 
\be
J_{\nu} (z) = \(\frac{z}{2}\)^{\nu}\int_{\mathcal{L}} \frac{d s}{2\pi i} \frac{1}{s^{\nu+1}}\, \exp\( s- \frac{z^2}{4 s}\)~,
\ee
where the contour $\mathcal{L}$ comes in  from $-\infty$, circles around the origin, and returns to $-\infty$.  With this representation of the Bessel function, we have $\langle \mO_{n_1}(\tau_1) \mO_{n_2}(\tau_2) \mO_{n_3}(\tau_3)\rangle$ in terms of six-contour integrals. In the limit of large $n_i$, we may evaluate these by saddle point analysis. We will only be interested in the dominant term, and will not compute the subleading corrections. 
Dropping all terms that are not exponential in the $n_i$, and not distinguishing between $n_i$ and $n_i -1$, we have,
\begin{multline} \label{threeInt}
\hspace{-1cm} \langle \mO_{n_1  }(\tau_1) \mO_{n_2 }(\tau_2) \mO_{n_3 }(\tau_3) \rangle\! \approx\!\frac{\sgn(\tau_{12}\tau_{13}\tau_{23})}{\Gamma(2\Delta)^3} 
\! \prod_{i =1}^3\! \int_{\mathcal{L}} \frac{d s_i}{2\pi i}\prod_{i=1}^3 \oint \!\! \frac{d x_i}{(2\pi i) } 
 \exp\(\! x_1 \frac{\tau_{23}}{\tau_{12} \tau_{13}}\! -\! x_2 \frac{\tau_{13}}{\tau_{12} \tau_{23}}\! +\! x_3 \frac{\tau_{12}}{\tau_{13}\tau_{23}}\) \\
\exp\( -\sum_{i=1}^3 n_i \log x_i + s_1 - \frac{ x_2 x_3}{\tau_{23}^2}\frac{1}{s_1} + s_2 + \frac{x_1 x_3}{\tau_{13}^2}\frac{1}{s_2} + s_3 - \frac{x_1 x_2}{\tau_{12}^2} \frac{1}{s_3}\)~.
\end{multline}
Note that, at this level of approximation, it makes no difference what $\Delta$ is. 
The saddle equations from varying the  $s_i$ are, 
\be \label{sisad}
s_1^2 = - \frac{ x_2 x_3}{\tau_{23}^2}~, \ \ \ \ \ 
s_2^2 = \frac{ x_1 x_3}{\tau_{13}^2}~, \ \ \ \ \\ 
s_3^2 = - \frac{x_1 x_2}{\tau_{12}^2}~.
\ee
The saddle equations from varying the $x_i$ are, 
\bea
- \frac{n_1}{x_1} + \frac{x_3}{\tau_{13}^2\,  s_2} - \frac{x_2}{\tau_{12}^2\, s_3} + \frac{\tau_{23}}{\tau_{12}\tau_{13}} &=& 0 \\
-\frac{n_2}{x_2} - \frac{x_3}{\tau_{23}^2\, s_1} - \frac{x_1}{\tau_{12}^2\, s_3} - \frac{\tau_{13}}{\tau_{12} \tau_{23}} &=& 0 \\
- \frac{n_3}{x_3} - \frac{x_2}{\tau_{23}^2\, s_1} + \frac{x_1}{\tau_{13}^2\, s_2} + \frac{\tau_{12}}{\tau_{13} \tau_{23}} &=& 0~.
\eea
We multiply the first equation by $x_1$, the second by $x_2$, and the third by $x_3$. We then apply (\ref{sisad}) to simplify the left-hand side. This gives, 
\be
s_2 + s_3 = n_1 - x_1 \frac{\tau_{23}}{\tau_{12}\tau_{13}}~, \ \ \ 
s_1 + s_3 = n_2 + x_2 \frac{\tau_{13}}{\tau_{12} \tau_{23}}~, \ \ \ 
s_1 + s_2 = n_3 - x_3 \frac{\tau_{12}}{\tau_{13} \tau_{23}}~.
\ee
We now trivially solve for the $x_i$, and insert into (\ref{sisad}) to get, 
\be \nonumber
s_1^2 = (n_2 - s_1 - s_3)(n_3 - s_1 - s_2)~, \ \ \ 
s_2^2 = (n_1 - s_2 - s_3)(n_3 - s_1 - s_2)~, \ \ \ 
s_3^2 = (n_1 - s_2 - s_3)(n_2 - s_1 - s_3)~.
\ee
Solving gives two  solutions. The first is,
\be \label{SadD}
s_1 = \frac{n_2 n_3}{n_1 + n_2 + n_3}~,  \ \ \ 
s_2 = \frac{n_1 n_3}{n_1 + n_2 + n_3}~, \ \ \ 
s_3 =  \frac{n_1 n_2}{n_1 + n_2 + n_3}~,
\ee
and is the dominant saddle, while the second is,
\be
s_1 = \frac{n_2 n_3}{-n_1 + n_2 + n_3}~, \ \ \ 
s_2 = \frac{n_1 n_3}{n_1 - n_2 + n_3}~, \ \ \ 
s_3 =  \frac{n_1 n_2}{n_1 + n_2- n_3}~.
\ee
Inserting the dominant  saddle into the integrand, we find the three-point function is, 
\be \label{3ptAnswer2}
\hspace{-.5cm} \langle \mO_{n_1}(\tau_1) \mO_{n_2}(\tau_2) \mO_{n_3}(\tau_3) \rangle\approx  \frac{1}{|\tau_{23}|^{n_2+n_3-n_1} |\tau_{13}|^{n_1+n_3 - n_2} |\tau_{12}|^{n_1 + n_2 - n_3}}\, \frac{(e\, N)^N}{n_1^{2 n_1} n_2^{2 n_2} n_3^{2 n_3}}~,\ \ \ \ n_1, n_2, n_3 \gg 1~,
\ee
where we defined $N = n_1 + n_2 +n_3$. 

In terms of $s_{n_1 n_2 n_3}$, comparing (\ref{3ptAnswer2}) with (\ref{OOOs}), we have that, 
\be
s_{n_1 n_2 n_3}^{(2)} \approx \frac{(2N)^{2N}}{(2n_1)^{2n_1} (2n_2)^{2n_2}(2n_3)^{2n_3}}\approx  \frac{(2N)!}{(2n_1)! (2n_2)! (2n_3)!}~,\ \ \ \ n_1, n_2, n_3 \gg 1~.
\ee
Equipped with the asymptotic limit of the three-point function, we can find the asymptotic limit of the cubic couplings of the dual bulk scalars $\phi_n$ dual to $\mO_{2n+1}$ \cite{GR2}. With the $\phi_n$ canonically normalized, we have,
\be \label{lambda}
\lambda_{n_1 n_2 n_3} \approx \frac{N!}{\Gamma(N-2n_1+\frac{1}{2}) \Gamma(N-2n_2+\frac{1}{2})\Gamma(N-2n_3+ \frac{1}{2})}~, \ \ \ \ n_1, n_2, n_3 \gg 1~, 
\ee 
where we have, for simplicity, dropped any order-one factors that may have appeared. 
One would ultimately like to have a string-like bulk interpretation of these couplings. 


\subsection{Asymptotic four-point function}\label{A4}
To find the behavior of the four-point function $\langle \mO_{n_1} (\tau_1) \cdots \mO_{n_4}(\tau_4)\rangle$ for large $n_i$ we perform an analogous analysis as with the three-point function. Representing the four-point function as a contour integral, and
dropping all terms that aren't exponential, we have,
\begin{multline} \label{4ptS2}
\langle \mO_{n_1}( \tau_1)\ldots \mO_{n_4}( \tau_4) \rangle \approx\frac{\sgn(\tau_{12}\tau_{23}\tau_{34}\tau_{41})}{\Gamma(2\Delta)^4} \prod_{i =1}^4\int_{\mathcal{L}}\frac{d s_i}{2\pi i}\prod_{i=1}^4 \oint \frac{d x_i}{(2\pi i) }\\ \exp\( - x_1 \frac{\tau_{24}}{\tau_{12} \tau_{14}} - x_2 \frac{\tau_{13}}{\tau_{12} \tau_{23}} - x_3\frac{\tau_{24}}{\tau_{23} \tau_{34}} - x_4 \frac{\tau_{13}}{\tau_{14}\tau_{34}}\)\,\\
\exp\Big( - \sum_{i = 1}^4 \, n_i \log x_i +  s_1 + \frac{x_1 x_2}{\tau_{12}^2\, s_1} + s_2 + \frac{x_2 x_3}{\tau_{23}^2\, s_2} +  s_3 + \frac{x_3 x_4}{\tau_{34}^2\, s_3} + s_4 + \frac{ x_4 x_1}{\tau_{41}^2\, s_4}\Big)~.
\end{multline}
At large $n_i$ we can approximate the integral by its saddle. 
Varying with respect to the $s_i$ gives the saddle equations,
\be \label{s111}
s_1^2 = \frac{x_1 x_2}{\tau_{12}^2}~,  \ \ \ 
s_2^2 = \frac{x_2 x_3}{\tau_{23}^2}~, \ \ \ 
s_3^2 = \frac{ x_3 x_4}{\tau_{34}^2}~, \ \ \ 
s_4^2 = \frac{x_4 x_1}{\tau_{41}^2}~.
\ee
Varying, in addition, with respect to $x_i$ gives the saddle equations, 
\bea
- \frac{n_1}{x_1} + \frac{x_2}{\tau_{12}^2\, s_1} + \frac{x_4}{\tau_{41}^2\, s_4} - \frac{\tau_{24}}{\tau_{12} \tau_{14}} &=& 0 \\
-\frac{n_2}{x_2} + \frac{x_1}{\tau_{12}^2\, s_1} + \frac{x_3}{\tau_{23}^2\, s_2} - \frac{\tau_{13}}{\tau_{12}\tau_{23}} &=& 0 \\
- \frac{n_3}{x_3} + \frac{x_2}{\tau_{23}^2\, s_2} + \frac{x_4}{\tau_{34}^2\, s_3} - \frac{\tau_{24}}{\tau_{23} \tau_{34}} &=& 0\\
- \frac{n_4}{x_4} + \frac{x_3}{\tau_{34}^2\, s_3} + \frac{x_1}{\tau_{41}^2\, s_4} - \frac{\tau_{13}}{\tau_{14} \tau_{34}} &=& 0~.
\eea
We multiply the first equation by $x_1$, the second by $x_2$, and so on, and use (\ref{s111}) to simplify, 
\bea \label{sSel}
s_4 + s_1 &= & n_1 + \frac{\tau_{24}}{\tau_{12} \tau_{14}} \, x_1~, \ \ \ \ \ \ \ \
s_1 + s_2  = n_2 + \frac{\tau_{13}}{\tau_{12}\tau_{23}}\, x_2~, \\ \nonumber
s_2 + s_3 &=& n_3 + \frac{\tau_{24}}{\tau_{23} \tau_{34}}\, x_3~,\ \ \ \ \ \ \ \
s_3 + s_4 = n_4 + \frac{\tau_{13}}{\tau_{14} \tau_{34}}\, x_4~.
\eea
Now, using the saddle equations, at the saddle we see that the four-point function is, 
\be \label{FourSEL}
\langle \mO_{n_1}( \tau_1)\ldots \mO_{n_4}( \tau_4) \rangle \approx \frac{1}{\prod_i x_i^{n_i}}\, \exp\(\sum n_i\)
\ee

Trivially solving (\ref{sSel}) for the $x_i$ and inserting into (\ref{s111}) gives, in terms of the cross-ratio $x = \frac{\tau_{12} \tau_{34}}{\tau_{13}\tau_{24}}$~, 
\bea \label{siEq}
s_1^2 &=& (1-x) (s_4 + s_1 - n_1) (s_1 + s_2 - n_2) \\ \nonumber
s_2^2 &=& x (s_1+ s_2 - n_2)(s_2 + s_3 - n_3) \\ \nonumber
s_3^2 &=& (1-x) (s_2 + s_3 - n_3)(s_3 + s_4- n_4) \\ \nonumber
s_4^2 &=& x( s_3 + s_4 - n_4)( s_4 + s_1 - n_1)~.
\eea
The solution to these equations for general $n_i$ is complicated. A simple case, which we focus on, is when all of the dimensions $n_i$ are equal. 
\subsubsection*{Equal $n_i$}
We set $n_1 = n_2 = n_3 = n_4$. 
In this case we can simplify (\ref{FourSEL}) to, 
\be
\langle \mO_{n_1}( \tau_1)\ldots \mO_{n_1}( \tau_4) \rangle \approx \exp(4 n) \frac{1}{(\tau_{12}^2 \tau_{23}^2 \tau_{34}^2 \tau_{41}^2 s_1^2 s_2^2 s_3^2 s_4^2)^{\frac{n}{2}}}~, 
\ \ \ \ n_1 \gg 1~.
\ee
The symmetric product of the times can alternatively be written as, 
\be
\tau_{12}^2 \tau_{23}^2 \tau_{34}^2 \tau_{41}^2= \tau_{12}^4  \tau_{34}^4 \(\frac{1-x}{x}\)^2~.
\ee
We define $\tilde{s_i} = s_i/n_1$. Then, since $(n/e)^n \approx n!$, 
\be \label{Osss}
\langle \mO_{n_1}( \tau_1)\ldots \mO_{n_1}( \tau_4) \rangle \approx \frac{1}{(n_1!)^4}\frac{1}{(\tau_{12}^2 \tau_{34}^2)^{n_1}} \frac{1}{\(\frac{(1-x)}{x} \ts_1 \ts_2 \ts_3 \ts_4\)^{n_1}}~, \ \ \ \ n_1 \gg 1~.
\ee
To complete the evaluation of the four-point function we need to solve (\ref{siEq}) for the $s_i$, and insert their product into (\ref{Osss}). There are eight solutions to (\ref{siEq}). In writing them, we assume that we have a time-ordered correlation function, so that cross-ratio of times is in the range $0<x<1$. The other time orderings can be worked out in a similar fashion. Of the eight saddle, two saddles give the product, 
\be
\ts_1 \ts_2 \ts_3 \ts_4 = \frac{ x-1}{16 x}~.
\ee
Another two saddles give, 
\be
\ts_1 \ts_2 \ts_3 \ts_4 = \frac{x}{16 (x-1)}~,
\ee
while the remaining four saddles give,
\be
\ts_1 \ts_2 \ts_3 \ts_4 = \frac{x(1-x)}{(1\pm\sqrt{x}\pm \sqrt{1-x})^4}~.
\ee
The dominant saddle, for all values of $0<x<1$, is clearly the one for which,
\be
\ts_1 \ts_2 \ts_3 \ts_4 = \frac{x(1-x)}{(1+\sqrt{x}+\sqrt{1-x})^4}~.
\ee
Inserting this into (\ref{Osss}), we have, 
\be \label{OOOOz}
\langle \mO_{n_1}( \tau_1)\ldots \mO_{n_1}( \tau_4) \rangle\approx  \frac{-1}{(\tau_{12}^2 \tau_{34}^2)^{n_1} }\Big(\frac{(\sqrt{x} + \sqrt{1-x} + 1)^4}{(1-x)^2}\Big)^{n_1} \frac{1}{(n_1!)^4}~, \ \ \ \  n_1 \gg1~.
\ee

As we cross the boundaries: $x=0$ or $x=1$, we observe the Stokes phenomenon: the dominant saddle changes. This means that if we want to consider the limit of $x\rightarrow 0$, or $x \rightarrow 1$,  we must  account for multiple  saddles. This can already be seen from (\ref{OOOOz}) since, by itself, it has  incorrect small $x$ behavior. In particular,  expanding around small $x$ gives rises to powers $x^{m/2}$, however the single-trace and double-trace operators appearing in  the OPE have integer dimension, so there should not be any terms with odd $m$. If we were to include one of the other saddles,
\be
\ts_1 \ts_2 \ts_3 \ts_4 = \frac{x(1-x)}{(1-\sqrt{x}+ \sqrt{1-x})^4}~,
\ee
and have it come with the same phase, then this would eliminate the odd $m$ in the expansion. Of course, to actually determine the phase one should compute fluctuations about the saddle, which we have not done. 

\subsubsection*{Mellin transform}
It is sometimes useful to study the four-point function in Mellin space, reviewed in Appendix~\ref{app:Mellin}. In terms of the variables $u=x^2$ and $v = (1-x)^2$, the four-point function (\ref{OOOOz})  is,
\be
\langle \mO_{n_1}( \tau_1)\ldots \mO_{n_1}( \tau_4) \rangle
 = \frac{1}{(n_1!)^4} \frac{-1}{(\tau_{13}^2 \tau_{24}^2)^{n_1} }\Big(\frac{(u^{1/4} + v^{1/4} + 1)^4}{ u v}\Big)^{ n_1}~.
\ee
Notice that, since  $u$ and $v$ are not independent, we could have written this in other ways. This ambiguity reflects the non-uniqueness of the Mellin amplitude for  CFT$_1$ four-point functions. However, the choice we made is natural because it is symmetric. Using the standard Mellin-Barnes representation, we can write,
\be
\Big(\frac{(u^{1/4} + v^{1/4} + 1)^4}{ u v}\Big)^{ n_1}=  \frac{1}{(u v)^{n_1}} \frac{1}{\Gamma(- 4 n_1)} \int \frac{d s}{2 \pi i} \int \frac{ d t}{2 \pi i}\, \Gamma( - 4 n_1 + s + t) \Gamma(- s) \Gamma(-t)\, (u^{\frac{1}{4}})^{s} (v^{\frac{1}{4}})^t~,  
\ee
and comparing with (\ref{4MeqD}), we find the Mellin amplitude is, 
\be \label{MelLarge}
M(\gamma_{12}, \gamma_{14}) =\frac{-1}{(n_1!)^4}  \frac{16}{\Gamma( - 4n_1)}\, \frac{\Gamma( 4n_1 - 4\gamma_{12} - 4 \gamma_{14}) \Gamma( 4 \gamma_{12} - 4 n_1) \Gamma( 4 \gamma_{14} - 4 n_1)}{\Gamma( n_1 - \gamma_{12} - \gamma_{14} )^2 \Gamma( \gamma_{12})^2 \Gamma(\gamma_{14})^2}~.
\ee

For a CFT$_1$ four-point function, it would seem more natural to consider a Mellin amplitude that is a function of only one variable. The reason for studying a two variable Mellin amplitude is because this is natural from the AdS$_2$ perspective, as will be discussed in the next section. 

\newpage
\section{Bulk} \label{sec:BULK}

\subsection{Constructing the bulk}
\begin{figure}[t]
\centering
\subfloat[]{
\includegraphics[width=.9in]{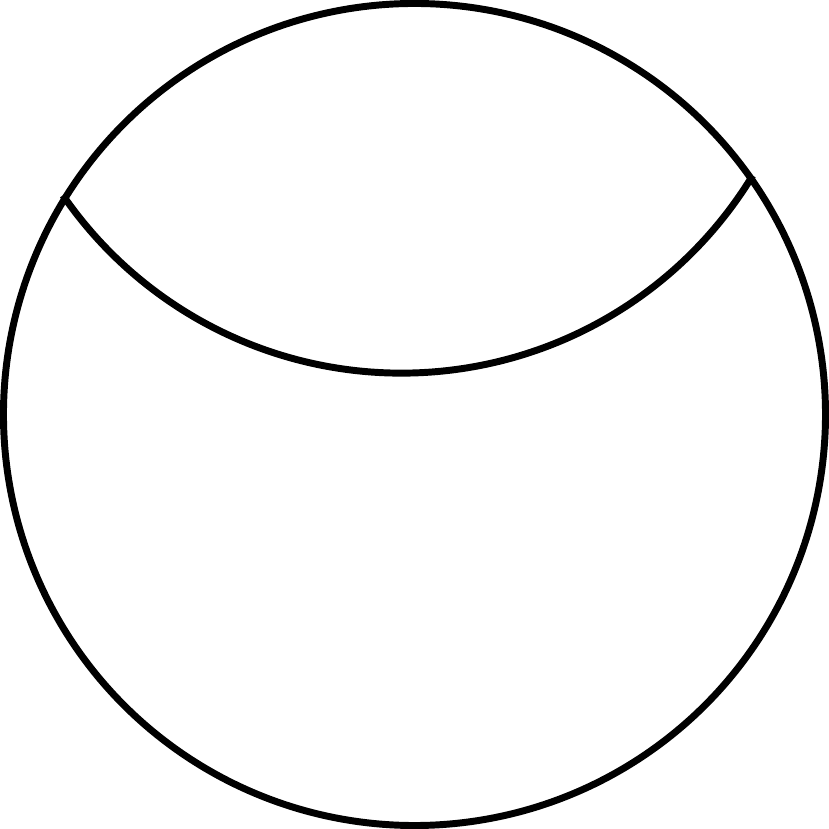}
} \hspace{3cm}
\subfloat[]{
\includegraphics[width=.9in]{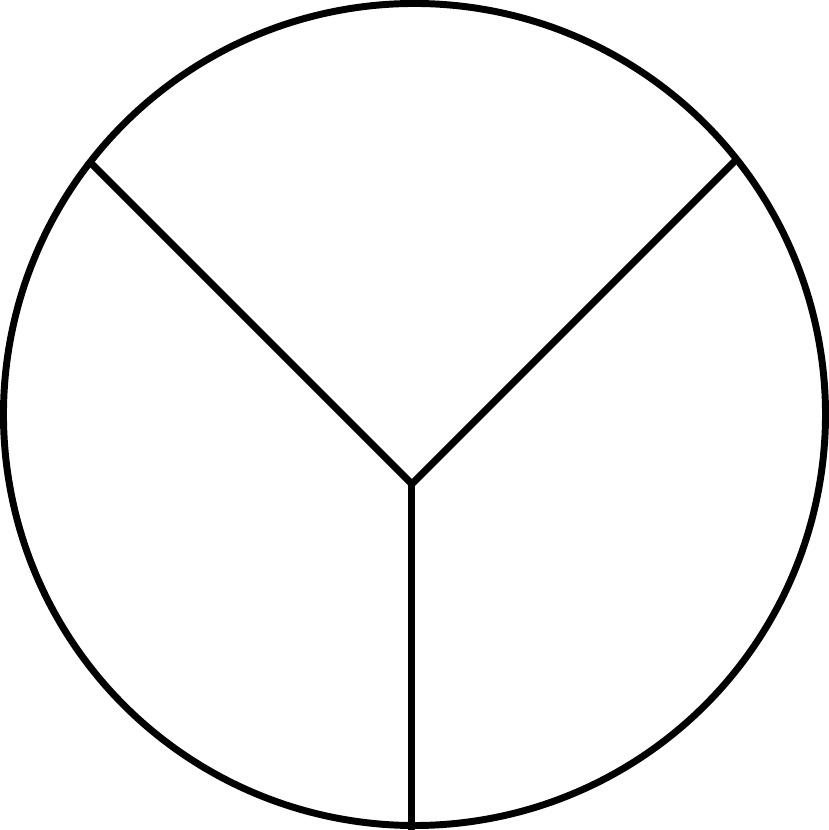}
} \ \ \ \ 
\subfloat[]{
\includegraphics[width=4.8in]{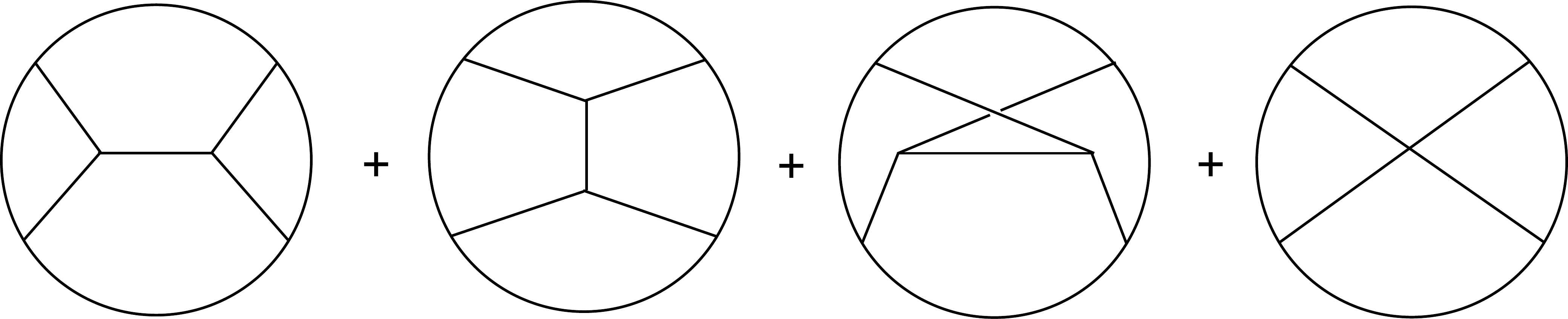}
}
\caption{The Witten diagrams for the AdS computation of  CFT correlation  functions of  $\mO$. The (a) two-point function (b) three-point function and (c) four-point function.   }  \label{Witten234}
\end{figure}

The bilinear, primary, $O(N)$ invariant singlets $\mO_n$
are, via the AdS/CFT dictionary, dual to scalar fields $\phi_n$ in AdS$_2$. Knowing all large $N$ connected correlation functions of the $\mO_n$,  in principle, fully fixes the classical bulk Lagrangian of the  AdS dual of SYK.

On general grounds, we expect the bulk Lagrangian, up to order $1/N$, to take the form,
\begin{multline}\label{Lbulk}
S_{bulk} = \int d^2 x\sqrt{g} \[ \frac{1}{2} (\partial\phi_n)^2 + \frac{1}{2} m_n^2 \phi_n^2 +\frac{1}{\sqrt{N}} \, \lambda_{n m k }\,  \phi_n \phi_m \phi_k\right.  \\
+ \left. \frac{1}{N}\(\lambda_{n m k l}^0\, \phi_n \phi_m \phi_k \phi_l + \lambda_{n m k l}^1\, \partial \phi_n \partial \phi_m \phi_k \phi_l + \ldots\) \]~.
\end{multline}
We have not included cubic interaction terms with derivatives, as these can be eliminated through field redefinitions \cite{GR2}. At the quartic level, it is no longer possible to eliminate derivatives, and indeed there should generically be an infinite number of independent quartic terms, with various combinations of derivatives. 

To establish the coefficients appearing in $S_{bulk}$, one should use this bulk action to compute CFT correlation functions, and fix the coefficients so as to match the SYK correlation functions. 
This is simple to do for the two-point and the three-point functions, as their functional form is fixed by conformal invariance. Evaluating the Witten diagram for the two-point function, Fig.~\ref{Witten234}(a), gives the standard relation between the mass of  $\phi_n$ and the dimension of  $\mO_n$,  $m_n^2 = h_n(h_n - 1)$. From the Witten diagram for the three-point function, Fig.~\ref{Witten234}(b), one obtains a simple relation between the cubic coupling $\lambda_{n m k}$ and the coefficient of the SYK three-point function, $c_{n m k}$.

Starting with the four-point function, the mapping is more involved. Conformal invariance restricts the four-point function to be a function of the cross-ratio, but is insufficient to fix the functional form. As result, matching between bulk and boundary requires matching two functions, rather than just two numbers.
 In particular, on the bulk side, computation of the four-point function involves summing over the exchange and contact Witten diagrams, shown in Fig.~\ref{Witten234}(c). One must sum over all exchange diagrams: in each of the three channels there is one for each exchanged $\phi_n$.  One must also sum over all contact Witten diagrams, accounting for the generically infinite number of quartic terms appearing in the bulk Lagrangian. 

One way of organizing the four-point function is by expanding each of the Witten diagrams as a sum of conformal blocks, and similarly for the SYK four-point function, and then adjusting the bulk couplings so as to make the  coefficients of all blocks match. The matching of the single-trace blocks is automatic, as these only depend on the cubic couplings. In particular, the $s$-channel Witten diagrams, expanded in terms of $s$-channel blocks, will contain single-trace blocks whose coefficients will match the coefficients of the single-trace blocks coming from the sum of $s$-channel SYK Feynman diagrams, that were computed in Sec.~\ref{Sec42}. The same holds for the $t$-channel and $u$-channel. The matching of coefficients of double-trace blocks is where the challenge lies: both the exchange and contact Witten diagrams will contain double-trace blocks, so one must adjust the quartic couplings in order for the total coefficients of the double-trace blocks to match the SYK result.  An approach of this type has been pursued in \cite{Bekaert:2014cea, Bekaert:2015tva, Sleight:2017fpc}, in the context of the duality between the free $O(N)$ model and Vasiliev theory.

A more tractable way of constructing the bulk at the quartic level, at least for local bulk theories of a few fields, is to study the four-point function in Mellin space. As discussed in \cite{Penedones:2010ue}, a contact Witten diagram has a Mellin amplitude that is a polynomial in the Mellin variables, whose order is set by the number of derivatives in the quartic interaction. In previous sections we wrote the SYK four-point function in Mellin space, so one could study it further in this context. The simplest limit is when all four operators have equal and large dimension, in which case the Mellin amplitude takes the form (\ref{MelLarge}). This does not have a natural interpretation as a polynomial, nor should we have expected it to, if the bulk Lagrangian has terms with an arbitrarily large number of derivatives, and moreover, no large gap. We leave an analysis of the bulk at the quartic level to future work: it is likely that the bulk theory should be regarded as a theory of extended objects, rather than local fields. So one should understand the CFT four-point function in this context instead. 

The only thing that we will do in the rest of the section is analyze further the exchange Witten diagrams and relate them to SYK exchange Feynman diagrams.

\subsection{Preliminaries}

We begin by collecting some relevant equations for AdS$_2$ computations of correlation functions. The discussion follows \cite{Penedones:2010ue}, with the notational exception that there $h$ denotes one-half of the boundary spacetime dimension, whereas for us the boundary spacetime dimension is one and $h$ denotes the operator dimension. 

Letting $X$ denote a bulk coordinate and $P$  a boundary coordinate, both in embedding space, the bulk-boundary propagator is,
\be
G_{h}(X, P) = \frac{\mC_{h}}{(- 2 P \cdot X)^{h}}~, \ \  \ \text{where} \ \ \ \ \  \mC_{h} = \frac{\Gamma(h)}{2\sqrt{\pi}\, \Gamma(h + \frac{1}{2})}~.
\ee
Correspondingly, this leads to a CFT two-point function,
\be
\langle \mO_{h}(P_1) \mO_{h}(P_2) \rangle = \frac{\mC_{h}}{(-2 P_1 \cdot P_2)^{h}}~,
\ee
where, upon converting from embedding space to physical space, $- 2 P_1 \cdot P_2 = (\tau_1 - \tau_2)^2$.

Consider a cubic bulk interaction  with coupling equal to one, $\phi_1 \phi_2 \phi_3$, involving fields $\phi_i$ dual to operators $\mO_i$ of dimension $h_i$. The corresponding tree-level Witten diagram determining the CFT three-point function involves a product of three bulk-boundary propagators, see Fig.~\ref{Witten234}(b),
\be \label{bulk3pt}
 \langle \mO_1(P_1) \mO_2(P_2) \mO_3(P_3)\rangle =  \int_{AdS} d X\, G_{h_1}(X, P_1)  G_{h_2}(X, P_2) G_{h_2}(X, P_3)~.
\ee
Evaluation of the integral gives,
\be
\langle \mO_1(P_1) \mO_2(P_2) \mO_3(P_3)\rangle =\frac{\lbb(h_1, h_2, h_3)}{(- 2P_2 \cdot P_3)^{\frac{h_2 + h_3 - h_1}{2}} (- 2P_1 \cdot P_3)^{\frac{h_1 + h_3 - h_2}{2}}(- 2P_1 \cdot P_2)^{\frac{h_1 + h_2 - h_3}{2}}}~,
\ee
where,
 \be \label{lbbD}
\lbb(h_1, h_2, h_3) =  \frac{ \Gamma\(\frac{h_1 + h_2 + h_3-1}{2}\) \Gamma\(\frac{h_2 + h_3 - h_1}{2}\) \Gamma\(\frac{h_1 + h_3 - h_2}{2}\) \Gamma\(\frac{h_1 + h_2 - h_3}{2}\)
}{16\pi \Gamma(h_1+ \frac{1}{2})\Gamma(h_2+ \frac{1}{2}) \Gamma(h_3+ \frac{1}{2})}~.
\ee
As a result,   the relation between the cubic couplings $\lambda_{123}$ and the  coefficients  $c_{123}$  of the CFT three-point function is, 
\be \label{lambdac}
\lambda_{123} = \frac{c_{123}}{\Lambda_{B \partial}(h_1, h_2, h_3)} \sqrt{\mC_1 \mC_2 \mC_3}~,
\ee
where the $\mC_i$ appear due to the CFT convention of two-point functions having norm equal to one.

For computing exchange Witten diagrams, we will need  the bulk propagator for a field dual to an operator of dimension $h$, 
\be \label{GBB}
G_{B B}^h(X,Y) = \frac{\Gamma(h)}{2\sqrt{\pi} \Gamma(h + \frac{1}{2})} u^{-h} {}_2 F_1(h, h, 2h, - \frac{4}{u})~,
\ee
where $u = (X-Y)^2$. One can verify the following representation of the propagator minus the ``shadow'' propagator, written in terms of two bulk-boundary propagators, 
\be  \label{GBBs}
\int_{\partial AdS} dP_0\,\, G_{h}(P_0, X) G_{1- h}(P_0, Y) =\frac{1}{(1- 2h)} \( G_{BB}^h(X,Y) - G_{BB}^{1-h} (X,Y) \)~.
\ee
One may notice the similarity between (\ref{GBB}) and the CFT$_1$ conformal blocks, and between (\ref{GBBs}) and the representation of the conformal block plus its shadow as a product of a three-point function involving $\mO_h$ and a three-point function involving its shadow, see Appendix~\ref{blocks}. This similarity will  be utilized later, in connecting boundary Feynman diagrams to bulk Witten diagrams. Performing a contour integral of (\ref{GBBs}) over $h$ gives the standard split-representation, 
\be \label{BBrho}
G_{BB}^h(X,Y) = \int_{}^{} \frac{ d h_c}{2\pi i}\,  \frac{2(h_c-\frac{1}{2})^2}{( h_c-h)(h_c-1 +h)}\int_{\partial AdS} dP_0\,\, G_{h_c}(X,P_0) G_{1-h_c}(Y,P_0)~,
\ee
where the $h_c$ integral runs parallel to the imaginary axis, $\frac{1}{2} - i\infty<h_c<\frac{1}{2} + i \infty$. 
Finally, a delta function in AdS can also be written in terms of a split-representation, with the same contour,
\be
\delta(X-Y) = -2 \int_{}^{} \frac{ d h_c}{2\pi i}\, (h_c - \frac{1}{2})^2 \int_{\partial AdS} dP_0\,\, G_{h_c}(X,P_0) G_{1-h_c}(Y,P_0)~.
\ee

\subsection{Exchange Witten diagrams}
\begin{figure}[t]
\centering
\includegraphics[width=3in]{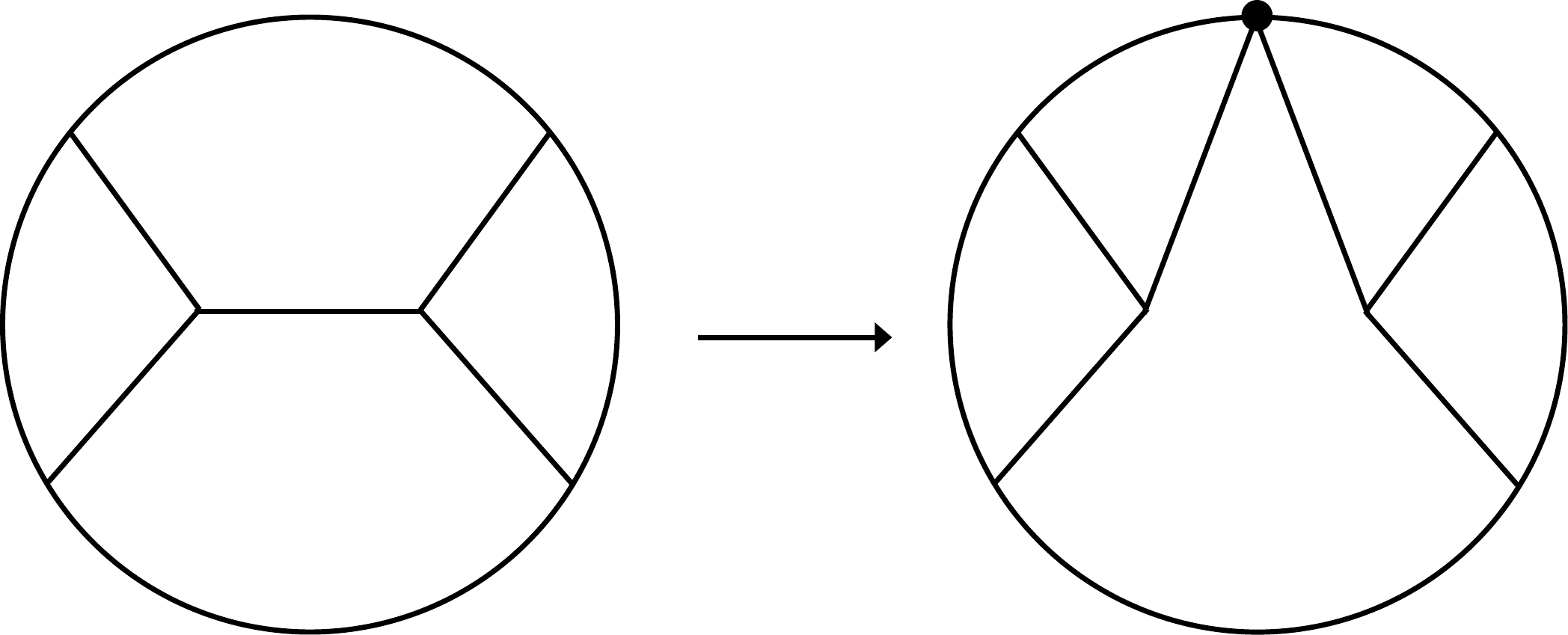}
\caption{A convenient way to evaluate an exchange Witten diagram is to make use of the split-representation of the bulk two-point function.} \label{BulkSplit}
\end{figure} 
Consider an $s$-channel exchange diagram, shown in Fig.~\ref{BulkSplit}, where a field dual to an operator of dimension $h$ is exchanged. This is given by, 
\be \label{WS}
\mW_s = \int d X d Y\, G_{h_1}(\tau_1, X) G_{h_2}(\tau_2, X) G_{BB}^h(X,Y) G_{h_3}(\tau_3, Y) G_{h_4}(\tau_4, Y)~. 
\ee
 In Appendix.~\ref{app:Witten} we evaluate this; using the split-representation for the bulk propagator gives a nice form in terms of single-trace and double-trace conformal blocks. 
For the bulk dual of SYK, since the bulk theory contains a whole tower of fields, we must sum over all the $\phi_{h}$, dual to $\mO_h$,  that can be exchanged, requiring us to evaluate,   
\begin{multline} \label{OOOOs1}
\langle \mO_1 \mO_2 \mO_3 \mO_4\rangle_s = \sum_{h=h_n} \frac{c_{12 h}}{\lbb(h_1, h_2, h)}\frac{c_{3 4 h}}{\lbb(h_3, h_4, h)} \, \mC_h \\
\int d X d Y\, G_{h_1}(\tau_1, X) G_{h_2}(\tau_2, X) G_{BB}^h(X,Y) G_{h_3}(\tau_3, Y) G_{h_4}(\tau_4, Y)~,
\end{multline}
where we have made use of the expression (\ref{lambdac}) relating the cubic couplings to the SYK  three-point function coefficients.  The sum is over all the physical $\mO_h$ in the theory; in particular the $h>0$ that satisfy the transcendental equation, $k_c(h) = 1$ given in 
(\ref{ghSYK}). 
The most direct way to evaluate this would be to use the expression in Appendix.~\ref{app:Witten} for a single Witten diagram expressed as a  single-trace conformal block and a sum of double-trace blocks, see Eq.~\ref{mWs}, and then evaluate the sum over all the operators $h_n$. This will yield a sum of single-trace and double-trace blocks. The coefficients of the single-trace blocks will clearly be the same as what was found for SYK $s$-channel exchange diagrams, $c_{1 2 h} c_{ 3 4 h}$, see Eq.~\ref{OOOOs3}. The coefficients of the double-trace blocks, however, will be complicated. In what follows, we will perform some manipulations to simplify them. 

An important step is to  start by replacing the sum  over the dimensions of the exchanged operators with a contour integral, 
\begin{multline} \label{OOOOs2}
\langle \mO_1 \mO_2 \mO_3 \mO_4\rangle_s =\int_{\mC} \frac{d h}{2\pi i } \frac{\rho(h)}{c_h^2} \frac{\Gamma(h)^2}{\Gamma(2h)}\frac{c_{12 h}}{\lbb(h_1, h_2, h)}\frac{c_{3 4 h}}{\lbb(h_3, h_4, h)} \mC_h \\
\int d X d Y\, G_{h_1}(\tau_1, X) G_{h_2}(\tau_2, X) G_{BB}^h(X,Y) G_{h_3}(\tau_3, Y) G_{h_4}(\tau_4, Y)~.
\end{multline}
To verify that this step is correct, we should check that if we close the contour in (\ref{OOOOs2}) we get back to (\ref{OOOOs1}). In particular, the integrand in (\ref{OOOOs2}) should not have any poles except at those $h$ equal to the physical dimensions, $k_c(h) = 1$, and moreover, for these $h$ the residue of the poles should agree with what is in (\ref{OOOOs1}). The latter property is clearly satisfied, due to the definition of $c_h^2$ in terms of the residue of $\rho(h)$, (\ref{OPEres}). To check the former, that there are no additional poles, recall the analytic structure of $c_{1 2 h}/c_h$, discussed at the end of Sec.~\ref{Sec42}. For $h$ in the right-half complex plane, the only poles we need to potentially be concerned  about are at $h = h_1 + h_2 + 2n$, however $\lbb(h_1, h_2, h)$ also has poles at these $h$, so the ratio $c_{1 2 h}/\lbb(h_1, h_2, h)$ is finite at  $h = h_1 + h_2 + 2n$. Thus, we are justified in going from (\ref{OOOOs1}) to (\ref{OOOOs2}). 

Proceeding, we make use of the property (\ref{c123Rel}) relating $c_{ 1 2 h}$ to $c_{ 12\, 1-h}$ to note that,
\be
\frac{c_h\, c_{1 2\, 1-h}}{c_{1-h}\, c_{12 h}} \frac{\lbb(h_1, h_2, h)}{\lbb(h_1,h_2,1-h)} = \frac{\Gamma(\frac{h}{2})^2 \Gamma(\frac{3}{2}- h)}{\Gamma(\frac{1- h}{2})^2 \Gamma( \frac{1}{2} +h)}~.
\ee
One can verify this implies the following identity,
\be \nonumber
\frac{1}{2}\frac{1}{(1-2h)}\(\frac{c_h\, c_{3 4 \, 1-h}}{c_{1-h}\, c_{3  4h}} \frac{\lbb(h_3 , h_4, h)}{\lbb(h_3 ,h_4,1-h)}  + \frac{\rho(1-h)}{\rho(h)} \frac{c_h\, c_{1 2\, 1-h}}{c_{1-h}\, c_{12 h}} \frac{\lbb(h_1, h_2, h)}{\lbb(h_1,h_2,1-h)}\) = \mC_h \frac{\Gamma(h)^2}{\Gamma(2h)}~.
\ee
Inserting this identity into  (\ref{OOOOs2}) we get, 
\begin{multline} \label{716}
\langle \mO_1 \mO_2 \mO_3 \mO_4\rangle_s =\int_{\mC} \frac{d h}{2\pi i } \frac{\rho(h)}{c_h^2} \frac{c_{12 h}}{\lbb(h_1, h_2, h)}\frac{c_{3 4 h}}{\lbb(h_3, h_4, h)} \\
\frac{1}{2}\frac{1}{(1-2h)}\(\frac{c_h\, c_{3 4 \, 1-h}}{c_{1-h}\, c_{3  4h}} \frac{\lbb(h_3 , h_4, h)}{\lbb(h_3 ,h_4,1-h)}  + \frac{\rho(1-h)}{\rho(h)} \frac{c_h\, c_{1 2\, 1-h}}{c_{1-h}\, c_{12 h}} \frac{\lbb(h_1, h_2, h)}{\lbb(h_1,h_2,1-h)}\) \\
\int d X d Y\, G_{h_1}(\tau_1, X) G_{h_2}(\tau_2, X) G_{BB}^h(X,Y) G_{h_3}(\tau_3, Y) G_{h_4}(\tau_4, Y)~.
\end{multline}
Recall that the contour $\mC$ has two pieces: a line parrallel to the imaginary axis, and circles around even integers,  as was shown in Fig.~\ref{figcontour}. Let us focus on the contribution of the line piece. For this,  we may change variables $h \rightarrow 1- h$ for the second term to get, 
\begin{multline} \nonumber
\langle \mO_1 \mO_2 \mO_3 \mO_4\rangle_s \supset \int d X d Y\, G_{h_1}(\tau_1, X) G_{h_2}(\tau_2, X) G_{h_3}(\tau_3, Y) G_{h_4}(\tau_4, Y)\\
\int_{h= \frac{1}{2}+is} \frac{d h}{2\pi i } \frac{\rho(h)}{c_h c_{1-h}} \frac{c_{12 h}}{\lbb(h_1, h_2, h)}\frac{c_{3 4\, 1- h}}{\lbb(h_3, h_4, 1-h)} \frac{1}{2}\frac{1}{(1-2h)} \( G_{BB}^{h}(X,Y) - G_{BB}^{1-h}(X,Y)\)~.
\end{multline}
Recalling the representation of the bulk two-point function minus its shadow, as given in (\ref{GBBs}), we  rewrite this as,
\begin{multline}
\langle \mO_1 \mO_2 \mO_3 \mO_4\rangle_s \supset
\frac{1}{2}\int_{h= \frac{1}{2}+is} \frac{d h}{2\pi i } \frac{\rho(h)}{c_h c_{1-h}} \frac{c_{12 h}}{\lbb(h_1, h_2, h)}\frac{c_{3 4\, 1- h}}{\lbb(h_3, h_4, 1-h)}\\
 \int d X d Y\,d\tau_0\,  G_{h_1}(\tau_1, X) G_{h_2}(\tau_2, X) G_{h_3}(\tau_3, Y) G_{h_4}(\tau_4, Y) G_{h}(\tau_0, X) G_{1- h}(\tau_0, Y)~.
\end{multline}
Recognizing that the integral over $X$ of three bulk-boundary propagators is what appears in the cubic Witten diagram, (\ref{bulk3pt}), and similarly for the integral over $Y$,
we finally have, 
\be \nonumber
\langle \mO_1 \mO_2 \mO_3 \mO_4\rangle_s \supset
\frac{1}{2}\int_{h= \frac{1}{2}+is} \frac{d h}{2\pi i }\frac{\rho(h)}{c_h c_{1-h}} \int d\tau_0   \langle \mO_1(\tau_1) \mO_2 (\tau_2) \mO_h(\tau_0)\rangle    \langle \mO_3(\tau_3) \mO_4 (\tau_4) \mO_{1-h}(\tau_0)\rangle~.
\ee
This precisely matches the analogous SYK answer, (\ref{exchangeF}), for the $s$-channel exchange Feynman diagrams,  for the contribution of the line piece of the contour.

 If we suppose, for the moment, that the contour $\mC$ necessary for a complete basis of conformal blocks in a one-dimensional CFT consisted only of the line parallel to the imaginary axis (and did not also require circles around positive even integers), then the above would have demonstrated that the sum over all $s$-channel exchange Feynman diagrams in SYK (Fig.~\ref{FigIntro3}) is equal to the sum over all $s$-channel exchange Witten diagrams. This would be a remarkably simple result. Furthermore, it would imply that the SYK Feynman diagrams with no exchanged melons, Fig.~\ref{FigIntro4}, are dual to the sum over all contact Witten diagrams. 

Unfortunately, we must also include the contribution of the contour in (\ref{OOOOs2}) coming from circles wrapping $h = 2n$. Here we can not change variables, so as to form the combination of the bulk propagator and its shadow needed to apply (\ref{GBBs}). We can of course use the expression for an individual Witten diagram as a sum of conformal blocks, and then sum over the $h= 2n$. This will give the same single-trace piece as in the SYK $s$-channel exchange diagrams answer,  as it must, but there will be additional double-trace terms (which, aside from simplicity, we had no reason not to expect).

\section{Discussion} \label{sec:dis}
The conformal symmetry of SYK at strong coupling fully fixes the functional form of the building blocks of all correlation functions. The dynamical information in the fermion two-point, four-point, and six-point functions is captured by $\Delta$, $\rho(h)$, and $c_{123}$, respectively. All higher-point functions are built out of these ingredients. 

The structure of $c_{123}$ is remarkable. Viewed as a function of, for instance,  $h_3$, it has poles at precisely $h_3 = h_1 + h_2 +2n$. These poles give rise to the double-trace blocks in the bilinear four-point function. This is presumably a general result for large $N$ conformal field theories: that the analytic continuation of a three-point function of scalar operators of dimensions $h_i$ has poles at $h_i= h_j+h_k+2n$, where $i, j, k $ are distinct and chosen from $1, 2, 3$.~\footnote{A simple CFT three-point function is the one obtained by computing the Witten diagram for scalars with cubic interaction $\phi_1 \phi_2 \phi_3$. This yields (\ref{lbbD}), which indeed has poles at  $h_i= h_j+h_k+2n$. } It is essential that there be no additional poles, with the exception of those  at the $h_3$ for which $\rho(h_3)=0$, or else the bilinear four-point function would have the wrong structure.  Furthermore, the $c_{123}$  are analytic functions of $h_i$ and $\Delta$, and lead to universality: to the extent that two theories in the SYK family have similar dimensions $h_i$ and $\Delta$, be they SYK at different $q$ or cSYK at different couplings, the $c_{123}$, and by extension, all higher-point functions (not accounting for the additional contact diagrams in Appendix.~\ref{sec:con}), will be similar. 
The large dimension bilinears have small anomalous dimensions. As a result, their correlators are well approximated by those of cSYK at weak coupling: generalized free field theory of fermions of dimension $\Delta$, in the singlet sector. 

Knowing all large-$N$ CFT correlation functions, in principle, determines the full tree-level AdS dual Lagrangian. However, thinking of the bulk as a collection of fields $\phi_n$, dual to the $\mO_n$, with some particular masses and couplings, is not the optimal language: there should be a string-like interpretation of the bulk, which still needs to be formulated. The place to start understanding the bulk is with the correlators of the large dimension operators: the interactions of the very massive bulk fields. We gave a simple expression for the cubic couplings of these. There is a  vague resemblance to string theory: a three-point function of vertex operators for massive string modes comes with combinatorial factors, as a result of the derivatives. The  four-point function of four equal and large dimension $\mO_n$ is also simple, and should have some string-like interpretation. We hope to report on this in future work. 

\bigskip

\section*{Acknowledgements} \noindent We thank O.~Aharony, A.~Gadde, T.~Hartman, M.~Isachenkov, K.~Jensen, J.~Kaplan, I.~Klebanov, J.~Maldacena, J.~Penedones, E.~Perlmutter, D.~Poland, and B.~van Rees for helpful discussions.  
This work was supported by NSF grant 1125915.

\appendix

\section{Conformal Blocks} \label{blocks}
Let $\mO_i$ be CFT$_1$ operators of dimensions $h_i$. Performing an OPE expansion, 
\be
\mO_1(\tau_1) \mO_2(\tau_2) = \sum_h c_{12 h} \mC_{12 h}(\tau_{12}, \partial_2) \mO_h~,
\ee
where $c_{1 2 h}$ are the OPE coefficients, and the function $\mC_{12 h}(\tau_{12}, \partial_2)$ is present in order to include the contributions of all the decedents of $\mO_h$. The functional form of $\mC_{12 h}(\tau_{12}, \partial_2)$ is fully fixed by conformal invariance: $SL(2,R)$ for a CFT$_1$. In particular, applying the OPE to the first two operators in  a three-point function gives,
\be
\langle \mO_1(\tau_1) \mO_2(\tau_2) \mO_3(\tau_3)\rangle =c_{123}\, \mC_{1 2 3}(\tau_{12}, \partial_2) \langle \mO_3(\tau_2) \mO_3(\tau_3)\rangle ~.
\ee
The function $\mC_{123}(\tau_{12}, \partial_2)$ can now be found in an explicit form, through Taylor expansion, in powers of $\tau_{12}$,  of the conformal three-point function on the left-hand side of the above,
\be \label{CC1}
\mC_{123}(\tau_{12}, \partial_2) \frac{1}{|\tau_{23}|^{2h_3}} = \frac{|\tau_{12}|^{h_3 - h_1 - h_2}}{ |\tau_{13}|^{h_1 + h_3 - h_2} |\tau_{23}|^{h_2 + h_3 - h_1}} =\frac{ |\tau_{12}|^{h_3 - h_1 - h_2}\, }{|\tau_{23}|^{2h_3}}\Big(1 + (h_2 - h_1- h_3)\frac{\tau_{12}}{\tau_{23}}+ \ldots\Big)~.
\ee
Equipped with $\mC_{123}(\tau_{12})$, through successive application of the OPE, the functional form of the building blocks of any correlation function is fixed. In particular, consider a four-point function, and apply the OPE either once or twice, 
\bea
\langle \mO_1(\tau_1) \mO_2(\tau_2) \mO_3(\tau_3) \mO_4(\tau_4)\rangle &=& \sum_h c_{1 2 h} \mC_{12 h}(\tau_{12}, \partial_2) \langle \mO_h(\tau_2) \mO_3(\tau_3) \mO_4(\tau_4)\rangle \\
&=&  \sum_h c_{12 h} c_{3 4 h} \mC_{1 2 h}(\tau_{12}, \partial_2)\, \mC_{3 4 h}(\tau_{34}, \partial_4)\, \frac{1}{|\tau_{24}|^{2 h}}~.
\eea
The conformal blocks are identified as the functions appearing in the latter expansion, 
\be
\mF_{1234}^h(x) \equiv  \mC_{1 2 h}(\tau_{12}, \partial_2)\, \mC_{3 4 h}(\tau_{34}, \partial_4)\, \frac{1}{|\tau_{24}|^{2 h}}~,
\ee
so that,
\be
\langle \mO_1(\tau_1) \mO_2(\tau_2) \mO_3(\tau_3) \mO_4(\tau_4)\rangle = \sum_h c_{1 2 h} c_{3 4 h}\, \mF_{1234}^h(x)~.
\ee
The explicit functional form of the conformal blocks is in terms of a hypergeometric function \cite{Dolan:2000ut},~\footnote{The notation, $\mF_{1234}^h(x)$, is somewhat inaccurate, because as a result of the prefactors, the conformal blocks are really functions of all the times, not just $x$. 
} 
\be \label{CPW}
\mF_{1234}^h(x) = \Big|\frac{\tau_{24}}{\tau_{14}}\Big|^{h_{12}}\Big|\frac{\tau_{14}}{\tau_{13}}\Big|^{h_{34}} \frac{1}{|\tau_{12}|^{h_1 +h_2}|\tau_{34}|^{h_3 + h_4}}x^{h}\, {}_2 F_1 (h- h_{12}, h+ h_{34}, 2h, x)~,
\ee
where $h_{ij} \equiv h_i - h_j$ and $x$ is the conformal cross-ratio, 
\be
x = \frac{\tau_{12} \tau_{34}}{\tau_{13}\tau_{24}}~.
\ee

A simple alternative way of deriving the conformal blocks is through the shadow formalism \cite{Ferrara1972, SimmonsDuffin:2012uy}, see also \cite{Murugan:2017eto, Bulycheva:2017uqj}. For an operator $\mO_h$ having dimension $h$, its shadow $\mO_{1-h}$ has dimension $1-h$. Consider the integral of a product of a three-point function involving $\mO_h$ and a three-point function involving $\mO_{1-h}$, 
\bea \nonumber
\mB_{1 2 3 4}^h &=&\int d\tau_0\,  \langle \mO_1(\tau_1) \mO_2(\tau_2) \mO_{h}(\tau_0)\rangle  \langle \mO_{1-h}(\tau_0)\mO_3(\tau_3) \mO_4(\tau_4)\rangle\\ \label{SbS}
& = &\int d\tau_0 \frac{|\tau_{12}|^{h- h_1 - h_2}}{|\tau_{10}|^{h+h_{12}} |\tau_{20}|^{h-h_{12}}}\frac{|\tau_{34}|^{1-h - h_3 - h_4}}{|\tau_{30}|^{1-h+h_{34} }|\tau_{40}|^{1-h - h_{34}}}~.
\eea
After a change of variables this becomes,
\be
\mB_{1234}^h =\frac{1}{|\tau_{12}|^{h_1 + h_2}}\frac{1}{|\tau_{34}|^{h_3 + h_4}}\Big|\frac{\tau_{24}}{\tau_{14}}\Big|^{h_{12}} \Big|\frac{\tau_{14}}{\tau_{13}}\Big|^{h_{34}} \int d\tau_0 \frac{|1-x|^{h_{12} - h_{34}} |x|^h}{|\tau_0|^{h- h_{12}}|\tau_0 - x|^{h+h_{12} } |\tau_0 - 1|^{1 - h - h_{34}}}~.
\ee
Evaluating the integral gives a sum of a conformal block of an exchanged $\mO_h$ and a conformal block of its shadow,
\be \label{IFF}
\mB_{1234}^h= \beta(h, h_{34}) \mF_{1234}^h (x) + \beta(1-h, h_{12}) \mF_{1234}^{1-h}(x)~, 
\ee
where we defined, 
\be \label{betaD}
\beta(h,\Delta) =\sqrt{\pi} \frac{\Gamma(\frac{h+\Delta}{2}) \Gamma(\frac{h-\Delta}{2})}{\Gamma(\frac{1-h+\Delta}{2})\Gamma(\frac{1-h-\Delta}{2})}\frac{\Gamma(\frac{1}{2}- h)}{\Gamma(h)}~.
\ee
In evaluating the integral, we have taken the cross-ratio to be in the range $0<x<1$. Through a simple change of variables, one can obtain $\mB_{1234}^h$ for other ranges of $x$ as well. 

In the special case that the four external operators are fermions with dimension $\Delta$, this kind of integral was encountered in the SYK fermion four-point function, see Eq.~\ref{PsiIntr}, in which case  $\Psi_h(x)$ was defined as, 
\be
\frac{2}{|\tau_{12}|^{2\Delta} |\tau_{34}|^{2\Delta}} \Psi_h(x) = \beta(h, 0)\mF_{\Delta}^{h}(x) + \beta(1-h,0) \mF_{\Delta}^{1-h}(x)~,
\ee
where $\mF_{\Delta}^h(x)$ denotes the conformal block (\ref{CPW}) with $h_i=\Delta$, 
\be \label{FermionBlock}
\mF_{\Delta}^{h} (x) = \frac{\sgn(\tau_{12})\, \sgn(\tau_{34})}{|\tau_{12}|^{2\Delta}|\tau_{34}|^{2\Delta}}\, x^h\, {}_2 F_1(h, h, 2h, x)~.
\ee
Since we are dealing with fermions, we have added an antisymmetry factor of  $\sgn(\tau_{12}) \sgn(\tau_{34})$ relative to the definition in (\ref{CPW}).

\subsection{Mellin space} \label{app:Mellin}
Mellin space is useful for large $N$ CFTs \cite{Mack:2009mi, Penedones:2010ue}. The Mellin amplitude $M(\gamma_{i j})$ for a four-point function  is defined by, 
\be \label{MelDEF}
\langle \mO_1(\tau_1) \cdots \mO_4(\tau_4)\rangle = \int \frac{[d \gamma]}{2\pi i}\, M(\gamma_{i j}) \prod_{i < j} \frac{\Gamma(\gamma_{i j})}{(\tau_{i j}^2)^{\gamma_{i j}}}~,
\ee
where the $\gamma_{i j}$ have the constraints, 
\be \label{Melrel}
\sum_{j=1}^4 \gamma_{i j} = 0~, \ \ \ \ \ \gamma_{i j} = \gamma_{j i }~, \ \ \ \ \ \ \gamma_{i i } = - h_i~,
\ee
and the integral $[d\gamma]$ is over  two independent $\gamma_{i j}$, which we will take to be $\gamma_{12}$ and $\gamma_{14}$. Solving the constraints for the others, 
\bea
\gamma_{13} &=& h_1 - \gamma_{12} - \gamma_{14}~, \hspace{3cm} \  \ \ 
\gamma_{23} = \frac{ - h_{12} + h_{34}}{2} + \gamma_{14} \\
\gamma_{24} & =& \frac{ h_1 + h_2 - h_3 + h_4}{2} - \gamma_{12} - \gamma_{14}~, \ \ \ \ 
\gamma_{34} = \frac{ - h_{13} - h_{24}}{2} + \gamma_{12}~.
\eea
The four-point function can therefore be written as, 
\be \label{OOOOmel}
\langle \mO_1(\tau_1) \cdots \mO_4(\tau_4)\rangle =  \frac{(\tau_{23}^2)^{ \frac{  h_{12} - h_{34}}{2}}}{(\tau_{13}^2)^{h_1} }\frac{(\tau_{34}^2)^{\frac{h_{13} + h_{24}}{2}}}{(\tau_{24}^2)^{\frac{h_1 + h_2 - h_3 +h_4}{2}} }\int \frac{ d\gamma_{12}}{2\pi i} \frac{d\gamma_{14}}{2\pi i} \, M(\gamma_{12}, \gamma_{14})\,   \prod_{i< j}\Gamma(\gamma_{ij})\,\, u^{- \gamma_{12}} v^{- \gamma_{14}}~,
\ee
where,
\be \label{uv}
u = \frac{\tau_{12}^2 \tau_{3 4}^2}{\tau_{13}^2 \tau_{24}^2}= x^2~, \ \ \ \ v = \frac{\tau_{14}^2 \tau_{23}^2}{\tau_{13}^2 \tau_{24}^2} = (1-x)^2~.
\ee
We see that, unlike in dimensions greater than one, in one dimension the cross-ratios $u$ and $v$ are not independent. This leads to the non-uniqueness of the four-point Mellin amplitude. In higher dimensions, the $p$-point Mellin amplitude is also not unique, for $p>d+2$.
In the case that all the operators have the same external dimension, $h_1 = h_2 = h_3 = h_4$, the four-point function simplifies to, 
\be \label{4MeqD}
\!\!\! \langle \mO_1(\tau_1) \cdots \mO_1(\tau_4)\rangle\! =\! \frac{1}{(\tau_{13}^2\tau_{24}^2)^{h_1}}\! \int \frac{d \gamma_{12}}{2\pi i} \frac{d\gamma_{14}}{2\pi i } \, M(\gamma_{12}, \gamma_{14})\, \Gamma(\gamma_{12})^2 \Gamma(\gamma_{14})^2 \Gamma( h_1 -\gamma_{12} - \gamma_{14})^2\, u^{- \gamma_{12}} v^{- \gamma_{14}}~.
\ee

In order to find the Mellin transform of a general four-point function, it is convenient to have the Mellin transform of a conformal block plus its shadow, $\mB_{1234}^h$. We will denote this by $\tilde{M}_{1234}^h(\gamma_{12}, \gamma_{14})$.   A simple way to find $\tilde{M}_{1234}^h$ is to start with the integral definition of $\mB_{1234}^h$ (\ref{SbS}) and evaluate it through the standard Mellin-Barnes technique. After an appropriate change of variables, it can be brought into the form (\ref{MelDEF}), with,
\be \label{Mtilde}
\tilde{M}_{1234}^h(\gamma_{12}, \gamma_{14}) = \frac{\pi^{\frac{1}{2}}}{\Gamma(\frac{h +  h_{12}}{2})\Gamma(\frac{h - h_{12}}{2}) \Gamma(\frac{1 - h + h_{34}}{2}) \Gamma(\frac{1 - h - h_{34}}{2})}\frac{\Gamma(\gamma_{12} + \frac{h - h_1 - h_2}{2})}{\Gamma(\gamma_{12})}\frac{\Gamma( \gamma_{12} + \frac{ 1- h - h_1 - h_2}{2})}{\Gamma(\gamma_{12} - \frac{h_{13} +h_{24}}{2})}~.
\ee

\section{Large $q$ Limit}  \label{largeQ}
In this appendix we study the large $q$ limit of the three-point and four-point functions. As a result of the fermions having a small anomalous dimension in the infrared, $\Delta = 1/q$, there are some simplifications. 

 For $q\gg 1$, the dimensions of the $\mO_n$ approach their free-field values, $2n+1$, 
\be \label{hnq}
h_n = 2 n + 1 + 2 \epsilon_{n}~, \ \ \ \  \epsilon_{n} = \frac{1}{q} \frac{2 n^2 +n + 1}{2 n^2 +n - 1}~, \  \ \ \ \ n\geq1~,\ \ \ q\gg 1~,
\ee
while the OPE coefficients in the large $q$ limit behave as,
\be \label{cnLarge}
c_n^2 = \eps_n^2 \frac{n(1+2n)}{\(n(1+2n)+1\)\(n(1+2n)-1\)}\frac{\sqrt{\pi} \Gamma(2n+1)}{\Gamma(2n+\frac{1}{2}) 2^{4n-2}}~, \ \ \ \ \ q \gg 1~.
\ee

\subsection*{Three-point function}
The contact  diagram contribution to the three-point function $\langle \mO_1 \mO_2 \mO_3\rangle$ has a coefficient that was denoted by $c_{123}^{(1)}$, given in (\ref{c1231}),   $c_{123}^{(1)} = c_1 c_2 c_3\, \mI_{123}^{(1)}$, where $\mI_{123}^{(1)}$ was given in (\ref{I1exact}). Inserting (\ref{hnq}) into $\mI_{123}^{(1)}$ and expanding to leading order in $\epsilon$ gives \cite{GR2}, 
\be \label{C1}
\mI_{123 }^{(1)} = 2 s_{123 }^{(1)} \frac{\eps_{n_1}+ \eps_{n_2} + \eps_{n_3}}{\eps_{n_1} \eps_{n_2} \eps_{n_3}}~, \ \ \ \ q \gg 1~,
\ee
where $s_{123}^{(1)}$ is,
\be \label{C1s}
\!\!s_{123 }^{(1)} =\!  (-4)^{n_1 + n_2 +n_3} \frac{\Gamma(\frac{1}{2}\!+n_2\!+\! n_3\! -\! n_1) \Gamma(\frac{1}{2}+n_1\! +\!n_3\!-\!n_2) \Gamma(\frac{1}{2}+ n_1\!+\!n_2\!-\!n_3) \Gamma(1\!+n_1\!+\!n_2\!+\!n_3)}{\pi^{\frac{3}{2}} \Gamma(1+2n_1 ) \Gamma(1+2n_2)\Gamma(1+2n_3)}  ~.
\ee
The planar diagram contribution to the three-point function has a coefficient that was denoted by $c_{123}^{(2)}$, given in (\ref{c1232}) as $c_{123}^{(2)} = c_1 c_2 c_3\, \xi (h_1)\xi(h_2) \xi(h_3)\, \mI_{123}^{(2)}$, where $\mI_{123}^{(2)}\equiv \mI_{123}^{(2)}(z=1)$ was given in (\ref{Ansatz}). Taking the large $q$ limit this simplifies.
 Defining 
\be
\eps^{\pm} = \eps\pm \Delta~,
\ee
the factor, $\xi (h_n)$ simplifies to, 
\be
\xi(h_n)  =\frac{\eps_n^-}{\eps_n} (n + \frac{1}{2})~,~\ \ \ \ \ \ q \gg 1~,
\ee
while the expression for $\mI_{123}^{(2)}$ simplifies to,
  \be \label{I1232q}
 \mI_{123}^{(2)} = s_{ 123}^{(2)} \(2\frac{(\eps_1^+ + \eps_2^-)(\eps_2^+ + \eps_3^-)(\eps_3^+ + \eps_1^-)}{\eps_1^+\eps_1^-\eps_2^+\eps_2^-\eps_3^+\eps_3^-} - \frac{1}{\eps_1^+ \eps_2^+\eps_3^+}- \frac{1}{\eps_1^- \eps_2^-\eps_3^-}\)~, \ \ \ \ q \gg 1~,
 \ee
 where we are using the short hand $\eps_{i} \equiv \eps_{n_i}$, and $s_{123}^{(2)}$ is,
  \be \label{s1232}
\!\!\! s_{123}^{(2)} =  \frac{(2n_1+2n_2-2n_3)!(2n_2 +2n_3-1)!}{(2n_1-1)!(2n_2-1)!(2n_3-1)! (1+2 n_2 - 2n_3)!}
\,\,\,\, \pFq{4}{3}{1\!-\!2n_1~, 2\! +\! 2n_1~, 1\! -\! 2n_3~, -\! 2n_3}{ 2~, 1\!-\!2n_2\! -\! 2n_3~, 2\! +\! 2n_2 \!-\! 2n_3}{1}~.
 \ee
 In writing it in this form we have assumed $n_1>n_2>n_3$. Using the definition of ${}_4 F_3$, this may be written as a single finite sum. Previously, in \cite{GR2}, we found (\ref{I1232q}) without taking the large $q$ limit of the exact answer, but rather by evaluating the integral $I_{123}^{(2)}$ to leading order in $1/q$. There we noted that $s_{123}^{(2)}$ is the same expression that appears in computing the three-point function in a generalized free field theory with fermions of dimension $\Delta$, in the limit $\Delta\rightarrow 0$. Specifically (see Eq.~\ref{pPoint}),  
 \be \label{mCSum}
\!\!\!\!\! s_{123}^{(2)} = -\!\!\!\sum_{p_1, p_2, p_3}\!\! \!\!\binom{2n_1}{p_1}\!\binom{2n_2}{p_2}\!\binom{2 n_3}{p_3}\!
 \binom{2n_1\!+\!p_2\!-\!p_1}{p_2+1}\!\binom{2n_2\!+\!p_3\!-\!p_2}{p_3+1}\!\binom{2 n_3 \!+\!p_1\!-\!p_3}{p_1+1}\!
 \frac{z^{p_1-p_2 + 2n_2 - 2 n_3}}{ (\!-1\!-\!z)^{p_3 - p_2+2n_1 - 2 n_3}} ~,
 \ee
where $z$ is a cross ratio of times; the answer is independent of $z$. While it is not manifest that (\ref{s1232}) and (\ref{mCSum}) are the same, one can verify that they are.

\subsection*{Four-point function}
The $s$-channel contribution to the four-point function was given in (\ref{OOOOs3}).
 The only term that was not  explicitly stated there is the residue of $c_{h 1 2}/c_h$ at $h = h_1 +h_2 + 2n$. This consists of two terms, coming from $c_{h 1 2} = c_{h 12}^{(1)} + c_{h 1 2}^{(1)}$. Using (\ref{c1231}) gives, 
 \be
\!\!\!\!\!\!\text{Res }\frac{ c _{h 1 2}^{(1)}}{c_h} \Big|_{h =h_1 + h_2 +  2n } 
\!\!\!=\!c_1 c_2 \frac{(-1)^n}{n!}\,\frac{\sqrt{\pi}\,  4^{h_1 + h_2 + n}\, \Gamma(1\! -\! h_1) \Gamma(1\!-\! h_2) \Gamma(1\!-\!h_1\! -\! h_2\!-\! 2n)}{\Gamma(1\!-\! h_1\! -\! n) \Gamma(1\!-\!h_2\! -\! n) \Gamma( \frac{3}{2}\! -\! h_1\! -\! h_2\! -\! n)} \!\[\frac{1}{\cos \pi( h_1\! +\! h_2)}\!-\!1\]~.
\ee
 For the other  contribution, using (\ref{c1232}), and noting that the term giving the residue comes from a gamma function in $\alpha_4$, we get, 
\begin{multline}
\text{Res}\,\frac{c^{(2)}_{h 1 2}}{c_h}\Big|_{h = h_1 + h_2 +2n} = c_1 c_2\, \, 2\frac{(-1)^n}{n!} \frac{\Gamma(\frac{2\Delta+1}{2})^3}{\Gamma(1-\Delta)^3}\frac{\Gamma(\frac{1-2 h_1}{2})}{\Gamma(h_1)}  \frac{\Gamma(\frac{1- h_1 - h_2 - 2n}{2})}{\Gamma(\frac{h_1 + h_2 + 2n}{2})}\frac{\Gamma(\frac{2 +h_1 - 2\Delta}{2})}{\Gamma(\frac{1-h_1 + 2\Delta}{2})}\frac{\Gamma(\frac{2 + h_2 - 2\Delta}{2})}{\Gamma(\frac{1-h_2 + 2\Delta}{2})}\\
\frac{\Gamma(\frac{1-h_1 - h_2}{2})}{\Gamma(\frac{h_1 + h_2}{2})}\frac{\Gamma(\frac{2h_1 + 2n}{2})}{\Gamma(\frac{1-2h_1 - 2n}{2})}\frac{\Gamma( \frac{2h_1 + 2h_2 + 2n-1}{2})}{\Gamma(\frac{2 - 2h_1 -2h_2 - 2n}{2})}\frac{\Gamma(\frac{3- h_1 - h_2 - 2\Delta}{2})}{\Gamma(\frac{h_1 + h_2 + 2\Delta}{2})}\frac{1}{\Gamma(\frac{2n+1}{2})} \\
\pFq{4}{3}{-2n~, 2h_1\! +\! 2h_2\!+\! 2n\! -\! 1~, h_1\! -\! 1\! +\! 2 \Delta~, h_1}{2 h_1~, h_1\! +\! h_2\! -\! 1\! +\! 2\Delta~, h_1\!+\! h_2}{1}~.
\end{multline}
The large $q$ limit of these expressions is not much simpler, so we won't write it.

Another term entering the four-point function is $c_{ 3 4 h}$ at $h = h_1 +h_2 + 2n$. In the large $q$ limit, $h_i$ are close to odd integers, and so $h$ is close to an even integer. This has a different large $q$ limit from the one we already studied, in which all three dimensions are close to odd integers. In the current case, two of the dimensions are odd, and one is even. Taking, more generally, 
\be
\tilde{h}_1 = 2 n_1 + 2 + 2\tilde{\eps}_1~, \ \ \ 
h_2 = 2 n_2 + 1 + 2\eps_2~, \ \ \ 
h_3 = 2 n_3 + 1 + 2\eps_3~,
\ee
and assuming $h_1> h_2> h_3$ (the other cases can be similarly worked out),
we find the large $q$ limit to be,
\begin{multline}
c_{ \tilde{1} 2 3}^{(2)} \rightarrow c_1 c_2 c_3 \frac{(\eps_2 + \eps_3)}{\eps_2 \eps_2^+\eps_3\eps_3^+}\frac{(2+2n_1)(1+2n_2)(1+2n_3)\Gamma(2+2 n_1+2n_2-2n_3)\Gamma(2n_2 + 2n_3)}{4 \Gamma(2n_1+1)\Gamma(2n_2) \Gamma(2n_3)\Gamma(2+2n_2-2n_3)}\\
\pFq{4}{3}{-2n_1~, 3+2n_1~, 1-2n_3~, -2n_3}{2~, 1-2n_2 - 2n_3~, 2+2n_2 - 2n_3}{1}
\end{multline}
It is also straightforward to take the large $q$ limit of the other piece, $c_{ \tilde{1} 2 3}^{(1)}$, but it does not simplify significantly. 

Finally, the expression for the four-point function contains a $\rho(h)$. Since the large $q$ limit of $k_c(h)$ is simple, $k_c(h) \rightarrow 2/(h(h-1))$, we have that, at large $q$,
\be
\rho(h)\rightarrow\frac{1}{\eps_1 + \eps_2}\,  \frac{1}{\pi^2}\frac{4h - 2}{h(h-1) - 2}~,
\ee
where we took $h = h_1 + h_2 + 2n$ and used the large $q$ expression for $h_i$ given in (\ref{hnq}).

\section{Free Field Theory} \label{FFT}
The cSYK model \cite{GR3} is a variant of SYK that is conformally invariant for all values of the coupling. All the results in the paper can be trivially generalized to cSYK for arbitrary coupling. In this appendix we study the particular limit of weak coupling, in which cSYK becomes a generalized free field theory of fermions of dimension $\Delta$.

\begin{figure}
\centering
\includegraphics[width=1.6in]{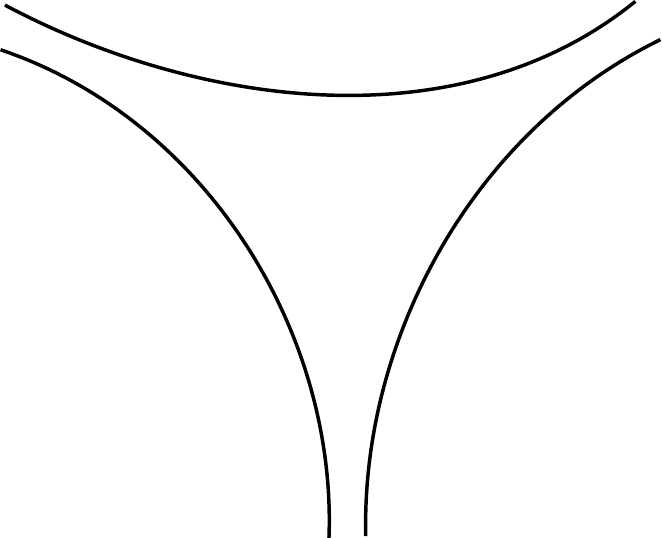}
\caption{In the limit that the coupling goes to zero, the only surviving Feynman diagram is the one without any melons.} \label{Fig6ptFree}
\end{figure}
The cSYK model \cite{GR3} has an action made up of the SYK interaction term (\ref{SSYK}) along with a bilocal kinetic term (\ref{S0}). The model has SL(2,R) symmetry for any value of the (marginal) coupling $J$, with a fermion two-point function, 
\be
G(\tau) = \bar{b} \frac{\sgn(\tau)}{|\tau|^{2\Delta}}~,
\ee
where $\bar{b}$ is given implicitly through, 
\be
\frac{\bar{b}^q}{1- 2 \bar{b}} = \frac{1}{2\pi J^2}(1- 2\Delta) \tan \pi \Delta~.
\ee
It is trivial to generalize the $J\gg1$ results in the body of the paper to any value of $J$. 

In the limit $J\rightarrow 0$, the action becomes that of a generalized free field theory, (\ref{S0}). 
In this limit the kernel $k_c(h)$ (\ref{ghSYK}) near the poles can be expanded as,
\be \label{ghpole}
k_c(h) = \frac{\gamma_n}{ h - (2n+2\Delta+1)}+ \ldots~, \ \ \ \ \ \ \ \gamma_n =2(1-\Delta)(1-2\Delta) \frac{\Gamma(2n+ 4\Delta)}{\Gamma(2n+2)\Gamma(2\Delta) \Gamma(1+2\Delta)}~.
\ee
 At leading order in $J^2$,  the bilinear dimensions $h_n$ are therefore, 
\be \label{hnWeak}
h_n = 2\Delta+ 2n+1 + (1-2 {\bar b}) \gamma_n ~.
\ee
Note that $(1-2{\bar b})$ scales like $J^2$ for small $J$. 
We can now take the limit of $J=0$, to find for the OPE coefficients $c_n$ \cite{GR3},
\be
c_n^2 = \frac{2}{ \Gamma(2\Delta)^2} \frac{ (4\Delta+1 + 4n )\Gamma( 2n +1 +2\Delta)^2 \, \Gamma(2n + 4\Delta)}{\Gamma(2n+2) \Gamma(4n + 2 +4\Delta) }~, \ \ \ \ \ J=0~.
\ee
Let us  look at the three-point function $\langle \mO_1 \mO_2 \mO_3\rangle$ in the limit of $J\rightarrow 0$. At zero-coupling, the only Feynman diagrams that appear are of the type shown in Fig.~\ref{Fig6ptFree}. Taking the general result for the coefficient $c_{123}^{(2)}$ given in (\ref{c1232}) for the  sum of the planar diagrams, and using the dimensions (\ref{hnWeak}), we find that $c_{123}^{(2)}$, now denoted as $c_{123}^{\text{free}}$ is, 
\begin{multline}
c_{123}^{\text{free}} = c_1 c_2 c_3 \frac{\sin^2 (2\pi \Delta)\Gamma(2\Delta)^2\Gamma(2n_2 + 2)\prod_{i=1}^3 \Gamma(-2n_i - 2\Delta)}{\pi^2 \Gamma(2n_2 - 2n_3 + 1)\Gamma(- 2n_2 - 2n_3 - 4 \Delta)\Gamma(2n_3 - 2n_1 - 2n_2 - 2\Delta)}\\
 \frac{\Gamma(2n_3 + 4\Delta)\Gamma(2n_2 + 2)\Gamma( 1-2n_2- 4\Delta)- \pi \csc(4\pi \Delta) \Gamma(2n_3+2)}{\Gamma(2n_3 + 2)\Gamma(2 n_2 + 4\Delta) - \Gamma(2n_2 + 2) \Gamma(2n_3 + 4\Delta)}\\
 \pFq{4}{3}{-1\!-\!2n_3~, -\! 2n_1 \!-\! 2\Delta~, -2n_3\! -\! 2\Delta~, 1\!+\! 2n_1\! +\! 2\Delta}{1\!+\!2n_2\! -\! 2n_3~, -2n_2\! -\! 2n_3\! -\! 4\Delta~, 2\Delta}{1}~.
\end{multline}
 In writing it in this form we  have assumed $n_1>n_2>n_3$.
  
In Sec.~\ref{FREE} we studied the generalized free field theory in a more direct way, computing the three-point function in terms of Wick contractions, see Eq.~\ref{pPOINT}, which instead gives the answer in the form  of a triple sum.

\section{Fermion Correlation Functions} \label{FermionP}
\begin{figure}[t]
\centering
\includegraphics[width=4.5in]{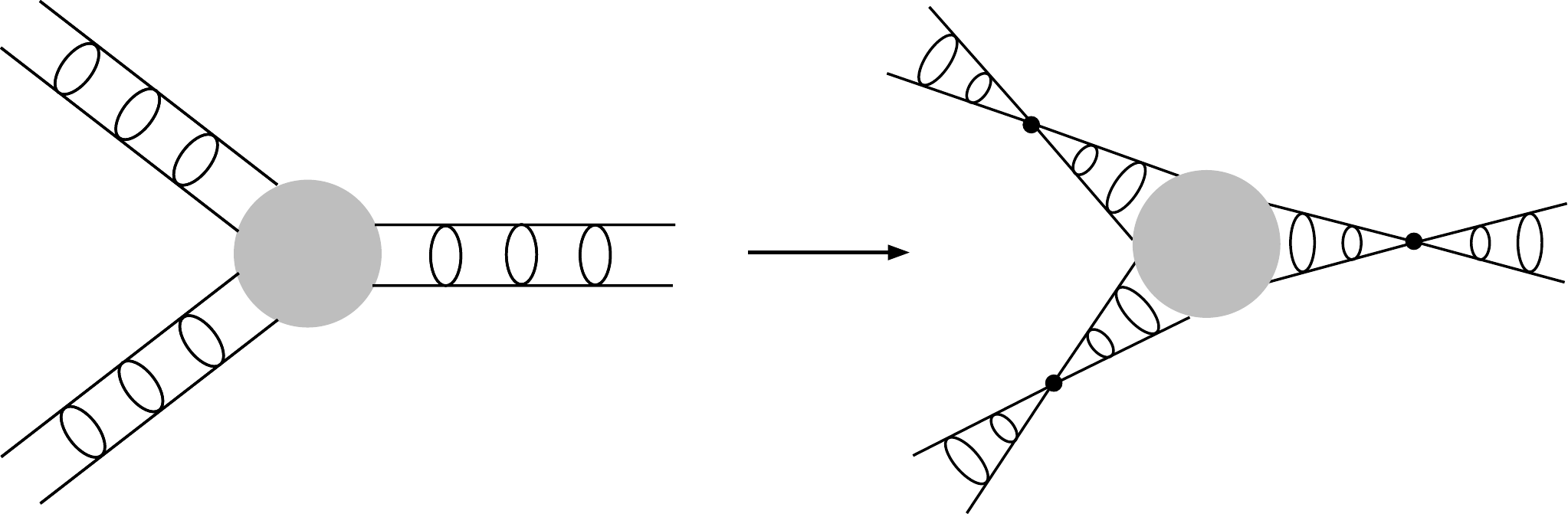}
\caption{The fermion six point function. Using the split representation of the four-point function, we can obtain the six-point function from the three-point functions of the bilinears.} \label{Fig6to3}
\end{figure}

In this appendix we show how to obtain a fermion $2 p$-point  correlation function from the $p$-point  correlation functions of the $\mO_i$.  

Let us start with the fermion six-point function. Since the $\mO_i$ are contained in the OPE of the fermions, it is clear that, due to conformal symmetry,  one can obtain the fermion six-point function from the three-point functions of the $\mO_i$. Indeed, such an expression was given in Eq.~\ref{mS_1}.~\footnote{ This expression is valid as long as the times in the six-point function are time-ordered, so that the OPE is valid.} 
Here we just give an alternative form of the expression.

The fermion six-point function is expressed in terms of three fermion four-point functions in (\ref{SFFF}). Making use of the fermion four-point function in the form (\ref{Fin2}), and then making use of  (\ref{OOO}) gives the desired expression, see Fig.~\ref{Fig6to3},
\begin{multline} \label{Sthree}
\mS =\prod_{i=1}^3\( \frac{1}{2}\int_{\mC}\frac{d h_i}{2\pi i}\,  \frac{\rho(h_i)}{ c_{h_i} c_{1-h_i}}\) \int d\tau_a d\tau_b d\tau_c   \langle \chi(\tau_1) \chi(\tau_2) \mO_{h_1}(\tau_a)\rangle
 \langle \chi(\tau_3) \chi(\tau_4) \mO_{h_2}(\tau_b)\rangle\\   \langle \chi(\tau_5) \chi(\tau_6) \mO_{h_3}(\tau_c)\rangle
 \langle\mO_{1-h_1}(\tau_a) \mO_{1-h_2}(\tau_b) \mO_{1-h_3}(\tau_c)\rangle~.
\end{multline}
Making use of (\ref{Ochichi}), this can be written explicitly as, 
\begin{multline} \label{SIX}
\mS =\prod_{i=1}^3\( \frac{1}{2}\int_{\mC}\frac{d h_i}{2\pi i}\, \rho(h_i)\)
G(\tau_{12})|\tau_{12}|^{h_1} G(\tau_{34}) |\tau_{34}|^{h_2} G(\tau_{56})|\tau_{56}|^{h_3} \\
\frac{c_{1-h_1\, 1-h_2\, 1-h_3}}{c_{1-h_1} c_{1-h_2} c_{1-h_3}}\int d\tau_a d\tau_b d\tau_c \frac{|\tau_{ab}|^{h_1 + h_2 - h_3 -1}|\tau_{ac}|^{h_1 + h_3 - h_2 -1} |\tau_{bc}|^{h_2 + h_3 - h_1 - 1}}{|\tau_{a1}\tau_{a 2}|^{h_1} |\tau_{b3} \tau_{b4}|^{h_2} |\tau_{c 5} \tau_{c 6}|^{h_3}}~.
\end{multline}

This expression  allows us to verify that   the argument given in Sec.~\ref{contact} for identifying the three-point function $\langle \mO_1 \mO_2 \mO_3\rangle$ from the fermion six-point function is correct. In particular, the three-point function is picked out as the coefficient of the term that has the correct scaling powers of $\tau_{12}, \tau_{34}, \tau_{56}$, in the limit that these become small, 
\be \label{goal}
\hspace{-.5cm} \mS = \!\!\! \sum_{h_1, h_2, h_3}\!\! c_{h_1} c_{h_2} c_{h_3} G(\tau_{12})|\tau_{12}|^{h_1} G(\tau_{34}) |\tau_{34}|^{h_2} G(\tau_{56})|\tau_{56}|^{h_3} \langle \mO_{h_1}(\tau_2) \mO_{h_2} (\tau_4) \mO_{h_3}(\tau_6)\rangle + \ldots
\ee
There are two contributions to this term coming from  (\ref{SIX}). The first involves setting, in the integrand,  $\tau_1 = \tau_2$, $\tau_3 = \tau_4$ and $\tau_5 = \tau_6$, and then performing the integral. The second involves doing a change of variables in the integral, $\tau_{a} \rightarrow \tau_a \tau_{12} + \tau_2$, $\tau_{b} \rightarrow  \tau_b \tau_{34} + \tau_4$, $\tau_c  \rightarrow \tau_c \tau_{56} + \tau_6$, 
 then taking $\tau_1 \rightarrow\tau_2$, $\tau_3 \rightarrow \tau_4$, 
$\tau_5 \rightarrow \tau_6$ and then performing the integral. We then relate $c_{1-h_1, 1-h_2, 1-h_3}$ to $c_{h_1, h_2, h_3}$ through repeated use of (\ref{c123Rel}). We  then  recover (\ref{goal}). 

There is a clear generalization of (\ref{Sthree}) to higher-point correlation functions. Specifically, the leading order in $1/N$ fully connected piece of a $2p$-point fermion correlation is given by, 
\begin{multline} \nonumber
\frac{1}{N^{2p}}\sum_{i_1, \ldots, i_p} \langle \chi_{i_1}(\tau_1) \chi_{i_2}(\tau_2) \cdots \chi_{i_p}(\tau_{2p-1}) \chi_{i_p}(\tau_{2p})\rangle 
\supset\\
  \prod_{i=1}^p\( \frac{1}{2}\int_{\mC}\frac{d h_i}{2\pi i}\,  \frac{\rho(h_i)}{ c_{h_i} c_{1-h_i}}\) \int d \tau_{a_1} \cdots d\tau_{a_p} \prod_{i=1}^p  \langle \chi(\tau_{2 i-1}) \chi(\tau_{2 i}) \mO_{h_i}(\tau_{a_i})\rangle\, 
 \langle\mO_{1-h_1}(\tau_{a_1}) \cdots \mO_{1- h_p}(\tau_{a_p})\rangle~.
\end{multline}

\section{Contact Diagrams} \label{sec:con}

\begin{figure}[t]
\centering
\includegraphics[width=2in]{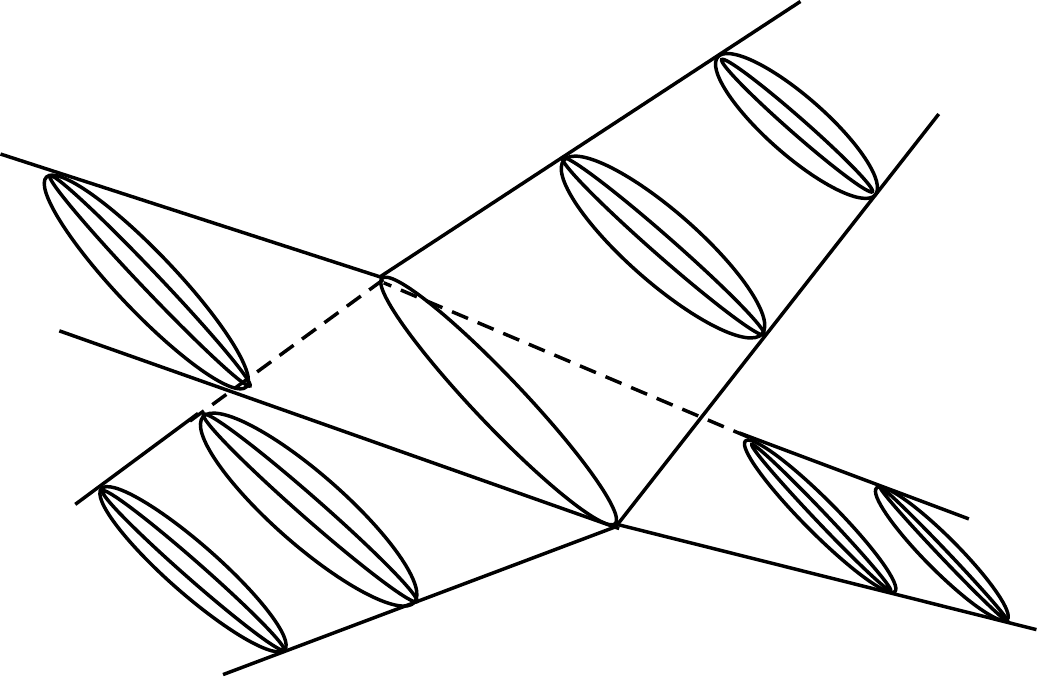}
\caption{There is an additional contribution to the eight-point function, involving four four-point functions meeting at a melon. This diagram is not planar, so it is difficult to draw. We have shown it for SYK with $q=6$, unlike the other diagrams  in the paper which are drawn for $q=4$. } \label{Fig8contact}
\end{figure} 

There is an additional contribution to the fermion eight-point function,  shown in Fig.~\ref{Fig8contact}, which has  four ladders glued to a single melon,
\be 
\mE_c = (q-1)(q-2)(q-3)J^2 \int d\tau_a d\tau_b\, G(\tau_{ab})^{q-4}\, \prod_{\text{i} = 1}^{4} \mF(\tau_{2\text{i}-1}, \tau_{2\text{i}}, \tau_a, \tau_b)~.
\ee
This gives rise to a  contribution to the bilinear four-point function  $\langle \mO_1(\tau_1) \cdots \mO_4(\tau_4) \rangle$ that is, 
\be \label{mC4}
\mathcal{C} = c_1 c_2 c_3 c_4 (q-1)(q-2)(q-3)b^q\, \int\frac{ d\tau_a d\tau_b }{|\tau_{ab}|^2} \prod_{i = 1}^4 \Big|\frac{\tau_{a b}}{\tau_{i a } \tau_{i b}}\Big|^{h_i}~.
\ee
This diagram is novel, and unlike the other contributions to the four-point function studied in the body of the paper, in the sense that it is not made up of fermion six-point functions glued together. 

For a $2p$-point fermion correlation function, there will be an analogous, novel, contact diagram, consisting of $p$ ladders glued to a melon, as long $q \geq p$. This term takes the form, 
\be
\frac{(q-1)!}{(q-p)!} J^2 \int d\tau_a d\tau_b\, G(\tau_{ab})^{q-p}\, \prod_{\text{i} = 1}^{p} \mF(\tau_{2\text{i}-1}, \tau_{2\text{i}}, \tau_a, \tau_b)~, 
\ee
and gives a contribution to the $p$-point function $\langle \mO_1(\tau_1) \cdots \mO_p(\tau_p)\rangle$ that is,
\be
c_1 \cdots c_p\,  \frac{(q-1)!}{(q-p)!}\,  b^q\, \int\frac{ d\tau_a d\tau_b }{|\tau_{ab}|^2} \prod_{i = 1}^p \Big|\frac{\tau_{a b}}{\tau_{i a } \tau_{i b}}\Big|^{h_i}~.
\ee
These integrals can be rewritten in terms of conformal cross-ratios. For instance, 
for the four-point function, we change variables to, 
\be
A= \frac{\tau_{a1} \tau_{23}}{\tau_{a 2} \tau_{13}}~, \ \ \ \ \ B = \frac{\tau_{b 2} \tau_{13}}{\tau_{b 1} \tau_{2 3}}~,
\ee
which gives,
\be
\mathcal{C} = c_1 c_2 c_3 c_4 (q-1)(q-2)(q-3)\, b^q\, \Big|\frac{\tau_{24}}{\tau_{14}}\Big|^{h_{12}}\Big|\frac{\tau_{14}}{\tau_{13}}\Big|^{h_{34}} \frac{1}{|\tau_{12}|^{h_1 +h_2}|\tau_{34}|^{h_3 + h_4}}\, \frac{|x|^{h_3 + h_4}}{|1-x|^{-h_1 + h_2 + h_3}} \mI_{C}~,
\ee
where,
\be \label{IC}
\mathcal{I}_C = \int d A\, d B \frac{| 1- A B|^{h_1 + h_2 + h_3 + h_4 - 2}}{|A|^{h_1}  |B|^{h_2} |(1-A)(1-B)|^{h_3} |(A+ x-1)(B+ \frac{1}{x-1})|^{h_4}}~, 
\ee
where $x$ is the cross-ratio of times. We will not proceed further with evaluating this integral, however one could evaluate it using similar methods as those employed in the paper: considering a restricted integration range and recognizing that portion of the integral as giving rise to a multivariable generalized hypergeometric function, finding all solutions to the differential equation defining this function, and then writing the result for the integral as a linear combination of solutions, and fixing the coefficients from different scaling limits.

\section{Witten Diagrams} \label{app:Witten}
In this appendix we recall the evaluation of exchange and contact Witten diagrams \cite{Liu:1998th, ElShowk:2011ag, Costa:2014kfa, Hijano:2015zsa}.
\subsubsection*{Exchange diagram}
Consider an $s$-channel exchange Witten diagram, $\mW_s$ defined by (\ref{WS}) and shown in Fig.~\ref{BulkSplit}. This involves external fields $\phi_i$ dual to operators $\mO_i$ of dimension $h_i$, and an exchanged  $\phi_h$, dual to an operator $\mO_h$ of dimension $h$.  Making use of the split representation of the bulk propagator, (\ref{BBrho}), we  write this as, 
\begin{multline}
\mW_s = \int \frac{ d h_c}{2\pi i}\, \rho_w(h, h_c) \int_{\partial AdS} dP_0\, \int_{AdS} d X dY \,  G_{h_c}(X,P_0) G_{1-h_c}(Y,P_0)\\ 
G_{h_1}(X,P_1) G_{h_2}(X, P_2)\,  G_{h_3}(Y, P_3) G_{h_4}(Y, P_4)~,
\end{multline}
where the integration contour is parallel to the imaginary axis $\frac{1}{2} - i\infty\! <\! h_c\! <\!\frac{1}{2}+ i \infty$, and we defined, 
\be
\rho_w(h, h_c) =  \frac{2(h_c-\frac{1}{2})^2}{( h_c-h)(h_c-1 +h)}~.
\ee
We see that $\mW_s$ involves a product of two three-point functions (\ref{bulk3pt}), integrated over $P_0$. The $P_0$ integral, done in (\ref{IFF}), gives a conformal block plus its shadow, leaving,
\be \nonumber
\!\!\!\! \mW_s =\!\! \int\! \frac{ d h_c}{2\pi i}\, \rho_w(h, h_c)  \lbb(h_1, h_2, h_c) \lbb(h_3, h_4, 1-h_c)\! \[\beta(h_c, h_{34}) \mF_{1234}^{h_c} (x) + \beta(1\!-\!h_c, h_{12}) \mF_{1234}^{1-h_c}(x)\]~.
\ee
We change integration variables $h_c \rightarrow 1- h_c$ for the second term, and use,
\be
\frac{\lbb(h_3, h_4, h_c)}{\lbb(h_3, h_4, 1-h_c)} =-  \beta(h_c, h_{34}) \frac{\Gamma(h_c)}{\sqrt{\pi}\,\Gamma(h_c- \frac{1}{2})}~,
\ee
in order to write this as,
\be  \label{mWs}
\mW_s = \int \frac{ d h_c}{2\pi i}\, \tilde{\rho}_w(h, h_c)  \lbb(h_1, h_2, h_c) \lbb(h_3, h_4, h_c) \mF_{1234}^{h_c} (x) ~,
\ee
where, 
\be
\tilde{\rho}_w(h, h_c) = - 2\sqrt{\pi}\frac{\Gamma(h_c - \frac{1}{2})}{\Gamma(h_c)} \rho_w(h, h_c)~.
\ee
If we wish,  we can, for $0<x<1$, close the contour to the right, writing the result as the expected sum of single-trace and double-trace conformal blocks. For simplicity, we assume none of the $h_i$ are equal and that $h_1 + h_2 >1/2$ and $h_3 + h_4>1/2$. Then, 
\be 
\mW_s = d(h_i,h) \mF_{1234}^h + \sum_{n=0}^{\infty} e_n(h_1, h_2, h_3, h_4, h) \mF_{1234}^{h_1+h_2 + 2n} + \sum_{n=0}^{\infty }e_n(h_3, h_4, h_1, h_2, h)\, \mF_{1234}^{h_3 + h_4 + 2 n}
\ee
where the coefficient of the single-trace block is,
\be
d(h_i,h) =2\sqrt{\pi} \frac{\Gamma(h+\frac{1}{2})}{\Gamma(h)}\lbb(h_1, h_2, h)\lbb(h_3, h_4, h)~,
\ee
while the coefficients of the double-trace blocks are, 
\be \nonumber
e_n(h_1, h_2, h_3, h_4, h)  = - \tilde{\rho}_w(h, h_1 + h_2 + 2n)\lbb(h_3, h_4, h_1+h_2+2n)\text{ Res } \lbb(h_1, h_2, h_c)\Big|_{h_c = h_1 + h_2 + 2n}\,  
\ee
where, 
\be\nonumber
\text{Res } \lbb(h_1, h_2, h_c)\Big|_{h_c = h_1 + h_2 + 2n} = - \frac{2 (-1)^n}{n!} \frac{\Gamma(h_1 + h_2 + n - \frac{1}{2})\Gamma(h_1 + n)\Gamma(h_2 + n)}{16\pi \Gamma(h_1 + \frac{1}{2})\Gamma(h_2 + \frac{1}{2}) \Gamma(h_1 + h_2 + 2n + \frac{1}{2})}~.
\ee

\subsubsection*{Contact diagram}
One can evaluate a contact Witten diagram using similar methods. 
Consider the contact diagram arising from the interaction, $
 \phi_1 \phi_2 \phi_3 \phi_4$. 
We need to evaluate,
\be\label{mFc}
\mW_c =  \int_{AdS} d X\, G_{h_1}(X,P_1) G_{h_2}(X,P_2) G_{h_3}(X,P_3) G_{h_4}(X,P_4)~.
\ee
We may trivially rewrite this as, 
\be
\mW_c = \int_{AdS} d X dY\, G_{h_1}(X,P_1) G_{h_2}(X,P_2) \delta(X-Y) G_{h_3}(Y, P_3) G_{h_4}(Y,P_4)~, 
\ee
and use the split-representation of the delta function \cite{Penedones:2010ue}, 
\be
\delta(X-Y) = \int \frac{ d h_c}{2\pi i}\, \rho_c(h_c) \int_{\partial AdS} dP_0\,\, G_{h_c}(X,P_0) G_{1 - h_c}(Y,P_0)~,
\ee
where the contour is as before, and,
\be
\rho_c(h_c) =- 2 (h_c-\frac{1}{2})^2~.
\ee
Following the same steps as in the calculation of the exchange Wittten diagram gives for the contact Witten diagram, 
\be 
\mW_c = \int \frac{ d h_c}{2\pi i}\, \tilde{\rho}_c(h_c)  \lbb(h_1, h_2, h_c) \lbb(h_3, h_4, h_c) \mF_{1234}^{h_c} (x) ~,
\ee
where, 
\be
\tilde{\rho}_c( h_c) = - 2\sqrt{\pi}\frac{\Gamma(h_c - \frac{1}{2})}{\Gamma(h_c)} \rho_c( h_c)~.
\ee
If we wish, we can, for $0<x<1$, close the contour to the right and write $\mW_c$ as a sum of double-trace conformal blocks. 

A general quartic interaction in the bulk will involve derivatives. For any specific set of derivatives it straightforward to write the contact Witten diagram as a sum of conformal blocks, as in the above case without derivatives, but it is difficult to write a general expression.

\section{An AdS$_2$ Brane in AdS$_3$} \label{KK}
The SYK model contains a tower of primary $O(N)$ invariant bilinears, $\mO_n$, with dimensions $h_n$. By the AdS/CFT dictionary, these are dual to a tower of massive fields $\phi_n$ with masses, 
\be
m_n^2 = h_n (h_n-1)~. 
\ee
For SYK at large $q$, the dimensions are, to leading order in $1/q$, given by $h_n = 2 n+1$. Therefore, for large $n$, the masses are approximates $m_n \approx 2n +1$. A natural way to approximately produce a spectrum of this type is to view it as arising from a Kaluza-Klein tower of a single scalar field in AdS$_2\times S^1$.~\footnote{We thank N.~Nekrasov for discussions on this.\\ \indent \, \,  See \cite{Das:2017pif} for work in a similar  direction.} To account for the bulk cubic couplings $\lambda_{n m k} \phi_n \phi_m \phi_k$, it is natural to introduce a cubic interaction $\phi^3$ in the AdS$_2\times S^1$ space.  One can then trivially compute the resulting cubic couplings: they are given by overlaps of the wavefunctions $e^{i m \theta}$ along the $S^1$. These couplings are, however, clearly a poor approximation to the true $\lambda_{n m k}$, given in (\ref{lambda}). For instance, these are of order-one, whereas the actual $\lambda_{n m k}$ grow exponentially as $n, m, k$ uniformly get large. 

The spectrum of the scalar in AdS$_2\times S^1$ is only approximately that of large-$q$ SYK for $n \gg 1$. In this appendix we will show that placing an AdS$_2$ brane inside of AdS$_3$, and considering a scalar in the AdS$_3$ spacetime, will exactly reproduce the large $q$ SYK spectrum. However,  the cubic couplings will still be completely off. This illustrates, perhaps unsurprisingly, that the spectrum is not by itself a strong enough clue as to the nature of the bulk theory. 

We write AdS$_3$ in coordinates, 
\be \label{AdSAdS}
ds^2 = d r^2 + \cosh^2 r\, d s_2^2~,
\ee
where $d s_2^2$ is the metric on AdS$_2$. Which coordinates one picks on the AdS$_2$ will not be relevant for us; a simple choice is global coordinates, 
\be
ds_2^2 = \frac{1}{\cos^2 \rho}\( - d t^2 + d \rho^2\)~.
\ee
The interpretation of (\ref{AdSAdS}) is that at each $r$ there is an AdS$_2$ space with radius $\cosh r$. The range of $r$ is from $-\infty$ to $\infty$. 

We place a brane  at some constant $r$, which without loss of generality, we take to be at $r=0$.  The tension of the brane is tuned so that it is static; for a general discussion, see \cite{Karch:2000ct}.  The wave equation for a scalar in AdS$_3$, $\( \Box - m^2\) \phi = 0$,  in terms of coordinates (\ref{AdSAdS}) is, 
\be
\frac{1}{\cosh^2 r}\partial_r \( \cosh^2 r\, \partial_r \phi\) + \frac{1}{\cosh^2 r} \Box_2 \phi = m^2 \phi~,
\ee
where $\Box_2$ is the AdS$_2$ Laplacian. 
Letting the solution be  of the form,
\be
\phi(r, \rho, t) = f(r)\, \psi(\rho, t)~, \ \ \ \ \ f(r) = \frac{u(r)}{\cosh r}~,
\ee
and letting $m_2^2$ denote the eigenvalue of the AdS$_2$ Laplacian, $\Box_2 \, \psi(\rho, t) = m_2^2\,  \psi(\rho, t)$,
we  get that the radial wavefunction satisfies, 
\be
- \frac{1}{2} u''(r)  - \frac{m_2^2}{2 \cosh^2 r} u(r) = - \frac{(m^2+1)}{2} u(r)~.
\ee
This is of the form of a Schr\"odinger equation for a particle of energy $- (m^2+1)/2$ in a potential $- m_2^2/(2 \cosh^2 r)$. Note that the mass $m$ is fixed: this is the mass of the scalar in AdS$_3$, which we choose at the beginning. On the other hand, $m_2$ is arbitrary and will only be constrained by quantization requirements. In particular, in order for $- (m^2+1)/2$ to be an eigenenergy of the potential, the values $m_2^2$ can not be arbitrary. This is a bit different from the scenario in which one compactifies along a compact manifold. 

In fact, this potential is the P\"oschl-Teller potential. Letting, 
\be
m_2^2 = n(n+1)~, \ \ \ \ \ \mu^2 = m^2 + 1~, 
\ee
where $n$ is a positive integer,  the eigenenergies are $\mu = 1, 2, \ldots, n$. The eigenfunctions are the associated Legendre functions, $
u(r)= P_{\lambda}^{\mu} \(\tanh(r)\)$. 

We will  choose the AdS$_3$ scalar to be massless, $m=0$. Then, from the point of view of the AdS$_2$ brane, a massless particle in AdS$_3$ looks like a tower of particles with masses $m_2^2 = n(n+1)$. This reproduces the large $q$ SYK spectrum, up to the fact that we should keep only odd $n$. This can be achieved through an appropriate choice of boundary conditions, as the Legendre polynomials $P_{n}^{1}\(\tanh(r)\)$
are odd under $r\rightarrow - r$ for even $n$, and even for odd $n$.

Let us now compute the cubic couplings. In the AdS$_3$ space we take the action, 
\be
S= \int dt d\rho d r \sqrt{-g} \(\phi \,\Box\, \phi + \lambda \phi^3\)
\ee
and insert
\be
\phi = \sum_{n=1}^{\infty} f_n(r)\, \psi_n(\rho, t)~, \ \ \ \ \ \ \ \ f_n(r) = \frac{1}{\cosh r} P_{n}^{1} \(\tanh(r)\)~.
\ee
We rewrite the action as, 
\begin{multline}
S = \int d\rho d t \sqrt{-g_2} \sum_{n, m} \(\int f_n(r) f_m(r)\)\, \psi_n(\rho, t) \Big(\, \Box_2 - n(n+1)\Big)\psi_m(\rho,t) \\+ \int d\rho d t \sqrt{-g_2} \sum_{n, m, k } \(\int dr \cosh^2 r f_n(r) f_m(r) f_k(r)\) \psi_n(\rho, t)\psi_m(\rho, t) \psi_k(\rho,t)~,
\end{multline}
where $g_2$ denotes the determinant of the AdS$_2$ metric. 
Rescaling the fields, 
\be
\tilde{\psi}_n =N_n\, \psi_n ~, \ \ \ \ \ N_n^2 \delta_{n m}= \int dr\, f_n(r) f_m(r)\, , \ \ \ N_n^2 = \frac{2n(1+n)}{1+2n}~,
\ee
we have,
\be
S = \int d\rho dt\sqrt{-g_2}\[ \sum_{n, m} \tilde{\psi}_n\Big(\, \Box_2 - n(n+1)\Big) \tilde{\psi}_n + \lambda_{n m k} \tilde{\psi}_n\tilde{ \psi }_m \tilde{\psi}_k\]
\ee
where, 
\be
\lambda_{n m k } = \frac{\lambda}{N_n N_m N_k}\int dr\, \cosh^2 r\, f_n(r) f_m(r) f_k(r)~.
\ee
Evaluating $\lambda_{n n n}$, the growth with $n$ is slow, and inconsistent with the required couplings of the bulk dual of SYK, (\ref{lambda}).

\vspace{.5cm}

\bibliographystyle{utphys}

\end{document}